\documentclass{article}
\usepackage[UTF8, scheme=plain, punct=plain, zihao=false]{ctex}
\usepackage{natbib} 
\usepackage{booktabs}
\usepackage{graphicx}
\usepackage {subcaption} 
\usepackage{color}
\usepackage{enumitem}
\usepackage{tcolorbox}
\usepackage{ulem}    

\tcbuselibrary{skins, breakable, theorems}
\usepackage{multirow}
\usepackage{makecell}
\usepackage[figuresright]{rotating}
\usepackage{stackengine}
\newcommand\xrowht[2][0]{\addstackgap[.5\dimexpr#2\relax]{\vphantom{#1}}}
\usepackage[title,titletoc]{appendix}

\usepackage{hyperref}
\hypersetup{colorlinks=true,linkcolor=blue,filecolor=red,urlcolor=blue,}

\title{Pulse in collapse: a game dynamics experiment}
\author{Yijia Wang  and Zhijian Wang  \\
Experimental social science laboratory, Zhejiang University, China }
\date{ }
\graphicspath{{figure/}} 

\begin{document}

\maketitle

\section*{Abstract}

The collapse process is a constitutive sub-process in the full finding Nash equilibrium process. We conducted laboratory game experiments with human subjects  to study this process.  We observed significant pulse signals in the collapse process. The observations from the data support the completeness and consistency of the game dynamics paradigm.

\noindent{\textbf{keyword~~}}
experimental economics; evolutionary game theory; game dynamic theory; iterated eliminating dominated strategy

\tableofcontents

\section{Introduction}

Game equilibrium finding is a process \cite{kandori2021,samuelson2016,dan1991,levine2016}.  
This process is also referred to as evolution, 
or adjustment process, or deviation from  equilibrium process, or the convergence process in game theory and experiment. It is a non-linear spontaneous evolution process driven by game participants' strategic interactions.

In the study of game theory, terms such as  adjustment dynamics \cite{kandori2021}, learning theory \cite{fudenberg1998},  population game dynamics, evolutionary game theory \cite{dan2016,sandholm2010} or market dynamics \cite{plott2008}, are related to this process. These terms are included in the game dynamics paradigm.  

As a scientific paradigm that moves beyond equilibrium theory (classical game theory, or game statics theory),  
the game dynamics paradigm is expected to be able to describe the regularity of the game dynamic process in terms of completeness and consistency, as well as reality and accuracy
\cite{samuelson2016,dan1991,levine2016,Newton2018}.  

Instead of using the statics theory to study the predictable equilibrium (stable statistical relationships 
among strategy), game dynamics theory studies the
predictable motion (temporary deviations from equilibrium). 
For example, in the rock-paper-scissors game \cite{zhijian2014rps,dan2014,zhijian2013,zhijian2011}, 
the persistent cycle is a typical motion pattern of \textit{predictable temporary deviation from equilibrium} (PTDE).
Such a pattern can be exploited \cite{mit2014}. In real life, most high-frequency trading strategies which are not fraudulent, but instead exploit the PTDE \cite{wiki:High-frequency_trading}.

Following this paradigm, we investigate 
an unignorable sub-process of 
the full finding Nash equilibrium process.  
In this section, we introduce the sub-process, namely collapse process; then, we introduce the research questions and offer a summary of the contents of this paper.  

\subsection{Collapse}
\begin{figure*}[ht!]
    \centering
    \begin{subfigure}[b]{0.4\textwidth}
    \includegraphics[scale=0.28]{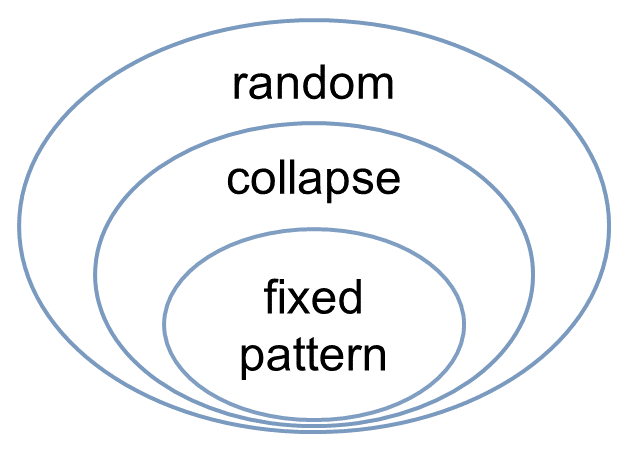}
    \caption{}~\\
    \centering
    \label{fig:main3periods}
    \end{subfigure}
    \begin{subfigure}[b]{0.4\textwidth}
    \includegraphics[scale=0.2]{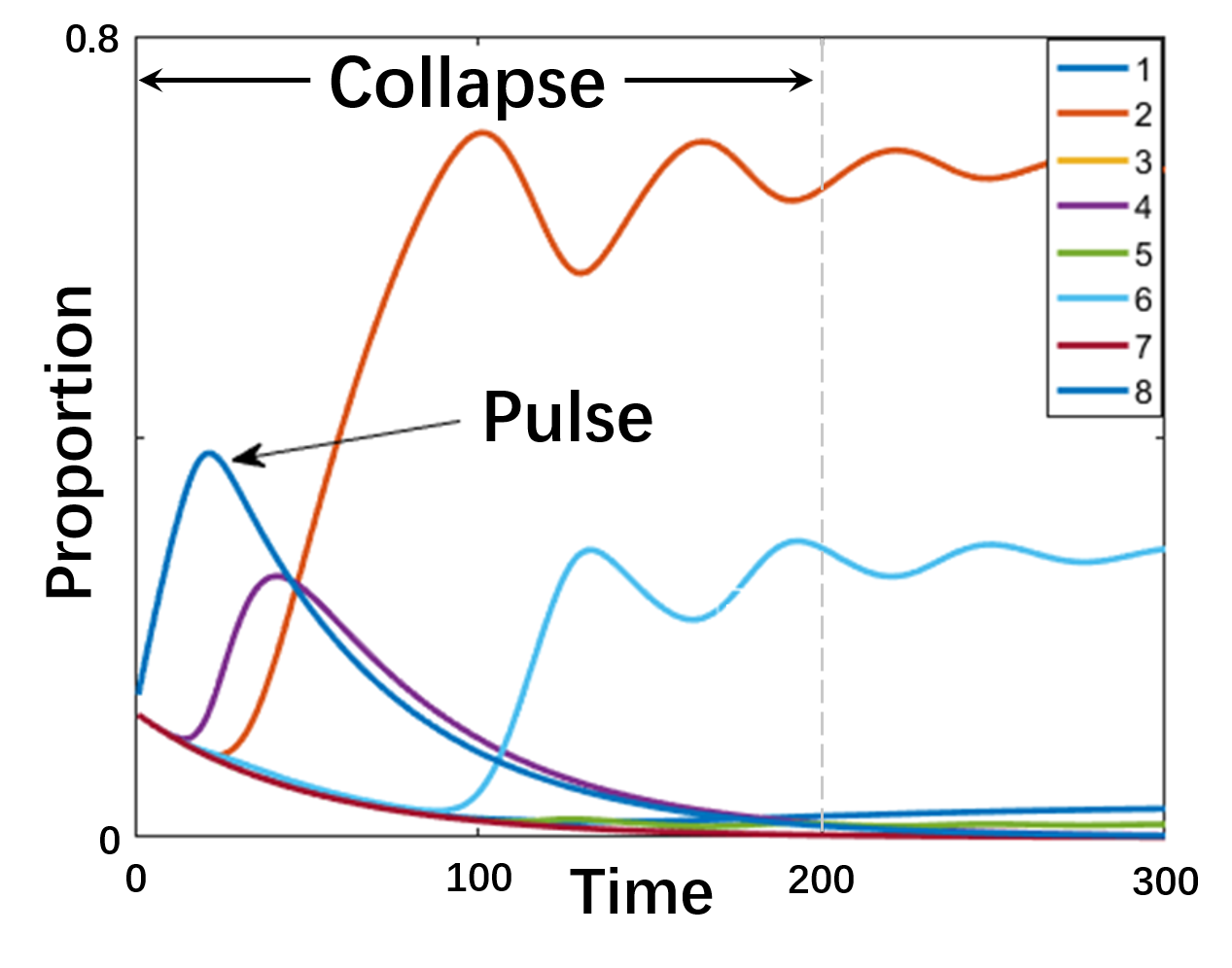}
    \caption{}~\\
    \centering
    \label{fig:main_pulse}
    \end{subfigure}
    \begin{subfigure}[b]{0.7\textwidth}
    \includegraphics[scale=0.55]{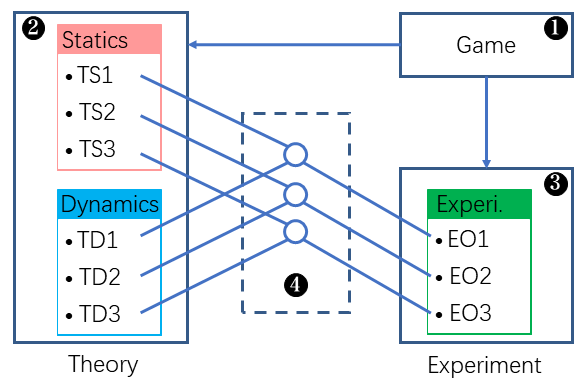}
    \caption{}
    \centering
    \label{fig:workflow}
    \end{subfigure}
    \caption{Conceptual figure. (a) The three periods in the full  Nash-equilibrium-finding process, moving from random to collapse to fixed pattern. This study focuses on the collapse process. (b) An illustration of a pulse in a collapse.(c) The workflow of this report: 
    The Game, \textbf{T}heory (\textbf{S}tatic and \textbf{D}ynamic) vs. \textbf{E}xperiment by \textbf{O}bservations (1-distribution; 2-cycle; 3-collapse). For the abbreviations, see Table \ref{tab:app_abbr}.}
    \label{fig:ConceptFigure4level}
\end{figure*}

A real game system can involve a large number of strategies, most of which would be dominated \textbf{during} the process of finding Nash equilibrium  \cite{binmore2007game,stratego2022}. 
As illustrated in Figure \ref{fig:main3periods},  the full equilibrium finding process can be classified into three stages: 
\begin{enumerate}
\item 
In the initial stages, having no information about the game, agents randomly choose a strategy vector from the strategy space for optimal payoff. During this period, the game's social evolution trajectory is highly stochastic, smoothly distributed throughout the full game space.  
\item 
 Later, in the period of infancy, each agent may start to identify their own nonprofit strategy by learning from strategy interactions. As a result, the dimensions of the game strategy space may collapse. In theory,`a dominated strategy can dominate a domination before being dominated', meaning that a phenomenon known as a pulse could exist. A conceptual rendering of a pulse in the collapse process is shown in Figure \ref{fig:main_pulse}. 
\item 
Finally, when there is no dominant strategy, the evolution trajectory will converge to a fixed pattern (a persistent cycle or fixed equilibrium point) of the game. 
When rendered as an image, the pattern is the same as condensed structures in astrophysics.
\end{enumerate}
The collapse appears in the second stage. 
The term `collapse', introduced to game dynamics here, is borrowed from the concept of `gravitational collapse' in astrophysics  \cite{wiki:Gravitational_collapse} 
which is a fundamental mechanism and 
a legitimate constituent part of the formation of structures in the universe. This study benefits from the physical pictures in astrophysics. 

\subsection{Research questions}

We investigate the collapse process 
by asking the following two questions:
\begin{itemize}
\item 
In controlled laboratory game experiments with human subjects,  
can significant new observations be obtained during a collapse process?
\item 
Can the existing game theory paradigm consistently and completely capture the collapse process?  
\end{itemize}

Answering these questions is not trivial, for the following reasons: 
\begin{itemize} 
\item 
From a scientific perspective, without evidence from the collapse process, our knowledge of the full equilibrium-finding process is obviously incomplete.
\item 
Within a scientific paradigm, without conducting experiments on collapses, there is no basis for the reality of the theory
\cite{Hofbauer1996Evolutionary,
Hofbauer2011Survival,
mertikopoulos2016learning}.

\end{itemize}
These issues relate to reality and accuracy, as well as to the completeness and consistency of the game dynamics paradigm, and are therefore not trivial. 

\subsection{Outline}
 
To answer our research questions, following  the experimental game theory paradigm \cite{plott2008,colin2003,dan2014,zhijian2013,stephenson2019,hans2021}, we carried out the workflow shown in Figure \ref{fig:workflow}:
\begin{itemize}
\item [\textbf{1}] 
We designed a parameterised three card poker game 
for  use in a laboratory study of  
the statics theory and the dynamics theory (see section \ref{sec:mt_game_design}).    

\item [\textbf{2}]
We derived the predictions for three variables (distribution, cycle, and pulse) by the  statics theory and the dynamics theory (see sections \ref{sec:mt_classical_predict} and \ref{sec:mt_dynamics_predict}).   
\item [\textbf{3}]
We conducted laboratory experiments which include 3-treatments with 72 human subjects, fixed-paired, and 1000 repeated games (see section \ref{sec:mt_Game_experiment}).   

\item  [\textbf{4}]
We outlined the experimental observations and their theoretical verification of the three variables (distribution, cycle, and pulse in sections \ref{sec:mt_exp_dis}, \ref{sec:mt_exp_eic} and \ref{sec:mt_exp_acc}, respectively). Additionally, we report the results of comparing the models (see section \ref{sec:mt_model_compare}).    
\end{itemize}
In the discussion section, we offer a summary, discuss the related literature, address the implications and applications, and introduce questions for future research.

\section{Game and the prediction}
This section explains the design of the game and the experiment, as well as the predictions made using the statics theory and the dynamics theory. A summary is shown in Table  
\ref{tab:method_sum}.  

\subsection{Game design}\label{sec:mt_game_design}
The original game is a simplified 3-card poker game. The game is introduced by Binmore, by replacing Von Neumann's numerical cards by a deck with only the 1, 2, 3 (page 93 in \cite{binmore2007game}). This game is a dynamic game with incomplete information. We generalised  this game to study the dynamics process. 

\begin{figure*}
    \centering
    \begin{subfigure}[b]{1\textwidth}
    \caption{The $(m,n)$-game}~\\
    \centering
    \includegraphics[scale=0.45]{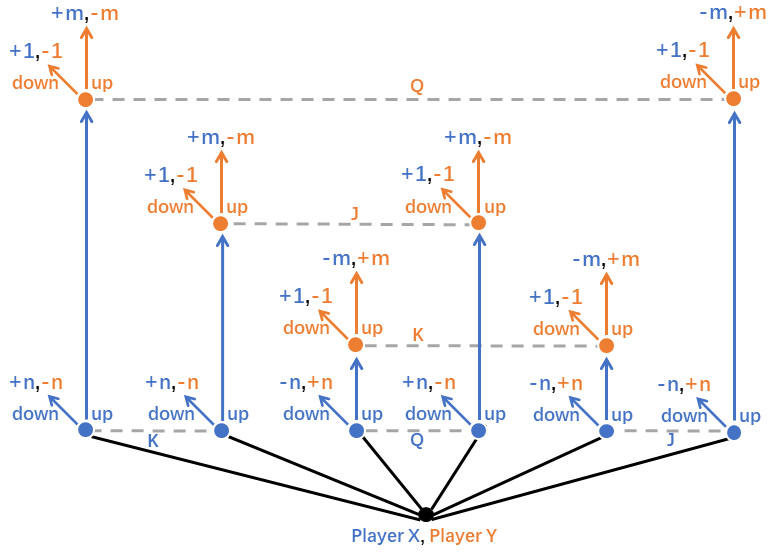}
    \end{subfigure}
   \footnotesize
 \begin{subfigure}[b]{1\textwidth}
    \caption{The payoff matrix of the (2,1)-game: Treatment A}\label{tab:matrix(2,1)} ~\\
\begin{tabular}{|c|c|c|c|c|c|c|c|c|}
  \hline
$(\textbf{m,n})\!=\!(2,1)$	&	$Y_1$ &	$Y_2$ &	$Y_3$ &	$Y_4$ &	$Y_5$ &	$Y_6$ &	$Y_7$ &	$Y_8$ \\
  \hline
$X_1$	&	0,~~0&0,~~0&0,~~0&0,~~0&0,~~0&0,~~0&0,~~0&0,~~0\\						$X_2$	&	0,~~0&0,~~0&1,~~-1&1,~~-1&1,~~-1&1,~~-1&2,~~-2&2,~~-2\\	 
$X_3$	&	2,~~-2&-1,~~1&2,~~-2&-1,~~1&3,~~-3&0,~~0&3,~~-3&0,~~0\\	 
$X_4$	&	2,~~-2&-1,~~1&3,~~-3&0,~~0&4,~~-4&1,~~-1&5,~~-5&2,~~-2\\ 
$X_5$	&	4,~~-4&1,~~-1&1,~~-1&-2,~~2&4,~~-4&1,~~-1&1,~~-1&-2,~~2\\ 
$X_6$	&	4,~~-4&1,~~-1&2,~~-2&-1,~~1&5,~~-5&2,~~-2&3,~~-3&0,~~0\\ 
$X_7$	&	6,~~-6&0,~~0&3,~~-3&-3,~~3&7,~~-7&1,~~-1&4,~~-4&-2,~~2\\ 
$X_8$	&	6,~~-6&0,~~0&4,~~-4&-2,~~2&~~8,~~-8~~&2,~~-2&	~~6,~~-6~~&0,~~0\\	 
\hline 
\end{tabular}
\end{subfigure}
   \footnotesize
 \begin{subfigure}[b]{1\textwidth}
    \caption{The payoff matrix of the (3,2)-game: Treatment B}\label{tab:matrix(3,2)} ~\\
\begin{tabular}{|c|c|c|c|c|c|c|c|c|}
  \hline
$(\textbf{m,n})\!=\!(3,2)$	&	$Y_1$ &	$Y_2$ &	$Y_3$ &	$Y_4$ &	$Y_5$ &	$Y_6$ &	$Y_7$ &	$Y_8$ \\
\hline
 $X_1$&	0,~~0&	0,~~0&	0,~~0&	0,~~0&	0,~~0&	0,~~0&	0,~~0&	0,~~0\\
$X_2$&	-2,~~2&	-2,~~2&	0,~~0&	0,~~0&	0,~~0&	0,~~0&	2,~~-2&	2,~~-2\\
$X_3$&	2,~~-2&	-2,~~2&	2,~~-2&	-2,~~2&	4,~~-4&	0,~~0&	4,~~-4&	0,~~0\\
$X_4$&	0,~~0&	-4,~~4&	2,~~-2&	-2,~~2&	4,~~-4&	0,~~0&	6,~~-6&	2,~~-2\\
$X_5$&	6,~~-6&	2,~~-2&	2,~~-2&	-2,~~2&	6,~~-6&	2,~~-2&	2,~~-2&	-2,~~2\\
$X_6$&	4,~~-4&	0,~~0&	2,~~-2&	-2,~~2&	6,~~-6&	2,~~-2&	4,~~-4&	0,~~0\\
$X_7$&	8,~~-8&	0,~~0&	4,~~-4&	-4,~~4&	10,~~-10&	2,~~-2&	6,~~-6&	-2,~~2\\
$X_8$&	6,~~-6&	-2,~~2&	4,~~-4&	-4,~~4&	10,~~-10&	2,~~-2&	~~8,~~-8~~&	0,~~0\\
 \hline
\end{tabular}
\end{subfigure}
   \footnotesize
 \begin{subfigure}[b]{1\textwidth}
    \caption{The payoff matrix of the (4,2)-game: Treatment C}\label{tab:matrix(4,2)} ~\\
\begin{tabular}{|c|c|c|c|c|c|c|c|c|}
  \hline
$(\textbf{m,n})\!=\!(4,2)$	&	$Y_1$ &	$Y_2$ &	$Y_3$ &	$Y_4$ &	$Y_5$ &	$Y_6$ &	$Y_7$ &	$Y_8$ \\
\hline
$X_1$	&	0,~~0&*0,~~0&0,~~0&*0,~~0&0,~~0&0,~~0&0,~~0&0,~~0\\							
$X_2$	&	-2,~~2&*-2,~~2&1,~~-1&1,~~-1&1,~~-1&1,~~-1&4,~~-4&4,~~-4\\							
$X_3$	&	2,~~-2&-3,~~3&2,~~-2&-3,~~3&5,~~-5&0,~~0&5,~~-5&0,~~0\\							
$X_4$	&	0,~~0&-5,~~5&3,~~-3&-2,~~2&6,~~-6&1,~~-1&9,~~-9&4,~~-4\\							
$X_5$	&	6,~~-6&*1,~~-1&1,~~-1&-4,~~4&6,~~-6&1,~~-1&1,~~-1&-4,~~4\\							
$X_6$&	4,~~-4&-1,~~1&2,~~-2&-3,~~3&7,~~-7&2,~~-2&5,~~-5&0,~~0\\							
$X_7$	&	8,~~-8&-2,~~2&3,~~-3&-7,~~7&11,~~-11&1,~~-1&6,~~-6&-4,~~4\\							
$X_8$	&	6,~~-6&-4,~~4&4,~~-4&-6,~~6&12,~~-12&2,~~-2&10,~-10&0,~~0\\\hline
\end{tabular}
\end{subfigure}
    \caption{The parameterized (m,n) game. (a) The extensive form \cite{binmore2007game}. (b,c,d) The normal form of the game (the matrix elements are multiplied by 6).} 
\label{fig:gametree_mn}
\end{figure*}

The extensive form of the generalised $(m,n)$ game is shown in Figure \ref{fig:gametree_mn}. 
When ($m,n$) = (2,1), the game is exactly the same as the original game (page 93 \cite{binmore2007game}).  
Following \cite{binmore2007game}, the normal form could be represented as the $8 \times 8$ bi-matrix shown in Figure \ref{tab:matrix(2,1)}. 
That is, each player (X or Y)  has eight strategies. 
This study mainly makes use of the bi-matrix form. 
The   parameters are assigned as $(m,n)$ = (2, 1), (3, 2) and (4, 2), which are referred to  as treatment A, B, and C, respectively.  
Details on the game design and the motivations are provided in SI section 1 (Game).   

\subsection{Classical game theory predictions}\label{sec:mt_classical_predict} 
 
\textbf{Method:} The concepts of static equilibrium and the iterated eliminating dominated strategy (IEDS) are useful when describing the full Nash-equilibrium-finding process, according to the classical game theory paradigm. For details of the method, see SI section 3 (Statics).
\begin{itemize}
\item For the three treatments, to solve the Nash equilibrium of the three bimatrix zero sum two person game, we can use quadratic programming method \cite{mangasarian1964two}. 
\item 
For the three treatments, the order of the eliminating dominated strategies, as well as the surviving  strategies, can be obtained by IEDS method.  
\end{itemize}

\textbf{Predictions:} The predictions and their verification methods are as follows:
\begin{itemize}
\item 
[TS1] Distribution. The equilibrium distribution results are shown in figure 2a. Following the methods of conventional experimental research  \cite{colin2003,selten2008,dan2014}, the predictions can be verified using the data, denoted as $\rho^E$ and shown in Table \ref{tab:dis_main}. 
\item 
[TS2] Cycle. 
IEDS provides the surviving strategy for a game, which is shown in Figure \ref{fig:survive_strategy} and in the row titled ‘Surviving Strategy’ in Table  \ref{tab:method_sum}. 
According to best-response analysis, the cycle's existence and relative strength in each treatment (A, B, and C) can be predicted as (Yes, No, Yes), referring to Figure \ref{fig:survive_strategy}.  These points can be verified  using the experimental cycle ($\vartheta^E$) in section \ref{sec:mt_exp_eic}.  
\item 
[TS3] Collapse. IEDS provides the eliminating order of a game. The results are shown in the row ' IEDS round' in Table \ref{tab:method_sum}. Mathematical analysis of the best response analysis shows that, among the dominated strategies, the strategy that is eliminated last could provide the pulse signal during the collapse. These points can be verified by the pulse ($\psi^E$) and the crossover points ($\chi^E$) using the experimental observations in section \ref{sec:mt_exp_acc}. 
\end{itemize} 
In addition, these three predictions can be applied to compare with those from dynamics theory. The results will be reported in section \ref{sec:mt_model_compare}.  
 \begin{figure*}[ht!]
    \centering
    \begin{subfigure}[b]{0.3\textwidth}
    \caption{}
    \centering
    \label{tab:(2,1)matrix_step3}
    \begin{tabular}{|c|c|c|c|c|c|c|c|c|}
  \hline
\multirow{2}*
{\makecell{A}}
&	 	\multirow{2}*{$Y_2$} &		\multirow{2}*{$Y_4$}  \\
&&\\
  \hline			
$X_2$	&	0,~~0&1,~~-1\\	
$X_6$	&	1,~~-1&-1,~~1\\
\hline
\end{tabular}
\end{subfigure}
\begin{subfigure}[b]{0.3\textwidth}
   \caption{}
    \centering
    \label{tab:(3,2)matrix_step5}
  \begin{tabular}{|c|c|}
  \hline
\multirow{2}*
{\makecell{B}}
&	\multirow{2}*{$Y_4$}   \\
&\\
  \hline
 $X_1$&	0,~~0\\
  \hline
 \end{tabular}
\end{subfigure}
\begin{subfigure}[b]{0.3\textwidth}
    \caption{}
    \centering
    \label{tab:(4,2)matrix_step3}
    \begin{tabular}{|c|c|c|}
  \hline
\multirow{2}*{\makecell{C}}	&	\multirow{2}*{$Y_2$} &		\multirow{2}*{$Y_4$ }\\
&&\\
\hline
$X_1$	&	0,~~0&0,~~0\\		
$X_2$	&	-2,~~2&1,~~-1\\		
$X_5$	&	1,~~-1&-4,~~4\\		
\hline
\end{tabular}
\end{subfigure}
    \centering
    \caption{The surviving strategy of the (A, B, C) game. 
    The expected fixed patterns of the (A, B, C) game are a
    fixed orbital cycle, pure Nash equilibrium (a fixed point or signal star), and a mixed pattern, respectively.}
    \label{fig:survive_strategy}
\end{figure*}
\begin{table}[ht!]
\small
\begin{center}
\caption{Equilibrium theory predictions and experiment protocol.}
\label{tab:method_sum}
    \begin{tabular}{|c|c|cc|cc|cc|}
      \hline\hline
   \multirow{3}*{ID} & Treatment ID     &	\multicolumn{2}{|c|}{A} 	  &   \multicolumn{2}{|c|}{B}  &  	\multicolumn{2}{|c|}{C}\\
        \cline{2-8}	
  ~& Game-$(m,n)$    &	\multicolumn{2}{|c|}{(2,1)}	  &   \multicolumn{2}{|c|}{(3,2)}  &  	\multicolumn{2}{|c|}{(4,2) }\\
        \cline{2-8}	
   ~&Player ID     & X & Y  &	     X & Y  &	          X & Y  \\
        \hline		
\multirow{6}*{\thead{Equili-\\brium \\Theory}}&  IEDS Round 1& 1,3,5,7 & $\alpha^*$	          & 3 & $\alpha$ &	            3& $\alpha $ \\
~&  IEDS Round 2& 4,8 & -&	            2,4,6,7,8 & -&	           4,6,7,8 & -\\
~&  IEDS Round 3&&&  - & 2&	       &   \\
~&  IEDS Round 4&&&  5 & -&	       &   \\
       \cline{2-8}	
~&Survive &\multirow{2}*{2,6}&\multirow{2}*{2,4} 	 &\multirow{2}*{1}&\multirow{2}*{4}	         &\multirow{2}*{1,2,5}&\multirow{2}*{2,4}\\
~&strategy && 	 &&	         &&\\	
        \hline		
\multirow{4}*{\thead{Experi-\\ment \\Protocol}}&Session number &	\multicolumn{2}{|c|}{12 }&	\multicolumn{2}{|c|}{12}&	\multicolumn{2}{|c|}{12}\\
  \cline{2-8}
~&Player in session &	\multicolumn{2}{|c|}{2}&	\multicolumn{2}{|c|}{2}&	\multicolumn{2}{|c|}{2}\\
  \cline{2-8}
~&Repeated round  &	\multicolumn{2}{|c|}{1000}&	\multicolumn{2}{|c|}{1000}&	\multicolumn{2}{|c|}{1000}\\ 
 \cline{2-8}
~&Matching   &	\multicolumn{2}{|c|}{~~Fix paired~~}&	\multicolumn{2}{|c|}{~~Fix paired~~}&	\multicolumn{2}{|c|}{~~Fix paired~~}\\
  \hline 
    \end{tabular} \\
    $^* \alpha$=1,3,5,6,7,8 means the strategies are eliminated simultaneously.  
\end{center}      
\end{table}

\subsection{Game dynamics theory prediction}\label{sec:mt_dynamics_predict}  
The game dynamics paradigm provides 
a set of concepts and methods on game evolution process \cite{dan2016,dan1991,huyck1999,nowak2015,zhijian2014rps}; 
This study adheres to this paradigm.

\textbf{Method:}  We utilize the logit dynamics equations system \cite{sandholm2010}; At the same time, referring to  \cite{dan2014,zhijian2013,nowak2015,dan2016}, 
this study makes use of the noise parameter
($\lambda$ = 50) and the adjustment time step 
($\triangle t$ = 0.02). 
Then,  with random initial conditions, we can generate the time series of 1000 rounds data as long as the 1000 round  experiment conducted in this research. 
Then, using the generated time series, we can calculate the predictions of the observations (distribution, cycle and collapse) simultaneously. 

It is worth noting that, based on the experiences of previous studies \cite{dan2014,zhijian2013}, we take the two parameters $(\lambda,\triangle t)$ simply by scanning two-dimensional parametric space, aiming for  $\rho^T_{\textbf{Dyn}(\lambda, \triangle t)}$ to be approximately close to $\rho^T_{\textbf{QP}}$. We do not fix the two parameters by the  experiment data.

\textbf{Predictions:} Following the methods, the results are predictions that can be experimentally verified. These results are as follows:  
\begin{itemize}
\item  [TD1] Distribution. 
The distribution of the treatments is shown in Figure \ref{fig:dyn_main} 
and Table \ref{tab:dis_main} as $\rho^T_{\text{Dyn}}$, which is verifiable by the experiment.  
\item [TD2] Cycle. 
The eigencycle spectra  of the treatments are shown in Figure  \ref{fig:eicycle_dyn_main} as $\vartheta^T$. 
For further details about eigencycle spectra see Appendix \ref{sec:app_cycle}. 
The refined predictions are that, for treatments (A,B,C), there exist (strong, no, weak) cycles, respectively.  
\item [TD3] Collapse. 
The theoretical accumulated curves ($\varrho^T$) are shown in Figure \ref{fig:dyn-X6}-\ref{fig:dyn-Y10}. The numerical results of the crossover points are shown in Table \ref{tab:dyn_cross}. The statistically significant theoretical pulse signals are shown in Table \ref{tab:dyn_Pulse_significant}. 
These predictions can be applied as the criterion to evaluate the consistency between the theory and the experiment.   

The crossover points ($\chi_-$) with crossover time $\tau >100$ are labelled with red arrows.  According to the  numerical results 
 shown in Table \ref{tab:dyn_cross}, there are (4,0,2) $\chi_-$  red arrows in treatments (A,B,C). 
\end{itemize}

\subsection{Game experiment}\label{sec:mt_Game_experiment}

The experiment was performed at Zhejiang University 
between  August and September 2018.  
A total of 72 undergraduate students from Zhejiang University volunteered to serve as the human subjects of this experiment.  

The main parameters and protocol are shown in the 'Experiment' rows in Table \ref{tab:method_sum}. 
There were three treatments, 
and each was assigned 24 human subjects. 
The 24 subjects were divided into 12 sessions 
of 2 participants. The two players played a game with 1000 repeated 
in this fixed pairing. 
The reasoning behind the protocol design  is explained in the SI Section 2.1.

In a session, a player can observe both sides' strategy IDs ($X_1, X_2, ..., X_8$) for player $X$ or ($Y_1, Y_2, ..., Y_8$) for player $Y$. But a player can only observe her/his own realised payoffs and the strategy used by their own opponent. We do not provide the subjects with the payoff matrix, and no discussion allowed.  

The experimental sessions lasted for an average of two hours. The payment of a subject, including CNY 20 for attending the session, as well as a payment based on the player's rank, averaged CNY 120.
The rank fee was determined by the rank of a participant’s total earning score in the 1000-round game, based on a comparison of players in the same role (player X or player Y) in the same experimental session.
For details on the design, procedures and data,  see SI section 2 (Experiment).

\section{Results}

In this section, we report the results of the experiment and the theoretical verification of the observations (distribution, cycle, and pulse), as well as the results of the model comparison.

\subsection{Distribution}\label{sec:mt_exp_dis}
 
 In order to illustrate the completeness and consistency, 
we report the distribution in brief.  For the method and details relating to the data,  see  Appendix---\ref{sec:app_cycle}. 

\textbf{Measurement.} The strategy distribution and its convergence are reported. The convergence is measured according to the Euclidean distance $\delta$ from the experimental distribution $\rho^E$ in respect to  two theoretical predictions (1)  the static (Nash) equilibrium $\rho^T_{\text{S}}$, and (2)  the time series produced by the dynamics model $\rho^T_{\text{D}}$. 

\textbf{Results.}  For the experimental distribution ($\rho^E$)
and its evolution, which is determined by the distribution values at various time intervals, as shown in Table \ref{tab:dis_main}, the main results are as follows:  
\begin{enumerate}
\item[EO1.1] 
~As expected, for all (A,B,C) treatments, over time, the observed distribution ($\rho^E$) moves  closer to
the static equilibrium ($\rho^T_{\text{S}}$) and to
the dynamics prediction ($\rho^T_{\text{D}}$).  
This is because the Euclidean distances decrease along time
($\triangle \delta_{\text{S}} < 0$  and $\delta_{\text{D}} < 0$)     for all of the treatments, as shown in Table \ref{tab:nash_distance}.

\item[EO1.2] ~In all (A,B,C) treatments, the dynamics prediction $\rho^T_{\text{D}}$ performs better than the statics prediction $\rho^T_{\text{S}}$. This finding is supported by comparing the Euclidean distance during the last rounds,  $\triangle \delta < 0$,   for all of the treatments, as shown in Table \ref{tab:nash_distance}.
\end{enumerate} 
To summarize, for the distributions, the experimental observations are consistent with the theoretical predictions and with the existing literature \cite{colin2003,plott2008,dan2014,dan2010,selten2008,nowak2015}. 
In other words, the full process shown  in Figure \ref{fig:main3periods} can be observed in our 1000 rounds of data.   

\subsection{Eigencycle}\label{sec:mt_exp_eic}
In order to illustrate the completeness and consistency, 
we report the cycle in brief. 

\textbf{Measurement.} Referring to \cite{zhijian2022,zhijian2020,2021Qinmei}, 
we used the eigencycle spectrum $\vartheta^E$
to present the cycle in the experiments. We measured the eigencycle value of the 120 subspace of the game state space for each treatment, respectively; then normalise $\vartheta^E_A, \vartheta^E_B, \vartheta^E_C$ for treatments (A,B,C) to attain  the maximum $|\vartheta^E| = 1$.  For more details on the measurements and the methods used for data presentation, see section 6 (Cycle). 

\begin{figure*}
    \centering
    \begin{subfigure}[b]{1\textwidth}
 \caption{Theoretical distributions ($\rho^T$)}~\\
    \centering
 \label{fig:dyn_main}
       \includegraphics[scale=0.50]{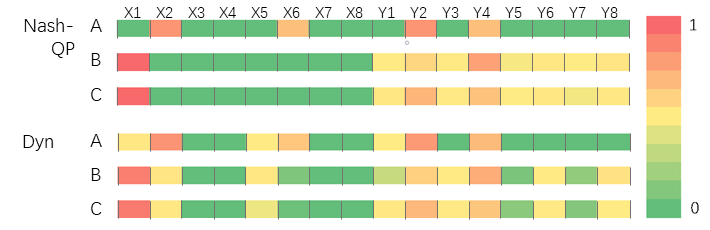}
    \end{subfigure}
    \begin{subfigure}[b]{1\textwidth}
    \centering
 \caption{Theoretical eigencycle spectrum ($\vartheta^T$)}~\\
 \label{fig:eicycle_dyn_main}
   \includegraphics[scale=0.63]{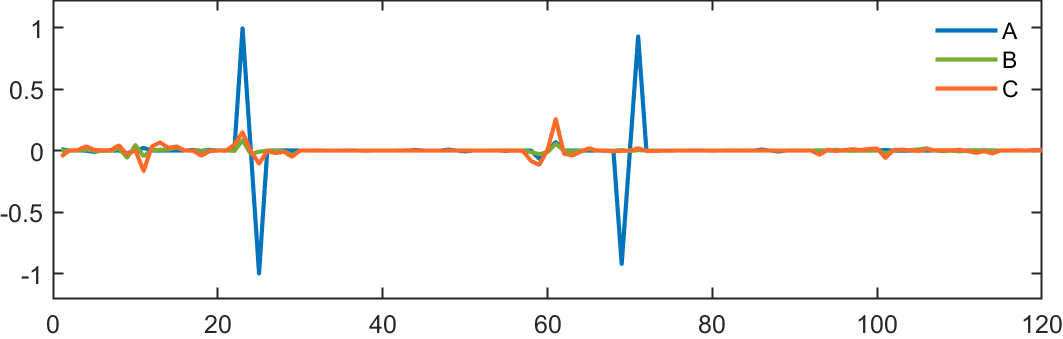}
    \end{subfigure}
    \begin{subfigure}[b]{0.3\textwidth}
    \caption{ $\varrho^T_\text{(X-A)}$}\label{fig:dyn-X6}~\\
       \includegraphics[scale=0.3]{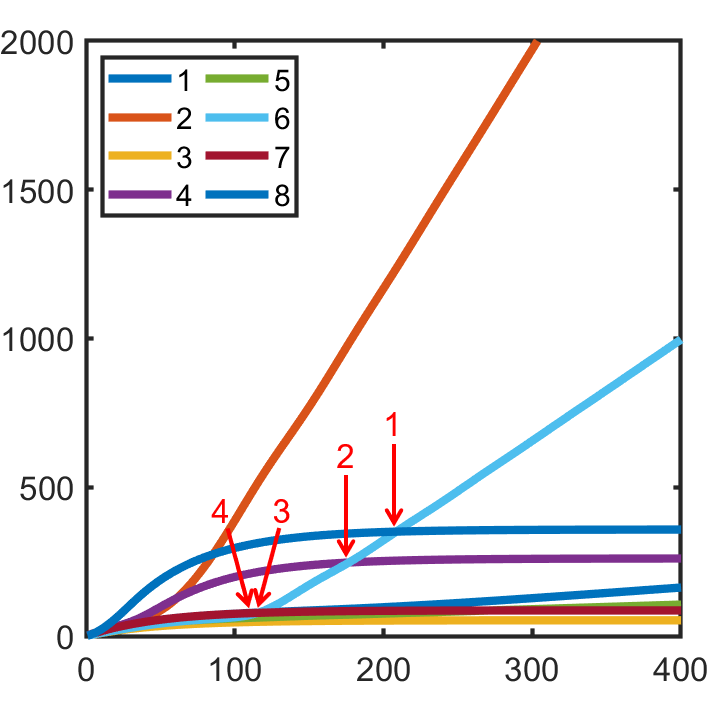}
        \end{subfigure}
    \begin{subfigure}[b]{0.3\textwidth}
    \caption{$\varrho^T_\text{(X-B)}$}\label{fig:dyn-X8}~\\
       \includegraphics[scale=0.3]{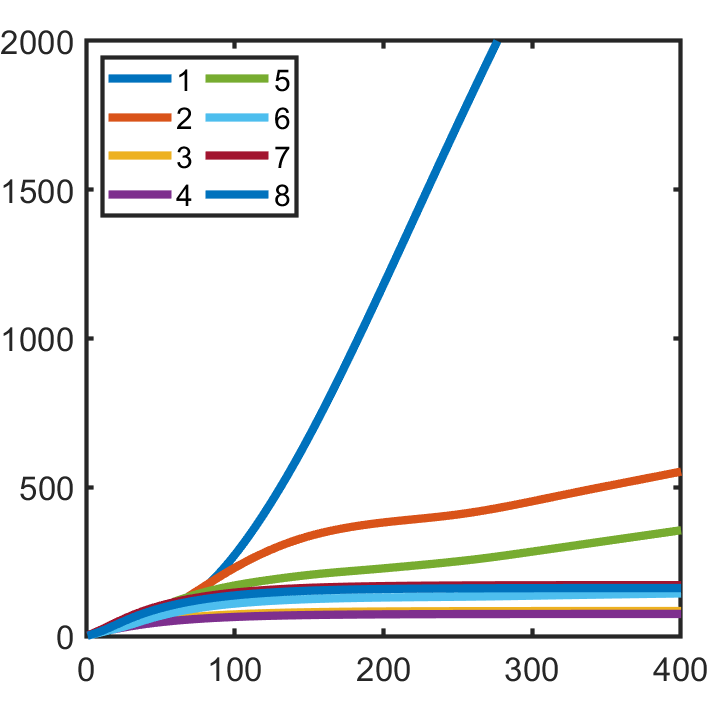}
        \end{subfigure}
    \begin{subfigure}[b]{0.3\textwidth}
         \caption{$\varrho^T_\text{(X-C)}$}\label{fig:dyn-X10}~\\
       \includegraphics[scale=0.3]{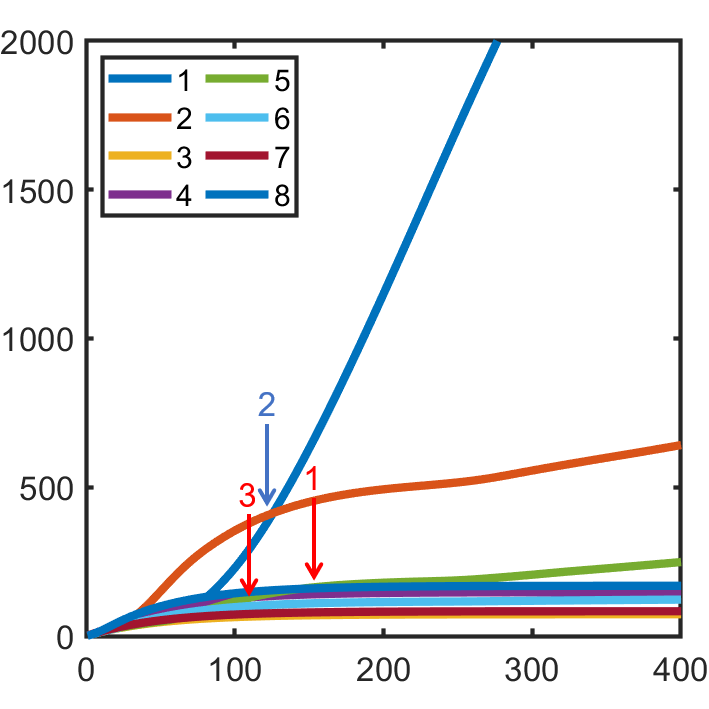}
    \end{subfigure}
     \begin{subfigure}[b]{0.3\textwidth}
    \caption{$\varrho^T_\text{(Y-A)}$}\label{fig:dyn-Y6}~\\
       \includegraphics[scale=0.3]{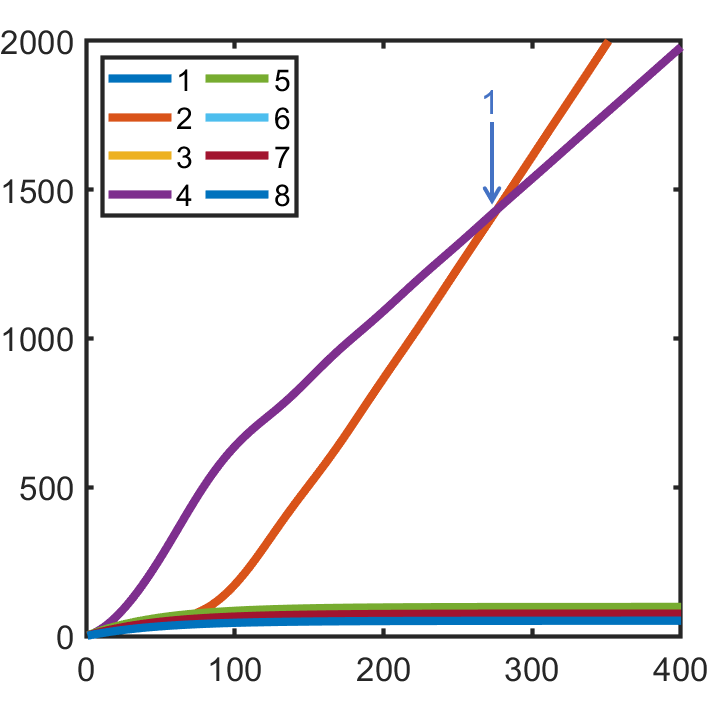}
        \end{subfigure}
    \begin{subfigure}[b]{0.3\textwidth}
    \caption{$\varrho^T_\text{(Y-B)}$}\label{fig:dyn-Y8}~\\
       \includegraphics[scale=0.3]{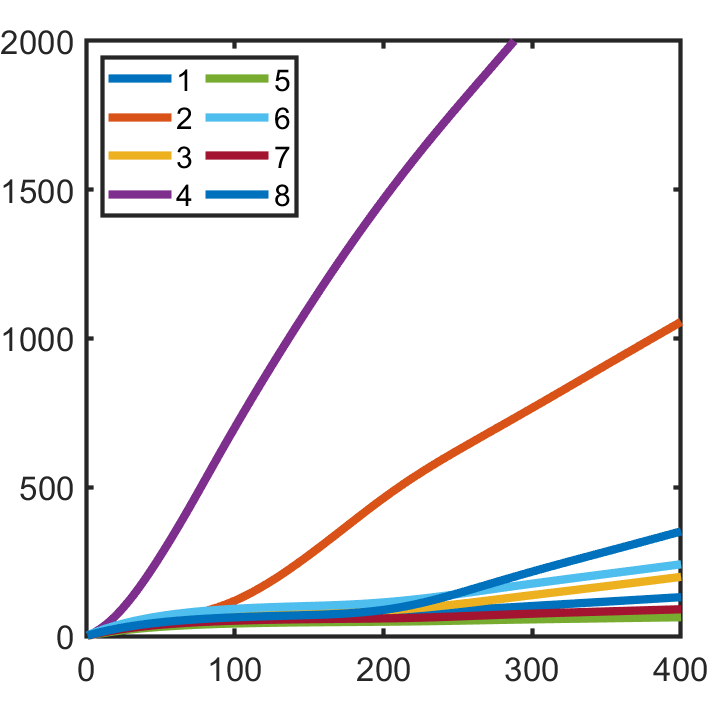}
        \end{subfigure}
    \begin{subfigure}[b]{0.3\textwidth}
         \caption{ $\varrho^T_\text{(Y-C)}$ }\label{fig:dyn-Y10}~\\
       \includegraphics[scale=0.3]{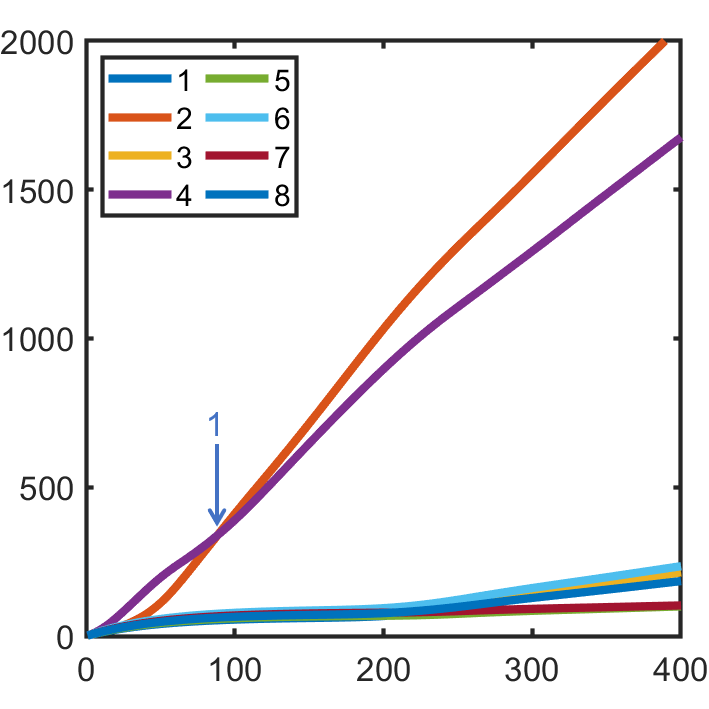}
    \end{subfigure}
    \caption{Theoretical results of  treatments A, B, and C. (a) Distributions $\rho^T$ from the two models; (b) eigencycle spectrum $\vartheta^T$; (c-h) accumulated curve ($\varrho^T$) of the player (X or Y). The crossover points ($\chi$) are labelled with arrows referring to the data in Table \ref{tab:dyn_cross}.}
    \label{fig:eig_str_}
\end{figure*}

\begin{figure*}
    \centering
    \begin{subfigure}[b]{1\textwidth}
 \caption{Experimental distributions ($\rho^E$)}~\\
 \label{fig:exp_main}
    \centering
       \includegraphics[scale=0.50]{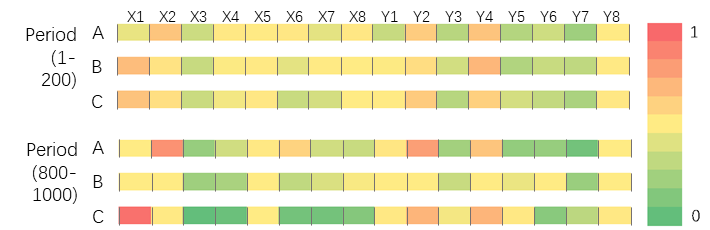}
    \end{subfigure}
    \begin{subfigure}[b]{1\textwidth}
    \centering
 \caption{Experimental eigencycle spectrum($\vartheta^E$)}~\\
 \label{fig:exp_3eigencycle}
       \includegraphics[scale=0.63]{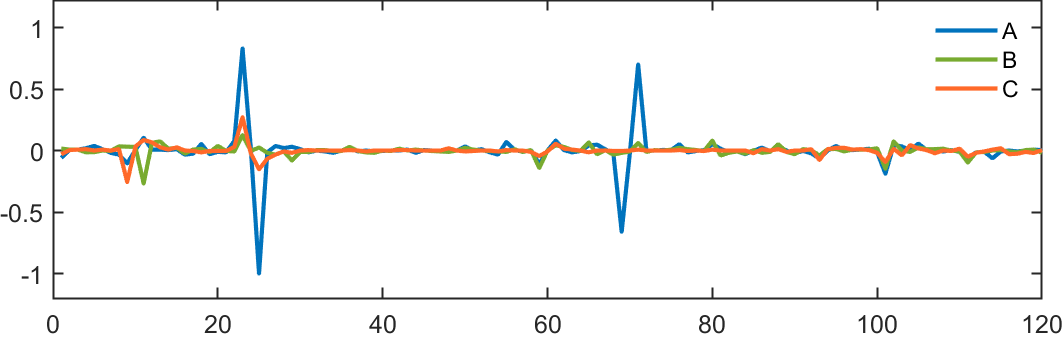}
    \end{subfigure}
    \begin{subfigure}[b]{0.3\textwidth}
    \caption{$\varrho^E_\text{(X-A)}$}\label{fig:exp-X6}~\\
       \includegraphics[scale=0.3]{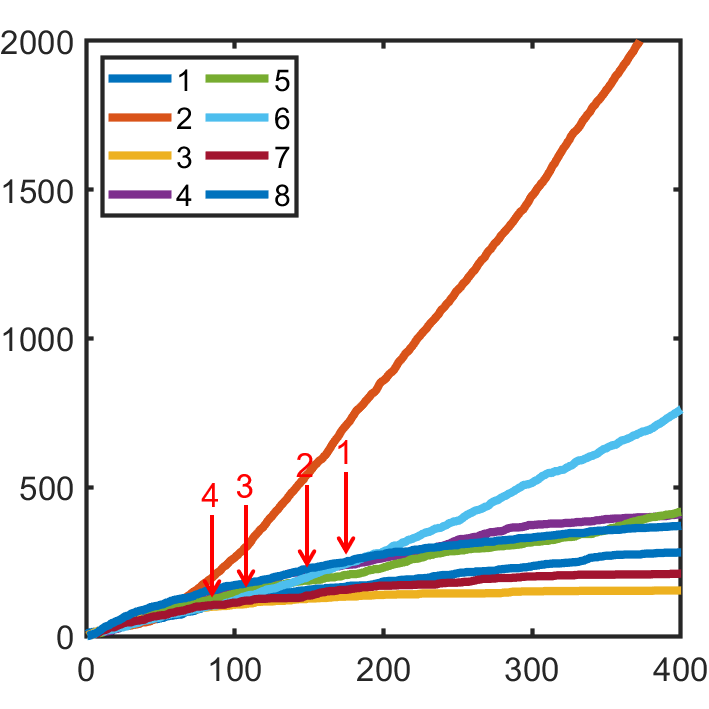}
        \end{subfigure}
    \begin{subfigure}[b]{0.3\textwidth}
    \caption{$\varrho^E_\text{(X-B)}$}\label{fig:exp-X8}~\\
       \includegraphics[scale=0.3]{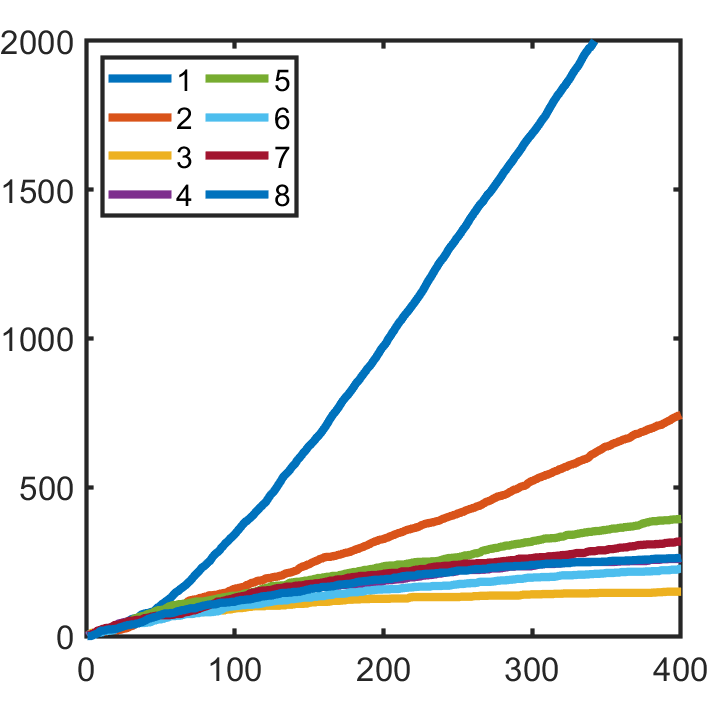}
        \end{subfigure}
    \begin{subfigure}[b]{0.3\textwidth}
         \caption{$\varrho^E_\text{(X-C)}$}\label{fig:exp-X10}~\\
       \includegraphics[scale=0.3]{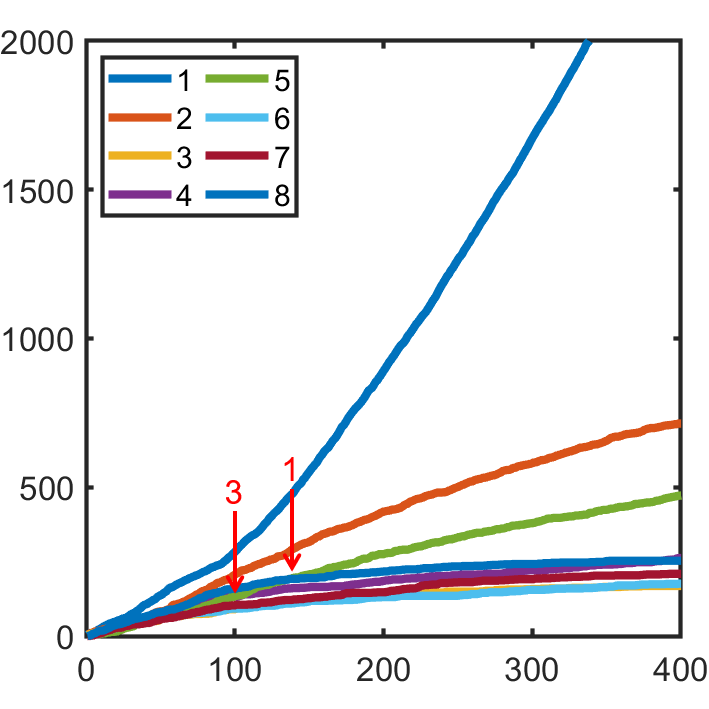}
    \end{subfigure}
     \begin{subfigure}[b]{0.3\textwidth}
    \caption{$\varrho^E_\text{(Y-A)}$}\label{fig:exp-Y6}~\\
       \includegraphics[scale=0.3]{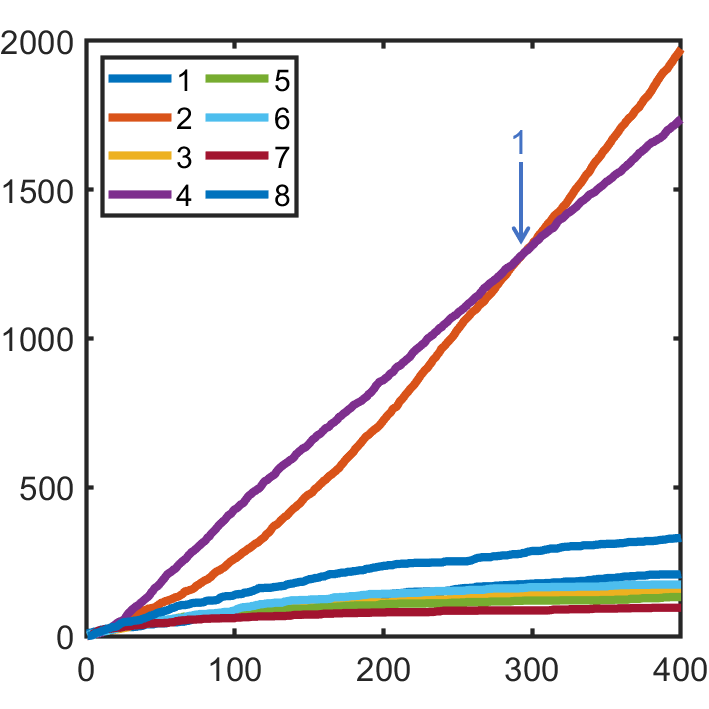}
        \end{subfigure}
    \begin{subfigure}[b]{0.3\textwidth}
    \caption{$\varrho^E_\text{(Y-B)}$}\label{fig:exp-Y8}~\\
       \includegraphics[scale=0.3]{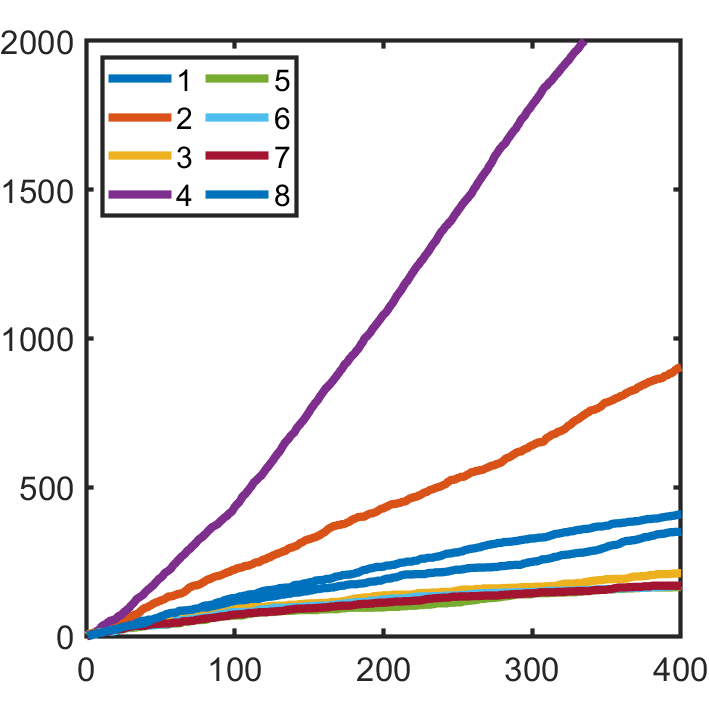}
        \end{subfigure}
    \begin{subfigure}[b]{0.3\textwidth}
         \caption{$\varrho^E_\text{(Y-C)}$}\label{fig:exp-Y10}~\\
       \includegraphics[scale=0.3]{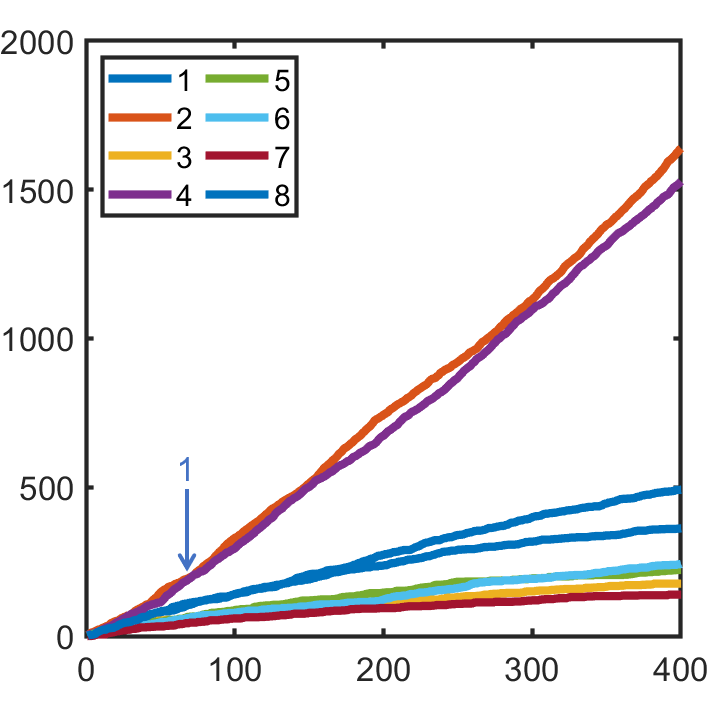}
    \end{subfigure}
    \caption{Experiment results of treatments A, B, and C. (a)Average distribution over time for the 1-200 period and the 800-1000 period; (b) eigencycle spectrum $\vartheta^E$; (c-h) accumulated curve ($\varrho^E$)  of the (X,Y) player over time. The crossover points ($\chi$) are labelled with arrows referring to the data in Table \ref{tab:exp_cross}.}
    \label{fig:eig_str_2}
\end{figure*}

\textbf{Results.} The experimental eigencycle spectrum $\vartheta^E$ is shown in Figure \ref{fig:exp_3eigencycle}. 
The main results are as follows: 
\begin{enumerate}
\item[EO2.1]
Observations on the (A,B,C) treatments, $\vartheta^E_A, \vartheta^E_B, \vartheta^E_C$ are consistent with  
the predictions made based on dynamics theory,
$\vartheta^T_A, \vartheta^T_B, \vartheta^T_C$, respectively, as shown in Figure \ref{fig:eicycle_dyn_main} and Figure \ref{fig:exp_3eigencycle}.This is confirmed by hypergeometric enrichment tests, which show significant enrichment of $\theta^T$ in the $\theta^E$ sets across all treatments (pooled $p = 1.69\times10^{-12}$). For details, see Table A5 and SI section 8.6.2.

\item[EO2.2]
Our observations were consistent with the predictions made based on the statics theory. 
According to the eigencycle spectra for the three treatments, the cycle $(X_2, X_6) \times (Y_2,Y_4)$ in treatment A is the strongest. This is consistent with the result of the best-response analysis, as shown in Figure \ref{fig:survive_strategy}. These findings are supported by the numerical analysis, as shown 
in Appendix \ref{sec:app_cycle_ieds}  

\end{enumerate} 
In summary, the cycles observed in the experiment are consistent with the theoretical predictions. The consistency of the results is in keeping with the existing literature on cycles \cite{zhijian2022,dan2014,wang2017,2021Qinmei,zhijian2020}.

\subsection{Collapse}\label{sec:mt_exp_acc}
The first research question requires us to identify the pulse in the collapse. To evaluate the consistency is to validate the dynamics paradigm during the collapse. Here, we offer the main results obtained regarding the collapse.   

 \textbf{Measurement.} 
For the collapse, there are three relevant measurements -- the accumulated curve ($\varrho$), the pulse signal ($\psi$) and the crossover points ($\chi$). For the definition of the measurements, see Appendix---\ref{sec:app_collapse_def}.

 \textbf{Results and explanation.} 
 The main experimental results are as follows: 

\begin{enumerate}
\item [EO3.1]
  \textbf{Accumulated curves ($\varrho^E$): }  The experimental accumulated curves $\varrho^E_\text{j - i}$ where $i \in $(A,B,C) indicates the treatment and  $j \in$ (X,Y) indicates the player, resulting in  six conditions. 
These  are  shown in 
Figure \ref{fig:exp-X6} - \ref{fig:exp-Y10}, which can be individually compared to the dynamics predictions  $\varrho^T_\text{j-i}$, as shown in Figure \ref{fig:dyn-X6} - \ref{fig:dyn-Y10}. The results of the visual comparison support the consistency between the dynamics theory and the experimental results. 

\item[EO3.2]
\textbf{Crossover point ($\chi^E$):} Following the dynamics prediction $\chi^T$ in Table \ref{tab:dyn_cross}, in the accumulated curve ($\varrho^E$),  
the $\chi^E_-$ are labelled using red arrows 
between  $\textbf{D}^-/\textbf{D}^+$;   
meanwhile the $\chi^E_+$ are labelled using blue arrows between the $\textbf{D}^+/\textbf{D}^+$. These are shown in Figure \ref{fig:exp-X6}-\ref{fig:exp-Y10}. The consistency between theory and experiment is visible.
\item[EO3.2.1]
The numerical results in Table \ref{tab:exp_cross} and Table \ref{tab:dyn_cross} support the consistency between the  theory and the experiment.  
For the dynamics predictions shown in Table \ref{tab:dyn_cross}, 
if $\tau^T \geq 89$ of $\psi ([Y_2,Y_4],\tau)$ is taken as the benchmark, there are nine predicted samples. 
The experiment showed that, 
following the nine that were theoretically predicted, 
 all eight of the $\chi^E$ labelled arrows 
 in experiment have the largest $\tau$ values 
 in terms of the related role (X,Y) 
 and the treatment (A,B,C).  
\item[EO3.2.2]
The crossover point is a consequence of the pulse in collapse. 
Referring to the relation between the pulse and the crossover time shown in section \ref{sec:app_collapse_def}, the observed crossover points are used to support the existence of pulse in [EO3.3.6] and  [EO3.3.7].
\item [EO3.3]
  \textbf{Pulse signals ($\psi^E$):} A pulse is a distinct observation made during the collapse.  
  For the definition of a pulse, see Equation \ref{eq:Pulse} in Appendix \ref{sec:app_collapse_def}. 
  By examining the data, we observed pulses to be significant.
  The main results regarding the pulses are as follows: 
\begin{itemize}
\item[(a)] Existence of the pulse:
\item [EO3.3.1]
There are a total of nine  pulses that are significant  ($p<0.05$), as shown in Table \ref{tab:exp_Pulse_significant}.  
\item [EO3.3.2]
Among these nine pulses, three pulses are of the strongest statistical significance ($p<0.010$), as shown in Table \ref{tab:exp_Pulse_sign3}. 
\item[(b)]	The pattern of the pulses' existence is consistent with the theoretical pattern: 
\item [EO3.3.3]
The number of pulses in the treatments does not deviate from the expectation of dynamics theory (see Table \ref{tab:dyn_Pulse_significant}). More pulses are observed in treatment A than in B, and in B than in C.    
\item [EO3.3.4]
All three of the pulses with the strongest significance ($p <0.010$) are included in the category of the most expected pulses,  according to dynamics theory (for details, see Table \ref{tab:dyn_Pulse_significant}). 
\item [EO3.3.5]
All four pulses with the strongest significance ($p\leq 0.011$) have the maximum surplus $\psi$ value among the nine pulse signals shown in Table \ref{tab:exp_Pulse_significant}.
\item[(c)] The pattern of the pulses' existence is consistent with the pattern of the crossover points.
\item [EO3.3.6]
Of the nine significant pulses ($p<0.05$), 
six are from treatment A and three are from C.  
More pulses are observed in treatment A than in C, 
as shown in Table \ref{tab:exp_Pulse_significant}. 
This is consistent with the $\chi^E$,
in that there are more significant crossover points 
observed in treatment A than in C, 
according to  the $\chi^E_-$ with 
the red arrows in Table \ref{tab:exp_cross}. 
\item [EO3.3.7]
As predicted in Table \ref{tab:dyn_cross},  
there are (4,0,2) red arrows in treatment (A,B,C), respectively. 
This is supported by the number of pulses and their significance in the experiment, 
as shown in Table \ref{tab:exp_Pulse_significant}. 
\item [EO3.3.8]
On the other hand, there is no significant pulse signal  ($p<0.05$) from the Y player in treatment A, B, or C. This is consistent with the dynamics predictions for pulses, as shown in Table \ref{tab:dyn_Pulse_significant}. 
\end{itemize}

Three exact binomial tests confirm the statistical robustness of the pulse signals. Using logit dynamics candidates, 5/9 pulses hit ($p=2.03\times10^{-11}$), referring to SI section 8.1.1; using parameter-free replicator dynamics, 5/9 hit ($p=4.85\times10^{-10}$), referring to SI section 8.3 ; and using IEDS-predicted surviving strategies, 8/9 hit at the strategy-pair level ($p=2.35\times10^{-4}$), referring to SI section 8.2. All results survive Bonferroni correction where applicable. These rule out multiple-testing artifacts and parameter-tuning concerns, while showing that the pulse signals observed in the collapse process are significant.

  \begin{table}[ht!]
    \centering
    \begin{tabular}{cccccccc}
    \hline
    Treat-&  Domin- & Domin- & Time  &  {paired-}  &   Surplus& Sample\\
   ments &  ated  &  ation &  block &~~ttest ($p$) &   & size  \\
         &   $s_j$   & $s_i$ &  $[t_0,t_1]$ &      & $\psi$  &  \textbf{N}    \\
    \hline
      A	& $X_8$ & $X_2$ & 11-20  & 0.007 & 17 & 120  \\
      A	&  $X_8$ & $X_6$ & 11-20  & 0.002 & 19 & 120  \\
     A	&  $X_4$ & $X_6$ & 81-90   & 0.007 & 15 & 120  \\ 
    \hline
    \end{tabular}
    \caption{All of the observed pulses $\psi^E$ were highly significant ($p <0.010$). The symbols are explained in Equation \ref{eq:Pulse} in Appendix \ref{sec:app_collapse_def} }
    \label{tab:exp_Pulse_sign3}
\end{table}   

 \end{enumerate} 

The main result from the collapse in experiment is that, the significant pulse signals are observed, which supported by the observed crossover points. In addition to existing literature \cite{zhijian2014rps,dan2014,dan2010,zhijian20142x2,nowak2015,stephenson2019,hans2021}, in the collapse process, the consistency between the  theory and the experiment is supported by new empirical evidence. 

\subsection{Model comparison}\label{sec:mt_model_compare}
Comparing the performances of the two theories under consideration is a key component of experiment research. Our main results are as follows:

\begin{enumerate} 
\item[ET1]  In terms of distribution, the dynamics prediction ($\rho^T_D$) outperformed statics prediction ($\rho^T_S$). This is supported by data ($\rho^E$) in section \ref{sec:mt_exp_dis}.  
 
\item[ET2]  For the  cycle, the dynamics prediction outperformed the statics prediction. This is  supported by comparing the results shown in EO2.1 and EO2.2 in section \ref{sec:mt_exp_eic}. 
Further analysis on cycle spectra can provide additional  evidence to enhance the results, see section \ref{sec:app_cycle}.  

\item[ET3] For the collapse, dynamics theory can quantitatively predict the pulse ($\psi$) as well as the crossover point ($\chi^E$). The predictions made using IEDS, a component of statics theory (TS3 in section \ref{sec:mt_classical_predict}), offer only limited explanation of the pulsing signals and the crossover point in terms of quantity.  As a result, the dynamics model outperforms the statics. 
\end{enumerate}
In summary,  game dynamics theory performs better than the static equilibrium   theory, according to our data. This finding is consistent with the existed literature \cite{dan2014,dan2010,nowak2015,zhijian20142x2,zhijian2014rps,stephenson2019,hans2021}.  
The summary of this section, see the first paragraph of the  Discussion.

\section{Discussion}
The main results of this study are the answers to the two key research questions posed above. 
\begin{enumerate}
\item Regarding the first question, we observed that the pulse signal, which is a distinct observation of the collapse predicted using the dynamic model, is significant.
\item Regarding the second question, in the full process of finding equilibrium, we showed that all of the significant observations were those most clearly predicted by the dynamic model. The completeness and the consistency of the dynamics paradigm are supported by our data.
\end{enumerate}
We posit that these results will assist the development of the game dynamics paradigm.

\subsection{Implications}\label{sec:mt_Implication}
To explain the implications of our results, we can refer to the scientific paradigm of game dynamics theory \cite{dan1991,dan1998,samuelson2016,kandori2021,dan2016,sandholm2010}. 
 This paradigm has a self-reinforcing closed feedback workflow loop 
 (for a brief introduction, 
see section \ref{sec:app_paradigm}), which enhances its distinct set of concepts 
 (e.g., natural selection, 
 the evolutionary stable strategy, and, potentially, cycle \cite{samuelson2016,dan2016,nowak2015,dan2014,zhijian2011,zhijian20142x2,zhijian2014rps,stephenson2019,hans2021}).

Since 1990 \cite{dan1991,samuelson2016},it has been expected that this paradigm would establish its own narrative of the game dynamics process, bringing about a paradigm shift with regard to the classical statics (Nash equilibrium) theory. However, the paradigm shift has not come to fruition. Those who are skeptical about using dynamics research to solve social and economic problems mainly focus on empirical observations \cite{kandori2021,samuelson2016,colin2003}.   
 
Since 2010,  the convenience of this method has been improved by empirical observations. The cycle is one example of this  \cite{dan2021,dan2014,2021Qinmei,zhijian20142x2,zhijian2013,zhijian2011,zhijian2012,stephenson2019,hans2021}, and it may become a constitute element in the paradigm’s distinct set of concepts. Because a cycle generally exists in mixed-strategy Nash equilibrium behaviours, it is a consequence of the engagement of strategy interactions; moreover, it demonstrates game evolution, and is an exploitable PTDE.

We posit that the pulse in collapse can be constantly developed, similarly to the cycle, because of the following:
\begin{enumerate}
\item 
The pulse fulfils the experimental observation gap in the collapse, which is a legitimate constituent part of the full equilibrium-finding process (as shown in  Figure \ref{fig:ConceptFigure4level}). 
\item
 The study on the pulse in collapse process can be helpful to the completeness and consistency of the game dynamics paradigm.  
\end{enumerate}

\subsection{Related works}
\textbf{On the elimination of a dominated strategy~~}
This study is not unique in noting the behaviours of the dominated strategy. For example: 

\begin{itemize}  

\item 
 Theoretically speaking, ‘A dominated strategy can dominate a domination before being dominated’; this phenomenon, namely, a pulse, is evident (e.g., in Gintis’ book \cite{gintis2000}, Figure 5.3, in section 5.8, the red mark during the infancy stage is the pulse signal). In finding the full Nash equilibrium process, the dominated strategies that are eliminated or survive have been extensively studied  \cite{Hofbauer1996Evolutionary, Hofbauer2011Survival, mertikopoulos2016learning}, but research on the pulse phenomenon is rare.
\item 
In experimental terms, research into dominated strategies is not uncommon
\cite{rapoport2000mixed,huyck1999,saran2016}. 
But 
these reports mainly focus on whether the dominated strategies are eliminated 
or attain equilibrium, 
but do not address what can be observed during the collapse.
\item \textbf{On why traditional literature has historically
paid little attention to the exact real-time trajectory of elimination} -- The difficulty in measuring the collapse process stems from a fundamental trade-off. In low-dimensional games with small strategy spaces, the time series can be frequently revisited, facilitating statistical testing; yet only a few strategies are eliminated, which fails to capture the diversity of the collapse. In high-dimensional systems, by contrast, the strategy space is favorable for observing the elimination of dominated strategies. However, within finite experimental time series, obtaining sufficient repeated measurements for each dominated strategy to yield statistically meaningful results is a demanding task, This requires both sophisticated experimental design and measurement, along with extensive data acquisition. This paper studies the dynamics in high-dimensional games.  In this direction,  research on the dynamics of high-dimensional strategic interactions is also developing rapidly \cite{Walunj2026,wang2026eigen,Dan2026,johnston2026}. \\

\end{itemize}
The uniqueness of this study in existing literature is that, 
it firstly provides the quantitative observation, 
in terms of the pulse, of the collapse process. 
By examining the collapse process, which has not been investigated before, the consistency and the completeness of the game dynamics paradigm are demonstrated.

~\\
\textbf{Further research}  
Further research  on the collapse and the pulse is needed.  Although the existence of the pulse in the collapse has been explored and pulse experiment is consistent with the theory, it is far from being considered a legitimate constituent part of the paradigm.  

The open question on how to select a dynamics model and its parameters when facing a real-life system is remains. In this study, following \cite{huyck2001,dan2014,zhijian2013}, we utilised logit dynamics, and fix its parameter grossly by the distribution close to the equilibrium distribution by QP. Although we can capture all of the various observations in various treatments without changing the parameters, admittedly, this work can not answer the open question. 

The collapse stage in games does not only move beyond the stationary concept of static equilibrium  \cite{selten2008,zhijian20142x2},but also departs from the concept of non-equilibrium stationary states (e.g., cycle) \cite{lax1960,zhijian2013,zhijian20142x2,Walunj2026,wang2026eigen}. 
By developing a better understanding of collapse, and of how to obtain the hierarchy of condensed structures or various dynamic equilibrium structures from a primordial soup or an existing state, we can expect solutions to appear in the coming decades.

~\\
\textbf{Potential applications} 
Exploiting the PTDE can be profitable, 
 e.g., in the arbitrage by high frequency trading (HFT) 
 \cite{wiki:High-frequency_trading}. 
The arbitrage of HFT comes from predictable 
non-equilibrium in the short-term, 
 rather than  equilibrium in the long-term.
The collapse is part of a process that is distant from equilibrium. 
It could therefore provide a helpful framework for understanding sudden shocks in a social system.
During such periods, many unconventional strategy behaviors might appear to dominate. The findings related to the pulse support the statement: ’A dominated can dominate a domination before dominated’. This means that non-equilibrium behaviors can be profitable in the short term.

This study has  real life applications in fields including artificial intelligence \cite{stratego2022} and the social systems addressed  in \cite{dan1998,samuelson2016}. 
the collapse is a legitimate constituent part of the process of game evolution. The Nash equilibrium is not necessarily the best strategy during the infancy period when playing a complex game, e.g., Stratego \cite{stratego2022}.

Regarding general decision making in real-life social systems, philosophically, linear thinking may be harmful. In periods of turmoil, the power of myopia, or the power of the best response, must be properly acknowledged.

\section{Conclusions}

Based on a dynamic game with incomplete information, this investigation illustrates the full equilibrium finding process shown in figure \ref{fig:ConceptFigure4level}, from random play to collapse to a fixed pattern (cycling and equilibrium). As a PTDE, the collapse constitutes fertile ground for the science of game dynamics, especially as it relates to the system of human social interactions.

\vspace{6pt}

\subsection*{Abbreviations} \label{sec:app_Abbreviations LX }
The abbreviations and mathematical symbols, shown in Table \ref{tab:app_abbr}, are used in the main text, this appendix and the supplementary information. 
 \begin{table}
    \centering
    \begin{tabular}{lll}
 \hline
PTDE && Predictable temporary deviation from equilibrium\\
    $\textbf{N}$ & & Sample size for statistical analysis  \\
    $\text{Dyn}(\lambda,\Delta)$ & & Logit Dynamics with noise $\lambda$ and time step $\Delta$,  \\
    & & a model in game dynamics theory \\
   IEDS  & &  Iterated
eliminating dominated strategy,    \\
    & & a method to find the Nash equilibrium \\
  QP & & Quadratic programming, an algorithm  \\
     & &  for Nash equilibrium distribution \\
  $\textbf{D}^\pm$   & & $+$, domination strategy, survival strategy; \\
  EO  & & Experimental observation   \\
  TS  & & Predictions from static equilibrium theory  \\
  TD  & & Predictions from dynamics theory  \\ 
  ET  & & Experiment data, used to compare the models'   \\ 
  Obs.  & & observations \\
  SI  & &  Supplementary information\\
     \hline
    $s_i$ & & The $i$-th optional strategy for a player\\
    $\rho$ & & The proportion vector of strategies used \\
    &$\rho(s_i)$&  The $i$-th component of $\rho$, \\
    &$\rho(s_i,t)$&   The $\rho(s_i)$ at time (round) $t$ \\
    &$\rho_{([t_1,t_2])}$& The time average of $\rho$ between $[t_1,t_2]$  \\
      &$\rho^+_*([s_i,s_j])$ & A surplus sample set \\
     $\delta$ & & The Euclidean distance between two distribution     \\
     && vectors,used to evaluate their difference \\ 
     \hline
   $ o $&  & A fixed point, rest point, or zero velocity point, equilibrium   \\
   $\lambda$ & & Eigenvalue   \\
   $\xi$ & & Eigenvector   \\
   $\eta_i$  & & The $i$-th component of a given eigenvector $\xi$  \\
    $\vartheta$ &  & The eigencycle set in the 2D subspace set of a game  \\
    & & space, describing the cyclic  motion strength    \\
& $\vartheta(m,n)$ & The $\vartheta$  at 
       dimension (1,2):=($\eta_m,\eta_n$)  \\
&$\vartheta(x)$ & Eigencycle spectrum, where $x$ is a natural number  \\
     \hline
   $\varrho$&  & Time-accumulated value curve of $\rho(t)$  \\
   &$\varrho(s_i)$ &   the $\varrho$ of the $s_i$ strategy  \\
   $\psi$   &  & A pulse signal   \\
      & $\psi([s_i,s_j],[t_1,t_2])$     & A pulse signal of strategy $j$ over $i$    \\
      &  &  during time interval $[t_1,t_2]$  \\
   $\chi$ &  & A crossover point, having two results  $t$ and $\varrho$  \\
         &$\chi([s_i,s_j],\tau)$& A crossover point of $\varrho(s_i)$ and $\varrho(s_j)$ at time \\
         &  & $\tau$, and when $t= \tau-0^+$, $s_j > s_i$  \\
        &$\chi_+(s_i,s_j)$&  $\chi$ if $s_i \in 
         \textbf{D}^+ \cap s_j \in 
         \textbf{D}^+$     \\
        &$\chi_-(s_i,s_j)$&   $\chi$ if $s_i \in 
         \textbf{D}^+ \cap s_j \in 
         \textbf{D}^-$\\
 \hline
    \end{tabular}
    \caption{Abbreviations and mathematical symbols.}
    \label{tab:app_abbr}
\end{table}

\clearpage
\begin{appendices}
\setcounter{table}{0}
\setcounter{figure}{0}
\renewcommand{\thetable}{A\arabic{table}}
\renewcommand{\thefigure}{A\arabic{figure}}
\renewcommand{\theequation}{A\arabic{equation}}
\renewcommand{\thesubsection}{\Alph{section}.\arabic{subsection}}

\section{}   
\subsection{Distribution} 
\label{sec:app_distribution}

This section reports the results on the distribution and its evolution, including the following elements: 
\begin{itemize}
\item
Theoretical and experimental numerical results of the average time distribution are shown Table \ref{tab:dis_main}.  
\item 
Numerical results relating to Euclidean distance, illustrating the evolution of the distribution, are shown in  \ref{tab:nash_distance}.   
\end{itemize}
For more details of the measurements and data, see SI section 5 (Distribution).   

\textbf{Explanation of support [EO1] and [EO2]. }
In Table \ref{tab:nash_distance}, the numerical Euclidean distance and its evolution are reported as follow:
\begin{enumerate}
\item The (1,2,4,5)-th row is the Euclidean distance of the given time interval.
\item The (3, 6)-th row is the difference in the Euclidean distance
\begin{equation}\label{QP}
\triangle \delta_{\textbf{QP}}=\delta_{\textbf{QP[801,1000]}}-\delta_{\textbf{QP[1,200]}}<0
\end{equation} 
\begin{equation}\label{Dyn}
\triangle \delta_{\textbf{Dyn}}=\delta_{\textbf{Dyn[801,1000]}}-\delta_{\textbf{Dyn[1,200]}}<0
\end{equation}
This means that $\rho^E$ is close to the predictions during the 1000 round game in all of the treatments (A,B,C). 
\item The (7)-th row is the difference between the two models. 
\begin{equation}\label{QP_Dyn}
\triangle \delta =\delta_{\textbf{Dyn[801,1000]}}-\delta_{\textbf{QP[801,1000]}} < 0 
\end{equation} 
This means that the $\rho^E_{800,1000}$ is closer to the expectation of the dynamics model than the Nash equilibrium according to the QP  algorithm in all of the treatments (A,B,C). 
\end{enumerate}

\begin{sidewaystable} 
    \centering
    \small
    \begin{tabular}{||c|c|c||cc||cc||cc||cc||cc||cc||}    \hline  
   \xrowht{13pt}
  \multirow{2}{*}{Treatment}&Period&\multirow{2}{*}{id}    &  \multicolumn{2}{c|}{$\rho^E_{(1,1000)}$} &\multicolumn{2}{c|}{$\rho^E_{(1,500)}$}	&	\multicolumn{2}{c|}{$\rho^E_{(500,1000)}$}	&	\multicolumn{2}{c|}{$\rho^E_{(800,1000)}$}	&	\multicolumn{2}{c|}{$\rho^T_{QP}$}	&	\multicolumn{2}{c|}{$\rho^T_{Dyn(0.02,50)}$}	\\ 
   \cline{2-2} \cline{4-15}\xrowht{8pt}
   & Game& &	\multicolumn{1}{|c|}{X}&	Y&	\multicolumn{1}{|c|}{X}&	Y&	\multicolumn{1}{|c|}{X}&	Y&	\multicolumn{1}{|c|}{X}&	Y&	\multicolumn{1}{|c|}{X}&	Y&	\multicolumn{1}{|c|}{X}&Y	\\
     	 \hline
A&(2,1)&1&	0.040&	0.050&	0.055&	0.042&	0.025&	0.058&	0.021&	0.057&	0.000&	0.000&	0.034&	0.001\\
A&(2,1)&2&	0.580&	0.497&	0.486&	{0.431}&	{0.673}&	0.563&	0.691&	0.595&	0.667&	0.667&	0.670&	0.629\\
A&(2,1)&3&	0.019&	0.017&	0.029&	0.028&	0.009&	0.007&	0.005&	0.007&	0.000&	0.000&	0.000&	0.000\\
A&(2,1)&4&	0.045&	0.346&	0.076&	0.367&	0.014&	0.324&	0.011&	{0.307}&	0.000&	0.333&	0.000&	0.370\\
A&(2,1)&5&	0.066&	0.015&	0.087&	0.023&	0.045&	{0.006}&	0.045&	0.005&	0.000&	0.000&	0.014&	0.000\\
A&(2,1)&6&	0.185&	0.019&	0.163&	0.030&	0.207&	0.007&	0.205&	0.005&	0.333&	0.000&	0.280&	0.000\\
A&(2,1)&7&	0.026&	0.011&	0.038&	0.018&	0.014&	0.005&	0.011&	0.002&	0.000&	0.000&	0.000&	0.000\\
A&(2,1)&8&	0.039&	0.045&	0.066&	0.061&	0.012&	0.029&	0.010&	0.022&	0.000&	0.000&	0.000&	0.000\\
 \hline										
 
B&(3,2)&1&	0.624&	0.061&	0.540&	0.071&	0.707&	0.051&	0.712&	0.044&	1.00&	0.033&	0.842&	0.025\\
B&(3,2)&2&	0.165&	0.188&	0.160&	0.189&	0.170&	0.186&	0.167&	0.186&	0.000&	0.186&	0.087&	0.239\\
B&(3,2)&3&	0.019&	0.030&	0.026&	0.040&	0.012&	0.020&	0.010&	0.017&	0.000&	0.041&	0.000&	0.053\\
B&(3,2)&4&	0.029&	0.579&	0.046&	{0.521}&	0.011&	0.636&	0.012&	0.650&	0.000&	0.573&	0.000&	0.501\\
B&(3,2)&5&	0.058&	0.030&	0.077&	0.034&	0.039&	0.026&	0.040&	{0.022}&	0.000&	0.021&	0.063&	0.006\\
B&(3,2)&6&	0.029&	{0.029}&	0.041&	0.034&	0.016&	0.025&	0.015&	0.027&	0.000&	0.049&	0.007&	0.053\\
B&(3,2)&7&	0.043&	0.020&	0.062&	0.030&	0.025&	0.011&	0.020&	0.009&	0.000&	0.023&	0.000&	0.012\\
B&(3,2)&8&	0.033&	0.063&	0.047&	0.081&	0.019&	0.045&	0.025&	0.045&	0.000&	0.074&	0.000&	0.111\\
 \hline												
C&(4,2)&1&	0.750&	0.051&	0.588&	0.064&	0.912&	0.039&	0.943&	0.038&	1.000&	0.058&	0.882&	0.078\\
C&(4,2)&2&	0.092&	0.386&	0.135&	0.362&	0.049&	0.411&	0.034&	0.426&	0.000&	0.422&	0.077&	0.413\\
C&(4,2)&3&	0.015&	0.027&	0.029&	0.033&	0.000&	0.022&	0.000&	0.014&	0.000&	0.043&	0.000&	0.061\\
C&(4,2)&4&	0.024&	0.363&	0.046&	0.325&	0.002&	0.400&	0.001&	0.425&	0.000&	0.330&	0.000&	0.321\\
C&(4,2)&5&	0.061&	0.043&	0.092&	0.044&	0.031&	0.042&	0.016&	0.040&	0.000&	0.025&	0.038&	0.011\\
C&(4,2)&6&	0.016&	0.040&	{0.030}&	0.051&	0.001&	0.030&	0.002&	0.004&	0.000&	0.055&	0.003&	0.061\\
C&(4,2)&7&	0.019&	0.018&	0.036&	{0.025}&	0.001&	0.010&	0.002&	0.009&	0.000&	0.021&	0.000&	0.009\\
C&(4,2)&8&	0.024&	0.071&	0.044&	0.096&	0.004&	0.047&	0.003&	0.044&	0.000&	0.046&	0.000&	0.047\\
\hline
       
\end{tabular}
  \caption{Distribution. Columns 1-3 show the treatment, game parameter, and strategy, respectively. The blocks   $\rho^E(t_0,t_1)$ are the experimental (X,Y) player's time average of $\rho$ between $(t_0,t_1)$. $\rho^T_{QP}$ and $\rho^T_{Dyn(0.02,50)}$ are the prediction from statics and dynamics theory, respectively. }
    \label{tab:dis_main}
\end{sidewaystable}

\begin{table}
    \centering
    \begin{tabular}{|l|c|c|c|}
    \hline
   &	 \multicolumn{3}{c|}{Treatment} \\\cline{2-4}
   &	A &	  B &	  C \\\hline
$\delta_{\textbf{QP[1,200]}}$&	~~0.4298&	~~0.5551&	~~0.5745\\\hline
$\delta_{\textbf{QP[801,1000]}}$&	~~0.3766&	~~0.4047&	~~0.4436\\\hline
~~~~~$\triangle \delta_{\textbf{QP}}$&	$-$0.0532&	$-$0.1504&	$-$0.1309\\\hline
$\delta_{\textbf{Dyn[1-200]}}$&	~~0.0781&	~~0.2088&	~~0.1442\\\hline
$\delta_{\textbf{Dyn[801-1000]}}$&	~~0.0059&	~~0.0097&	~~0.0072\\\hline
~~~~~$\triangle \delta_{\textbf{Dyn}}$&	$-$0.0722&	$-$0.1991&	$-$0.1370\\\hline
~~~~~~~$\triangle \delta$&	$-$0.3707&	$-$0.3950&	$-$0.4364\\\hline
    \end{tabular}
    \caption{ The Euclidean distance and its evolution. For details,  see the paragraph `Explanation of support [EO1] and [EO2]'.}
    \label{tab:nash_distance}
\end{table}

\begin{table}
    \centering

\clearpage
     	$$\begin{array}{|c|c|cc|r|}
       \hline\xrowht{8pt}
     	\text{Treatment}    & x  &  m & n  & \vartheta^E~~~ \\
       \hline
     	 \text{A} & 25 & 2 & 12 & -1.000 \\
     	 \text{A} & 23 & 2 & 10 & 0.830 \\
     	 \text{A} & 71 & 6 & 12 & 0.700 \\
     	 \text{A} & 69 & 6 & 10 & -0.660 \\
     	 \text{A} & 101 & 10 & 12 & -0.189 \\
     	 \text{A} & 59 & 5 & 10 & -0.110 \\
     	 \text{A} & 9 & 1 & 10 & -0.104 \\
     	 \text{A} & 11 & 1 & 12 & 0.104 \\
     	 \text{A} & 61 & 5 & 12 & 0.081 \\
     	 \text{A} & 22 & 2 & 9 & 0.078 \\
     	 \text{A} & 55 & 5 & 6 & 0.070 \\
     	 \text{A} & 93 & 9 & 10 & -0.067 \\
       \hline
     	 \text{B} & 11 & 1 & 12 & -0.269 \\
     	 \text{B} & 101 & 10 & 12 & -0.148 \\
     	 \text{B} & 59 & 5 & 10 & -0.143 \\
     	 \text{B} & 23 & 2 & 10 & 0.126 \\
     	 \text{B} & 111 & 12 & 13 & -0.098 \\
     	 \text{B} & 80 & 7 & 12 & 0.081 \\
     	 \text{B} & 29 & 2 & 16 & -0.081 \\
     	 \text{B} & 102 & 10 & 13 & 0.075 \\
     	 \text{B} & 13 & 1 & 14 & 0.074 \\
     	 \text{B} & 65 & 5 & 16 & 0.067 \\
     	 \text{B} & 12 & 1 & 13 & 0.063 \\
     	 \text{B} & 71 & 6 & 12 & 0.061 \\
       \hline
     	 \text{C} & 23 & 2 & 10 & 0.272 \\
     	 \text{C} & 9 & 1 & 10 & -0.257 \\
     	 \text{C} & 25 & 2 & 12 & -0.153 \\
    	 \text{C} & 101 & 10 & 12 & -0.094 \\
     	 \text{C} & 11 & 1 & 12 & 0.086 \\
     	 \text{C} & 93 & 9 & 10 & -0.077 \\
     	 \text{C} & 26 & 2 & 13 & -0.066 \\
     	 \text{C} & 12 & 1 & 13 & 0.066 \\
     	 \text{C} & 61 & 5 & 12 & 0.056 \\
     	 \text{C} & 111 & 12 & 13 & -0.051 \\
     	 \text{C} & 59 & 5 & 10 & -0.045 \\
     	 \text{C} & 104 & 10 & 15 & 0.044 \\
       \hline
     	\end{array} ~~~~
           	\begin{array}{|c|c|cc|r|}
       \hline\xrowht{8pt}
     	\text{Treatment}   & x  & m  & n  & \vartheta^T ~~~\\
       \hline
     	 \text{A} & 25 & 2 & 12 & -1.000 \\
     	 \text{A} & 23 & 2 & 10 & 0.995 \\
     	 \text{A} & 71 & 6 & 12 & 0.928 \\
     	 \text{A} & 69 & 6 & 10 & -0.923 \\
     	 \text{A} & 61 & 5 & 12 & 0.068 \\
     	 \text{A} & 59 & 5 & 10 & -0.068 \\
     	 \text{A} & 9 & 1 & 10 & -0.021 \\
     	 \text{A} & 11 & 1 & 12 & 0.021 \\
     	 \text{A} & 1 & 1 & 2 & 0.011 \\
     	 \text{A} & 5 & 1 & 6 & -0.011 \\
     	 \text{A} & 86 & 8 & 10 & 0.008 \\
     	 \text{A} & 88 & 8 & 12 & -0.008 \\
       \hline
     	 \text{B} & 23 & 2 & 10 & 0.084 \\
     	 \text{B} & 9 & 1 & 10 & -0.058 \\
     	 \text{B} & 61 & 5 & 12 & 0.056 \\
     	 \text{B} & 11 & 1 & 12 & -0.046 \\
     	 \text{B} & 10 & 1 & 11 & 0.044 \\
     	 \text{B} & 29 & 2 & 16 & -0.037 \\
     	 \text{B} & 59 & 5 & 10 & -0.030 \\
     	 \text{B} & 24 & 2 & 11 & -0.028 \\
     	 \text{B} & 15 & 1 & 16 & 0.025 \\
     	 \text{B} & 101 & 10 & 12 & -0.024 \\
     	 \text{B} & 106 & 11 & 12 & 0.018 \\
     	 \text{B} & 8 & 1 & 9 & 0.017 \\
       \hline
     	 \text{C} & 61 & 5 & 12 & 0.256 \\
         \text{C} & 11 & 1 & 12 & -0.168 \\
     	 \text{C} & 23 & 2 & 10 & 0.150 \\
    	 \text{C} & 59 & 5 & 10 & -0.117 \\
     	 \text{C} & 25 & 2 & 12 & -0.107 \\
     	 \text{C} & 58 & 5 & 9 & -0.084 \\
     	 \text{C} & 13 & 1 & 14 & 0.067 \\
    	 \text{C} & 101 & 10 & 12 & -0.060 \\
     	 \text{C} & 29 & 2 & 16 & -0.051 \\
     	 \text{C} & 1 & 1 & 2 & -0.049 \\
     	 \text{C} & 22 & 2 & 9 & 0.048 \\
     	 \text{C} & 8 & 1 & 9 & 0.042 \\
       \hline
     	\end{array}$$ 
    \caption{Experimental cycle $\vartheta^E$ and the theoretical cycle $\vartheta^T$. The top 12 maximum strengths in the eigencycles  of the spectrum from treatments (A,B,C). $x$ is the $x$-axis value in the eigencycle spectrum, and ($m,n$) are the strategy IDs in the unified dimension of the game, which are defined in SI section 6 (Cycle). }
    \label{tab:exp_dyn_eic12}
\end{table}

\subsection{Cycle}
As the main focus of this study is the collapse, we report the cycle only briefly.
The definition of the eigencycle spectrum, as well as the details of the numerical results of the eigencycle spectrum, are shown in SI section 6 (Cycle). This section includes the following contents: 
\label{sec:app_cycle} 

\begin{enumerate}
\item Results on the maximum eigencycles in the eigencycle spectrum, in the experiment and in theory, respectively, are shown in Table \ref{tab:exp_dyn_eic12}. 
\item The verification of the theories by experiment. 
\end{enumerate}

\subsubsection{Verify the statics}
\label{sec:app_cycle_ieds}
In the main text section \ref{sec:mt_classical_predict}  [TS2], we state that the cycle's existence in treatment-(A, B, C) can be predicted as (Yes, No, Yes), respectively.  
This prediction is based on the IEDS approach in the static equilibrium theory. 

  From Table \ref{tab:exp_dyn_eic12}, we obtain the following results. 
\begin{itemize}
\item In treatment A, the closed loops are  
\begin{equation}\label{eq:cycle_in_A1}
   2  \rightarrow 10 \rightarrow 6 \rightarrow 12 \rightarrow 2 
\end{equation}   
\item  In treatment C, the closed loops are 
\begin{equation}\label{eq:cycle_in_C1}
 2\rightarrow 10 \rightarrow 5\rightarrow 12 \rightarrow 2
\end{equation}  
\begin{equation}\label{eq:cycle_in_C2}
 2\rightarrow 10 \rightarrow 1 \rightarrow 12 \rightarrow 2
\end{equation}      
\end{itemize}   
the IEDS predictions for the cycle are supported. We observed a weak cycle in treatment B, which is not regarded as evidence against IEDS because of its weakness, referring to equation \ref{eq:eic_loop_3strength}. 

\subsubsection{Verification of the dynamics theory }\label{sec:app_cycle_dyn}
The predictions given in the main text section 2.3 [TD2] are verified here.
In treatments A and C, the predictions match the experimental data:

\begin{itemize}
\item In the cycle loop, there exist cycles for equation \ref{eq:cycle_in_A1}  in treatment A, and for equation \ref{eq:cycle_in_C1} 
 and \ref{eq:cycle_in_C2} in treatment C;
\item For the strength of the cycle loop, referring to Table \ref{tab:exp_dyn_eic12}, comparing the average of the strength of $\vartheta^E_{\text{Treatment}}$ in the loops, the result is
\begin{equation}\label{eq:eic_loop_3strength}
    \vartheta^E_{\text{A}} > \vartheta^E_{\text{C}} > \vartheta^E_{\text{B}.}
\end{equation} 
\end{itemize}
For treatment B, the dynamics predicts little cycle. 
Considering equation \ref{eq:eic_loop_3strength}, 
we do not regard the loop of $\vartheta_B^E$ as significant evidence  against the prediction.  
As a result, the consistency between the dynamics and the experiment is supported.
\subsection{Collapse}
\label{sec:app_collapse}

This section includes the measurement definition for collapse, as well as the experimental data. For more details on this subsection, see SI section 7 (Collapse). 
\subsubsection{Measurement definition}\label{sec:app_collapse_def}

\begin{figure}[h!]
\centering 

\includegraphics[angle=0,width=.5\textwidth]{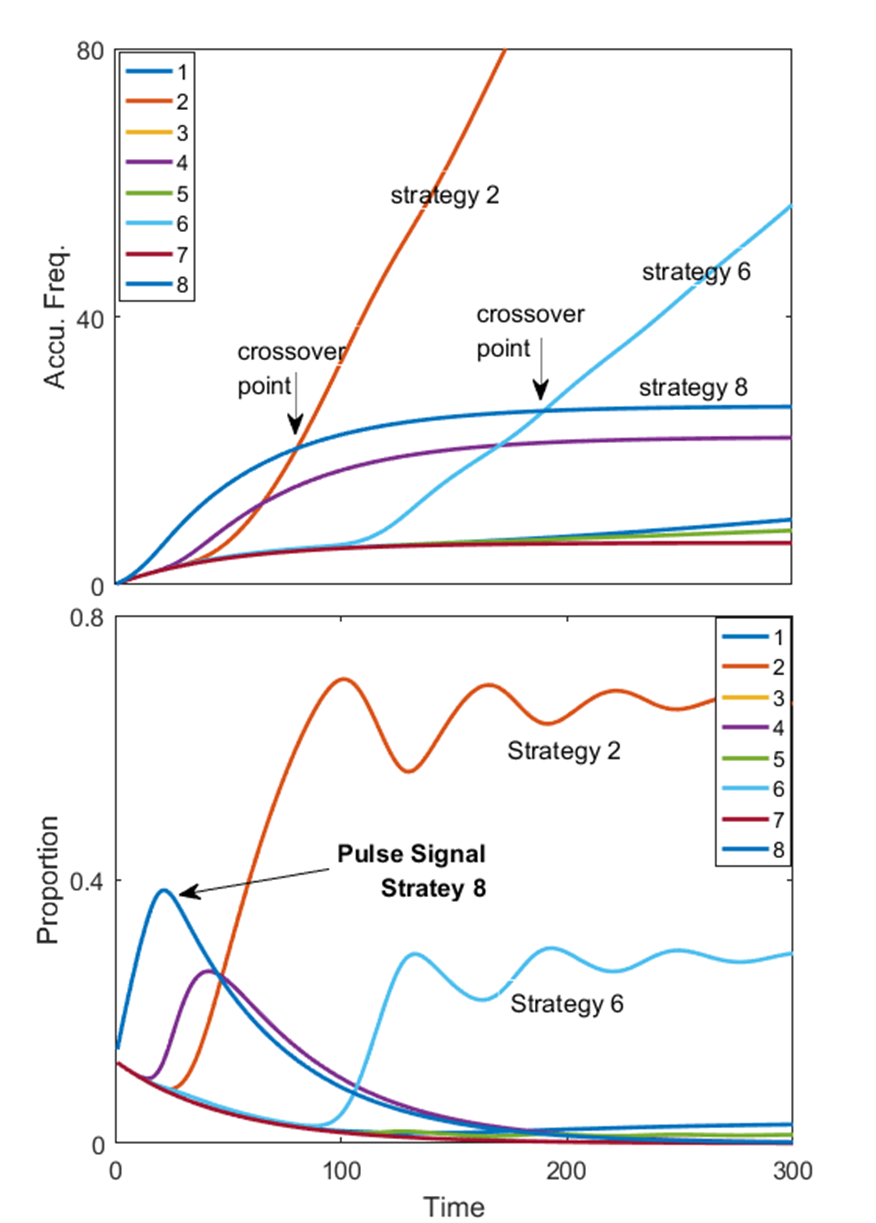}

\caption{\textbf{Conceptual figure for the pulse signal and the crossover point during a space collapse process.} (a) Illustrates a pulse of  strategy 8 in the strategy proportion time series $\rho(x_8,t)$, which dominates when $t \in [0, 30]$  and is eliminated (at $t \approx 130$). (b) Illustrates a crossover point $\chi([x_2,x_8], 70)$ (see the red arrows) in the accumulated curves $\varrho(x_2,t)$ and $\varrho(x_8,t)$.}
\label{fig:Concept_cross_pulse}
\end{figure}

\textbf{Conceptual Example} 
Figure \ref{fig:Concept_cross_pulse} offers an example of the pulse signal, accumulated curve and the crossover points. It is derived from the theoretical results of the strategy evolution over time of the X population in treatment A.  

~\\
\textbf{Accumulated curve ($\varrho$)}
~\\
We denote the accumulated usage of strategy $s_k$ over interval 
$[t_0,t_1]$  as
 $\varrho(x_k, t)$, which is 
\begin{equation}\label{eq:accu_cont}
\varrho(s_k, t_0, t_1) = \int_{t_0}^{t_1} \rho(s_k,t)dt,    
\end{equation} 
where $\rho(x_k,t)$ is the proportion of the observed $s_k$ at time $t$. 
In this study, We use discrete time version and set  $ dt = 1$.   
Without additional specification, in this study we set $t_0 = 0$, and the discrete version of the Equation \ref{eq:accu_cont} can be rewrites as 
\begin{equation}\label{eq:accu_disc}
\varrho(s_k, t) = \sum_{t'=0}^{t} \rho(s_k,t'),    
\end{equation} 
which is the \textbf{accumulated curve} in this study.\\
\textbf{Pulse signal}
A pulse signal, or pulse, can be defined following these steps:
\begin{enumerate}
\item 
\textbf{A sample of surplus} in time series, denoted as $\rho^+\big([s_i, s_j],t\big)$, is the proportion difference between two strategies $[s_i, s_j]$ at $t$,   
\begin{equation}\label{eq:surplus}
    \rho^+\big([s_i, s_j],t\big) = \rho(s_j,t) -  \rho(s_i,t).
\end{equation}

\item 
\textbf{A surplus sample set} $\rho^+_*$ refers to  the surplus samples from a given successive time interval $[t_0,t_1]$ (time block),  
\begin{equation}\label{eq:surpluset}
    \rho^+_*\big([s_i, s_j],[t_0,t_1]\big) = \{ \rho^+\big([s_i, s_j],t_k\big), ~ \forall t_k \in [t_0\!\!+\!\!1,t_0\!\!+\!\!2,..., t_1] \}.
\end{equation} 

\item Sample size $\textbf{N}$. For statistical analysis, the sample set size is 
\begin{equation}\label{eq:samplesize}
    \textbf{N}= (t_1 - t_0) \times \text{Number of experiment session} 
\end{equation} 
For example, assuming that a time block 
is set as 10 rounds,  
from the 81$^{st}$ round to 90$^{th}$ round in a 1000 round 
repeated game session, 
the element of the set is 10. 
We have 12 sessions for a given treatment, 
so the  sample size $\textbf{N}$ is 120. 
Under the null hypothesis from classical game theory (i.e., no learning effects and a fixed strategy distribution), the samples are  independent.

\item \textbf{Pulse ($\psi$)} is the total surplus 
of $s_j$ over $s_i$ in a time block $[t_0,t_1]$:  
\begin{equation}\label{eq:Pulse}
     \psi \big([s_i, s_j],[t_0,t_1]\big) = \sum_{t_k=t_0 + 1}^{t_1} \rho^+\big([s_i, s_j],t_k\big);   
\end{equation}   
For example, in the top line in Table \ref{tab:exp_Pulse_significant}, 
the $\psi = 17$. The explanation for this is that: 
\begin{itemize}
\item $X_8 \in \textbf{D}^-$ is a dominated, and $X_2 \in \textbf{D}^+$ is a domination. 
\item There are 120 samples from the X population of treatment A and in the $t \in [11-20]$ rounds
of the $\text{ses} \in [1,12] $ sessions. 
 \item  The surplus observed is the sum over the sample set shown in 
 equation \ref{eq:surpluset}. That is 
 \begin{equation}
\sum_{\text{ses}=1}^{12} \sum_{t=11}^{20}  \big(\rho(X_8, t, \text{ses}) - \rho(X_2, t, \text{ses})\big) = 17
 \end{equation} 
\end{itemize}

\item \textbf{The statistical significance  of a pulse $\psi$} is reported by the \textbf{ttest} over the surplus sample set $\rho^+_*$ shown in Equation \ref{eq:surpluset}. On the sample size \textbf{N} for \textbf{ttest},  see the example following the Equation \ref{eq:samplesize}. 
 
\item \textbf{Strong significant} When $p<0.010$, the statistic result is reported to be `in strongly significant'. When $0.010 \leq p < 0.050$, the statistic result is reported as `in significant` but not as `strongly significant`. 

\end{enumerate}

\textbf{Crossover point: $\chi(x_i,x_j,\tau)$} ~~The definition  of the crossover point depends on the accumulated curve ($\varrho$) shown in Equation \ref{eq:accu_disc}.  Assuming that: 
\begin{equation}\label{eq:tau_0}
    \varrho(s_i,t) - \varrho(s_j,t) = 0
\end{equation}   
the crossover point is defined as $\chi(s_i,s_j,\tau)$ in which 
\begin{itemize}
\item $\tau$ is the solution of the $t$ 
\item $\chi(s_i,s_j,\tau)$ := $\varrho(s_i,t) = \varrho(s_j,t)$
\end{itemize}

\textbf{The two classes of crossover points}
\begin{itemize}
\item $\chi_+$ --- when the paired comparison is between two  ($\textbf{D}^+$), for example, the blue arrows in Figure \ref{fig:exp-X6} - \ref{fig:exp-Y10} are the crossover points $\chi_+$.

\item $\chi_-$ --- when the paired comparison is between the   ($\textbf{D}^-$) and   ($\textbf{D}^+$), for example, the red arrows in Figure \ref{fig:exp-X6} - \ref{fig:exp-Y10} are the crossover points $\chi_-$.

\end{itemize}

 \textbf{The relation between the pulse and the crossover time.}
Based on the results reported in [E32.6] in section \ref{sec:mt_exp_acc}, 
we hypothesize that, if a crossover $\chi_- ([x_i,x_j], \tau)$ 
( $s_j \in \textbf{D}^-$ and $s_i \in \textbf{D}^+$) has larger $\tau$,
$s_j$ is more strongly expected to provide the pulse signal.  
 
Assuming that $\rho(s_i)$ and $\rho(s_j)$ are equal at $t=0$,  if a long run dominated ($s_j \in \textbf{D}^-$) 
has a pulse signal  referring to a long run domination 
($ s_i \in \textbf{D}^+$), that is 
\begin{equation}\label{eq:pulse_cross}
\rho(x_j,t) > \rho(x_i,t) ~~~~~~\forall  t \in [0, T],
\end{equation} 
As a mathematical result, there must be a crossover point in its accumulated curve $\varrho(x_j,\tau)$ referring to $\varrho(x_i,\tau)$ in which $\tau > T$. 
This is one explanation for the hypothesis.

For example, in experimental pulse $\psi^E$ (Table \ref{tab:exp_Pulse_sign3}), 
all the strongest significant pulses belong to the largest $\tau$ in $\chi^E$ (Table \ref{tab:exp_cross}). The 
result reported in [E32.6] refers to this mathematical property.   
 
\subsubsection
{Data}\label{sec:app_collapse_data}
Here, we present the data that support the results shown in  section \ref{sec:mt_exp_acc} in main text.    
\begin{enumerate}
\item 
1.	The experimental pulse signal observed is statistically significant (p<0.05):  see 
 Table \ref{tab:exp_Pulse_significant} 
\item
 The theoretical pulse signal in statistical significant; Listed are the most significant samples (top 18) ordered by the time block surplus values of treatments (A,B,C): 
 see Table \ref{tab:dyn_Pulse_significant}. Note: the time series from the dynamics model are smooth, meaning that more significant pulse signals can be obtained from this model, when compared with those from the highly stochastic human subject experimental time series.
\item 
Experimental crossover point: see 
 Table \ref{tab:exp_cross}
\item 
Theoretical crossover point:  see 
 Table \ref{tab:dyn_cross}
\end{enumerate}

\begin{table}[ht!]
    \centering
    \begin{tabular}{ccccccc}
    \hline
    Treat- & Domin- & Domin- & time   & paired- &   Surplus& Sample  \\
    ments & ated &  ation   &  block  &  ttest($p$) &    & size    \\
          & $s_j$   & $s_i$ &  $[t_0,t_1]$ &        & $\psi^E$  &  \textbf{N}    \\
    \hline
     	 A & $X_8$ & $X_2$ & 11-20 & 0.007 & 17 & 120  \\
     	 A & $X_8$ & $X_6$ & 11-20 & 0.002 & 19 & 120  \\
     	 A & $X_8$ & $X_2$ & 21-30 & 0.011 & 15 & 120  \\
     	 A & $X_4$ & $X_6$ & 41-50 & 0.027 & 13 & 120  \\
     	 A & $X_1$ & $X_6$ & 81-90 & 0.023 & 12 & 120 \\
     	 A & $X_4$ & $X_6$ & 81-90 & 0.007 & 15 & 120 \\ 
     	 C & $X_8$ & $X_5$ & 1-10 & 0.027 & 11 & 120  \\
     	 C & $X_7$ & $X_5$ & 11-20 & 0.049 & 10 & 120  \\
     	 C & $X_8$ & $X_5$ & 11-20 & 0.049 & 10 & 120\\
    \hline
    \end{tabular}
     \caption{Statistically significant experimental pulses ($\psi^E$) ($p$<0.05).}
    \label{tab:exp_Pulse_significant}
\end{table}  

\begin{table}[ht!]
    \centering
    \begin{tabular}{ccccccc}
    \hline 
    Treat- & Domin- & Domin- & time   & paired- &   Surplus& Sample  \\
    ments & ated &  ation   &  block  &  ttest($p$) &    & size    \\
          & $s_j$   & $s_i$ &  $[t_0,t_1]$ &        & $\psi^T$  &  \textbf{N}    \\
    \hline 
     	 A & $X_8$ & $X_6$ & 21-30 & 0.000 & 35.04 & 120 \\
     	 A & $X_8$ & $X_2$ & 21-30 & 0.000 & 34.75 & 120 \\
     	 A & $X_8$ & $X_6$ & 31-40 & 0.000 & 29.97 & 120 \\
     	 A & $X_8$ & $X_2$ & 11-20 & 0.000 & 29.20 & 120 \\
     	 A & $X_8$ & $X_6$ & 11-20 & 0.000 & 29.08 & 120 \\
     	 A & $X_8$ & $X_6$ & 41-50 & 0.000 & 24.47 & 120 \\
     	 A & $X_4$ & $X_6$ & 41-50 & 0.000 & 23.87 & 120 \\
     	 A & $X_8$ & $X_2$ & 31-40 & 0.000 & 23.84 & 120 \\
     	 A & $X_4$ & $X_6$ & 51-60 & 0.000 & 22.01 & 120 \\
     	 A & $X_4$ & $X_6$ & 31-40 & 0.000 & 20.47 & 120 \\
     	 A & $X_8$ & $X_6$ & 51-60 & 0.000 & 19.98 & 120 \\
     	 A & $X_4$ & $X_6$ & 61-70 & 0.000 & 18.84 & 120 \\ \hline
     	 B & $X_7$ & $X_1$ & 21-30 & 0.000 & 18.05 & 120 \\ \hline
     	 C & $X_8$ & $X_5$ & 11-20 & 0.000 & 21.64 & 120 \\
     	 C & $X_8$ & $X_1$ & 11-20 & 0.000 & 21.64 & 120 \\
     	 C & $X_8$ & $X_2$ & 11-20 & 0.000 & 18.75 & 120 \\
     	 C & $X_8$ & $X_5$ & 21-30 & 0.000 & 18.39 & 120 \\
     	 C & $X_8$ & $X_1$ & 21-30 & 0.000 & 18.39 & 120 \\
    \hline
    \end{tabular}
    \caption{ Theoretical pulse ($\psi^T$).  The most strongly expected top 18 samples (twice the $\psi^E$ is significant) ordered by the $\psi^T$ values of all the treatments (A,B,C). Note: the time series from the dynamics model are smooth, meaning that more significant pulse signals can be obtained from this model, when compared with those from the highly stochastic experimental time series.}
    \label{tab:dyn_Pulse_significant}
\end{table}  

\clearpage
 
\begin{table}
    \centering
    \begin{tabular}{c|c|c|c|c|c|c}
       \hline
     		 Treat- & Domin- & Domin- & Crossover   &\multicolumn{3}{c}{Arrow}\\
       \cline{5-7}\xrowht{9pt}
     	  ment &     ation &  ated &  Time ($\tau$) &Color & ID & Fig ID\\
       \hline
       \hline
 	$A$& 	$X_2$&	$X_1$&	47&-	&-	&Fig \ref{fig:exp-X6}\\
 	$A$&	$X_2$&	$X_6$&	47&-	&-	\\
 	$A$&	$X_2$&	$X_7$&	52&-	&-	\\
  	$A$&	$X_2$&	$X_3$&	55&-	&-	\\
	  $A$&	  $X_2$&	$X_4$&	58&-	&-	\\
	  $A$&	$X_2$&	$X_5$&	58&-	&-	\\
	  $A$&	$X_2$&	$X_8$&	67&-	&-	\\
 	$A$&	$X_6$&	$X_1$&	44&-	&-	\\
	  $A$&	$X_6$&	$X_3$&	74&-	&-	\\
	  $A$&	$X_6$&	$X_3$&	79&-	&-	\\
	  $A$&	$X_6$&	$X_7$&	92&	    {red}&	4\\
	  $A$&	$X_6$&	$X_1$&	106&	{red}&	3\\
	  $A$&	$X_6$&	$X_5$&	139&-	&-	\\
	  $A$&	$X_6$&	$X_4$&	169&	{red}&	2\\
	  $A$&	$X_6$&	$X_8$&	186&	{red}&	1\\
       \hline							
	$A$&	$Y_2$&	$Y_4$&	297&	blue&	1&Fig \ref{fig:exp-Y6}\\
       \hline							
       \hline							
	$C$&	$X_2$&	$X_8$&	45&-	&-	&Fig \ref{fig:exp-X10}\\
	$C$&	$X_5$&	$X_6$&	49&-	&-	\\
	$C$&	$X_5$&	$X_6$&	54&-	&-	\\
	$C$&	$X_5$&	$X_3$&	56&-	&-	\\
	$C$&	$X_5$&	$X_4$&	100&	{red}&	3\\
	$C$&	$X_5$&	$X_8$&	138&	{red}&	1\\
							
       \hline							
	$C$&	$Y_2$&	$Y_4$&	74&	blue&	1&Fig \ref{fig:exp-Y10}\\

       \hline
       \hline
    \end{tabular}
    \caption{Experimental crossover point $\chi^E$ at $\tau^E > 40$.
    The arrow columns refer to the arrows in Figure \ref{fig:exp-X6}-\ref{fig:exp-Y10}, based on  the theoretical arrows in Table \ref{tab:dyn_cross}.
    For the definition of $\tau$, see Equation  \ref{eq:tau_0}.}
    \label{tab:exp_cross}
\end{table}

\begin{table}
    \centering
    \begin{tabular}{c|c|c|c|c|c|c}
       \hline
       \hline
     	 Treat- & Domin- & Domin- & Crossover   &\multicolumn{3}{c}{Arrow}\\
       \cline{5-7}\xrowht{9pt}
     	  ment &     ation &  ated &  Time ($\tau$) &Color & ID & Fig ID\\
       \hline
       \hline
		$A$&	$X_2$&	$X_4$&	62&-	&-	&Fig \ref{fig:dyn-X6}\\
	$A$&		$X_2$&	$X_8$&	86&-	&-	\\
	$A$&		$X_6$&	$X_5$&	65&-	&-	\\
	$A$&		$X_6$&	$X_7$&	116&	{red}&	4\\
	$A$&		$X_6$&	$X_1$&	117&	{red}&	3\\
	$A$&		$X_6$&	$X_4$&	177&	{red}&	2\\
	$A$&		$X_6$&	$X_8$&	209&	{red}&	1\\
       \hline							
     	$A$&	$Y_2$&	$Y_1$&	71&-	&-	&Fig \ref{fig:dyn-Y6}\\
     	$A$&	$Y_2$&	$Y_5$&	74&-	&-	\\
     	$A$&	$Y_2$&	$Y_4$&	277&	blue&	1\\
       \hline							
       \hline							
     	$B$&	$X_1$&	$X_3$&	42&-	&-	\\
     	$B$&	$X_1$&	$X_6$&	56&-	&-	\\
     	$B$&	$X_1$&	$X_8$&	69&-	&-	\\
     	$B$&	$X_1$&	$X_7$&	72&-	&-	\\
     	$B$&	$X_1$&	$X_5$&	75&-	&-	\\
     	$B$&	$X_1$&	$X_2$&	84&-	&-	\\
       \hline							
       \hline							
	$C$&	$X_1$&	$X_7$&	52&-	&-	&Fig \ref{fig:dyn-X10}\\
	$C$&	$X_1$&	$X_6$&	66&-	&-	\\
	$C$&	$X_1$&	$X_4$&	77&-	&-	\\
	$C$&	$X_1$&	$X_8$&	82&-	&-	\\
	$C$&	$X_1$&	$X_2$&	126&	blue&	2\\
	$C$&	$X_5$&	$X_7$&	61&-	&-	\\
	$C$&	$X_5$&	$X_6$&	82&-	&-	\\
	$C$&	$X_5$&	$X_4$&	112&	{red}&	3\\
	$C$&	$X_5$&	$X_8$&	144&	{red}&	1\\

 \hline
	$C$ & $Y_2$ & $Y_4$  & 89 & blue & 1 &Fig \ref{fig:dyn-Y10}\\

       \hline
       \hline
    \end{tabular}
    \caption{Theoretical crossover point $\chi^T$ at $\tau^T > 40$. The arrow columns relate to the arrows ($\tau > 89$) in Figure \ref{fig:dyn-X6}-\ref{fig:dyn-Y10}. For the definition of $\tau$, see Equation  \ref{eq:tau_0}.}
    \label{tab:dyn_cross}
\end{table}

\subsection{The game dynamics paradigm and its workflow}
\label{sec:app_paradigm} 
Similarly to the paradigms found in natural science, the game dynamics paradigm is also \textbf{a distinct set of concepts}, including theories, research methods, postulates, and standards for what constitutes legitimate contributions to a field. The reality and accuracy, as well as the completeness and consistency, of this distinct set of concepts must be proven. Key to developing a paradigm is a closed-loop of workflow. The workflow includes the following steps:
\begin{enumerate} 
\item   Describe a game (G) and its playing protocol (P); 
\item  Take a proper dynamics equation system  T (e.g., replicator dynamics or logit dynamics) and specify the parameters of T by (G,P). Then, derive the theoretical observation 
O$^T$(G,P) by solving T(G,P), e.g., the reset point, the eigen system, or by time series analysis among others technologies;
\item Conduct experiments or collect empirical data to measure the observation  O$^E$(G,P)
\item Evaluate the O$^T$(G,P) by empirical observation O$^E$(G,P).  
\item Iterate Step 1 to Step 4 and feedback to step 1 as a loop, by trial and error, to find the best fit between the theory and the empirical system in general.    
\end{enumerate}
This iterated closed-loop workflow is widely applied in various scientific fields, and can constantly improve the  establishment of a paradigm.  In addition to the reality and the accuracy, the paradigm is endorsed by the completeness and the consistency between: 
$$O^T(P,G) ~~\text{vs}~~ O^E(P,G)$$
endorse the paradigm. Social science follows this workflow, as does natural science\cite{dan2016,colin2003,plott2008}.   

\end{appendices}

\clearpage
\part{SI --- Pulse in collapse: a game dynamics experiment}
\setcounter{section}{0}  
\renewcommand{\thesection}{\arabic{section}}

\section{SI --- Game}

\subsection{The $(m,n)$ poker game design}
The experimental design refers to a simplified Von Neumann's poker game, as set out on page 93 in Ref. \cite{binmore2007game}. 
It replaces Von Neumann's numerical cards with a deck of only J, Q and K cards (denoted as 1, 2 and 3 in the main text). It is a dynamic game with incomplete information. We strongly suggest the reader refer to the text book cited, especially the referenced part (near page 93 \cite{binmore2007game}), in order to understand the concept of the game design used in this study.

\paragraph{Game rule}
The game is composed of two stages of betting. The game can be visualized as in Figure   \ref{fig:game_generized_pres}(a).
The top card is dealt to player X, and the second card to player Y. The names on the nodes indicate who makes the decision in the stage, and the outstretched arrow shows different strategies that can be adopted. When the arrow points to a different node, the game moves on to the next stage.

The number at the end of an arrow indicates the payoff to be received by player X; the payoff to be received by player Y is indicated by the opposite number, since this is a zero–sum game. This finishes a round.

In order to observe various phenomena in a collapse process, and verify the    consistency and completeness of the game dynamics paradigm, we extrapolate the original game to  the ($m,n$)-game.

\paragraph{Generalize the game} 
The generalization of the game is the process of adapting the original game shown in Figure  \ref{fig:game_generized_pres}(a) to the generalized game shown in Figure  \ref{fig:game_generized_pres}(e).  
The procedure of generalizing the game includes the following steps, as shown in Figure  \ref{fig:game_generized_pres}(a)-(e). 
\begin{enumerate}
\item 
[(a)] The original game tree. There are six invariant upward-facing branches; 
\item 
[(b)] An invariant branch of the original game tree is selected. 
\item 
[(c)] The location of the parameter is selected. The location is labeled with a square in sub-figure (c).  
\item 
[(d)] The selected branch is parameterized to an $(m, n)$ parameterized branch; 
\item 
[(e)] The $(m, n)$ parameterized game tree is rebuilt in the form of the (m, n) parameterized branch. A full $(m, n)$ game tree is thus obtained.
\end{enumerate}

\subsection{Extensive form presentation}
\begin{figure}[!ht]
    \centering
    \includegraphics[scale=0.55]{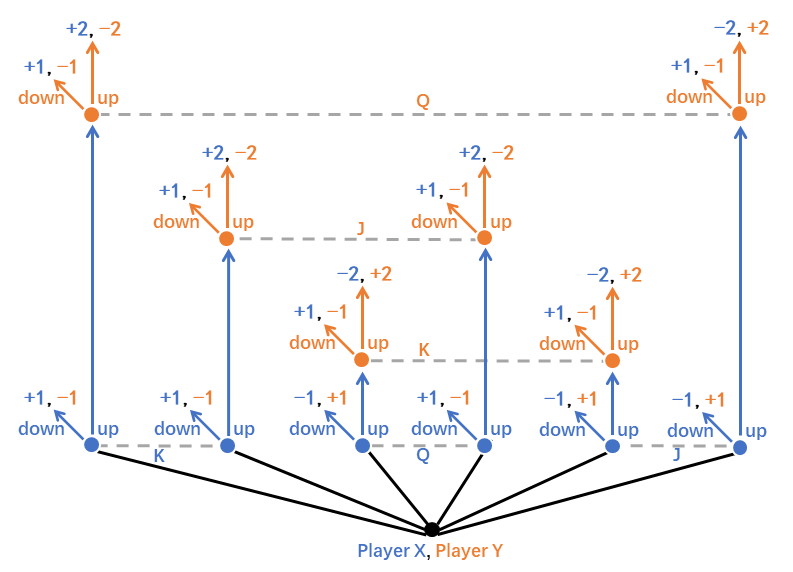}  
    \includegraphics[scale=0.55]{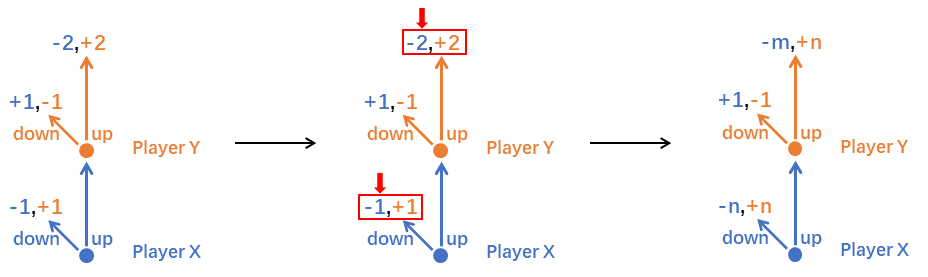} 
    \includegraphics[scale=0.55]{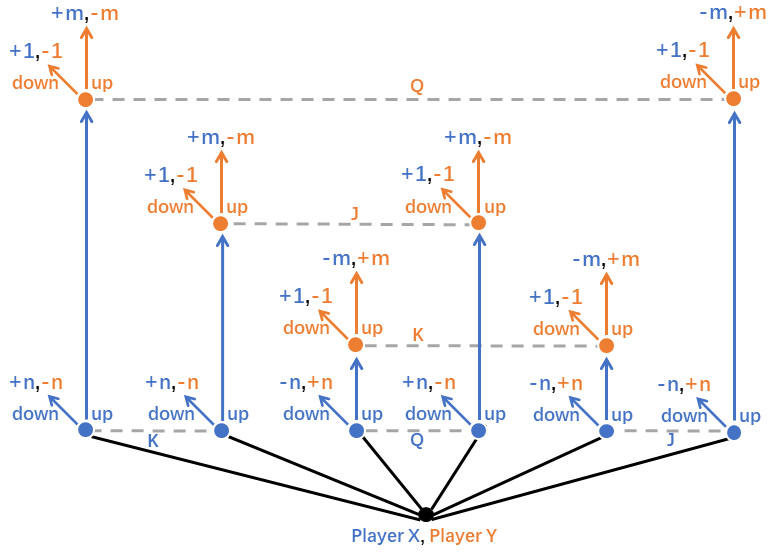} \\
     \caption{The protocol to generalize the von Neumann 3-card poker game to $(m,n)$-game. (a) The original game tree. There are 6 invariant upward-facing branches.  (b) An invariant branch of the original game tree. (c) The labels of the locations to be parameterized. (d) A ($m,n$)-parameterized branch. (e) The ($m,n$)-game tree.}
    \label{fig:game_generized_pres}
\end{figure}

The extensive form is shown in Figure \ref{fig:game_generized_pres}(e). 

 %
 %
\subsection{Normal form presentation}
This game is a dynamic game that lacks complete information. The strategy of each player is illustrated by eight elements \cite{binmore2007game}, 
$$\{s_1,s_2,s_3\} \text{~~and~~} s_1, s_2,s_3  \in \text{[down, up]}:=[0,1]$$ 
where $s_i$  indicates the strategy assigned when a player receives card $i \in [1,2,3]:=[J,Q,K]$. We assign 
\begin{eqnarray}
   &&[0,0,0],[0,0,1],[0,1,0],[0,1,1],[1,0,0],[1,0,1],[1,1,0],[1,1,1] \\
   &\longrightarrow& \{1,2,3,4,5,6,7,8\}
\end{eqnarray} 
The result of this is the following:  
\begin{itemize}
\item The strategy set of Player X is $(X_1, X_2, X_3, ..., X_8)$
\item The strategy set of Player Y is $(Y_1, Y_2, Y_3, ..., Y_8)$
\end{itemize}
From this point, the game can be presented in a bi-matrix form. The payoff matrices of the games, where $(m,n) = (2,1),~ (3,2),  ~\text{and}~ (4, 2)$  are shown in the main text, and replicated in the SI in Table \ref{tab:(2,1)_matrix_2}, Table \ref{tab:(3,2)_matrix_2}, and Table \ref{tab:(4,2)_matrix_2}.

\subsection{$(m,n)$: A Stress Test for Dynamic Theory}

The design of the three parameter sets is characterised by its systematic variation of the static structure of a baseline game through parameterisation---altering equilibrium types, strategy elimination orders, and the resulting dynamic patterns (such as cycle strength and pulse signals) in a theoretically predictable manner. 

This creates a controlled experimental platform in which the three treatments serve as distinct test cases for the dynamic theory.

In essence, the design functions as a \emph{stress test} for the game dynamics paradigm. The theory must account for whether the predicted dynamic phenomena---most notably, the pulse signals during collapse---emerge as expected across qualitatively different static environments (Treatments A, B, and C). 

This multi-treatment design strengthens the empirical conclusions considerably: it demonstrates not only that the theory works in one specific setting, but also that it exhibits \textbf{consistency} and \textbf{completeness} across a range of theoretically meaningful variations.

The fact that the observed pulse patterns (A: strong, B: none, C: weak) align with the theoretical predictions across all three treatments provides compelling evidence that the dynamic framework captures the underlying regularities of the collapse process in a coherent and generalisable manner.

\begin{table}[ht!]
\centering
\label{tab:treatment_design}
\begin{tabular}{l|p{0.2\linewidth}|p{0.2\linewidth}|p{0.2\linewidth}}
\toprule
\textbf{Attribute} & \textbf{Treatment A} & \textbf{Treatment B} & \textbf{Treatment C} \\
\midrule
Parameters & \((2,1)\) & \((3,2)\) & \((4,2)\) \\
Equilibrium type & Fully mixed NE & Pure-strategy NE & Continuum NE \\
IEDS elimination &  elimination (simple) & Cross-elimination (complex) & Multiple-step elimination (moderate) \\
Cycle expectation & Strong & None & Weak \\
Pulse expectation & Strong pulse & No pulse & Weak pulse \\
\bottomrule
\end{tabular}
\caption{Three-dimensional cross-design of the experimental treatments}
\end{table}


\section{SI --- Experiment}\label{sec:app_experiment}

\subsection{Protocol design and considerations}
The main parameters are shown in Table 1 in the main text. 
The experiment protocol is as follows: 
\begin{itemize}
\item 
Treatment: There are three treatments, namely the (2,1) game as treatment A, (3,2) game as treatment B, and (4,2) game as treatment C.   
\item 
Sessions: Each treatment involves  12 sessions. 
\item 
Players: Each session has two players.  
There are a total of 72 student subjects involved in all the experiments. 
\item 
Matching: When there are two players in the game, this is a fixed pair match.
\item 
Rounds: The game is repeated for 1000 rounds, known by each player.  
\item 
Role: Each player knows that their role, which is either 'X' or 'Y'.  
\item 
Strategies: Each player knows they have eight strategies to choose from in each round.
\item 
Information 1: The participants do not know the payoff matrix, and they make their decisions by trial and error.
\item 
Information 2: The feedback of each round includes the following: 
    \begin{itemize}
    \item –	The point added to the player’s points total accumulated from the previous rounds;
    \item –	The strategy they used in the previous round, as well as that used by their opponent.  
    \end{itemize} 
\item 
Information 3: The participants know the game comprises a fixed two-player pair
\item 
Information 4: No communication is allowed.  
\item 
Payment:  The rank is obtained by two aspects: 
    \begin{itemize}
    \item The points the players accumulated throughout the game;
    \item There are always several players (N) playing as “X” or “Y” within the same experiment. These players form a share group;
    \item  The rank is given in the order of high to low regarding the accumulated points in the share group; 
    \item –	The participants are paid in relation to their rank;
    \item The payment to players is determined by an arithmetic sequence, in which the highest pay is CNY 175 and the lowest is CNY 25;
    \item –	In addition, each is paid CNY 20 for attending the session; 
    \item The average of the payment to each  individual is a constant, CNY 120, within each share group.
    \end{itemize} 
\end{itemize}

\paragraph{The design considerations}
The background of game dynamics theory is the population game model (evolutionary game theory), which comprises an infinite number of players in continuous time. Usually, when studying game dynamics, it is necessary to imitate the environment required by evolutionary game theory. That is the main protocol, besides a long period of continuous game playing, using many players, and random matching or mean matching. These are the conventions of evolutionary game experimental research expected \cite{plott2008,dan2016}.

However, in this study, we use the two-player fixed paired protocol.  This is explained below.

\begin{figure*}[!ht]
    \begin{subfigure}[b]{0.32\textwidth}
     \begin{center}
     \caption{Game-1}
       \includegraphics[scale=0.15]{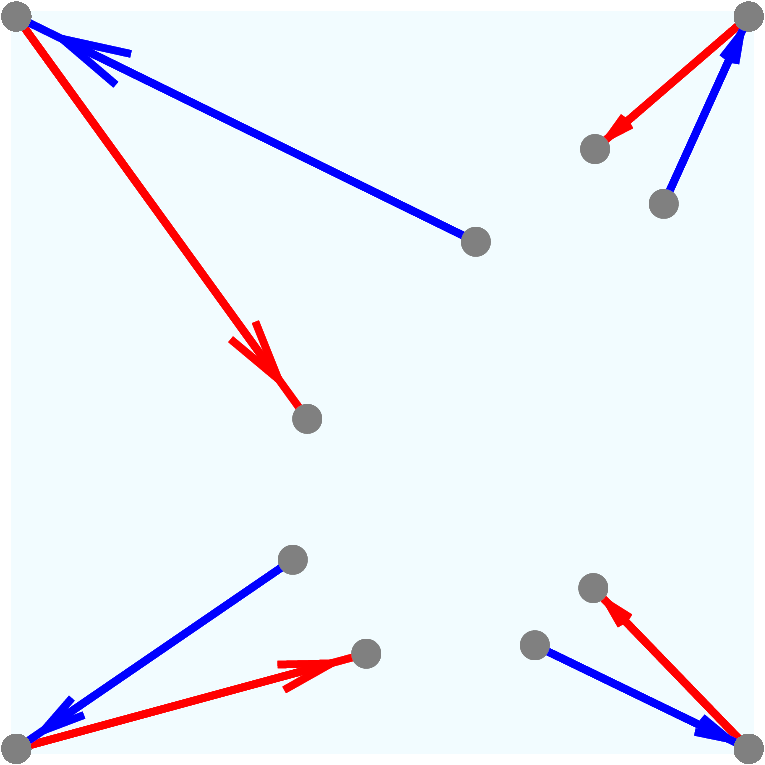}
       \end{center}
    \end{subfigure}
    \begin{subfigure}[b]{0.32\textwidth}
    \begin{center}
    \caption{Game-9}
       \includegraphics[scale=0.15]{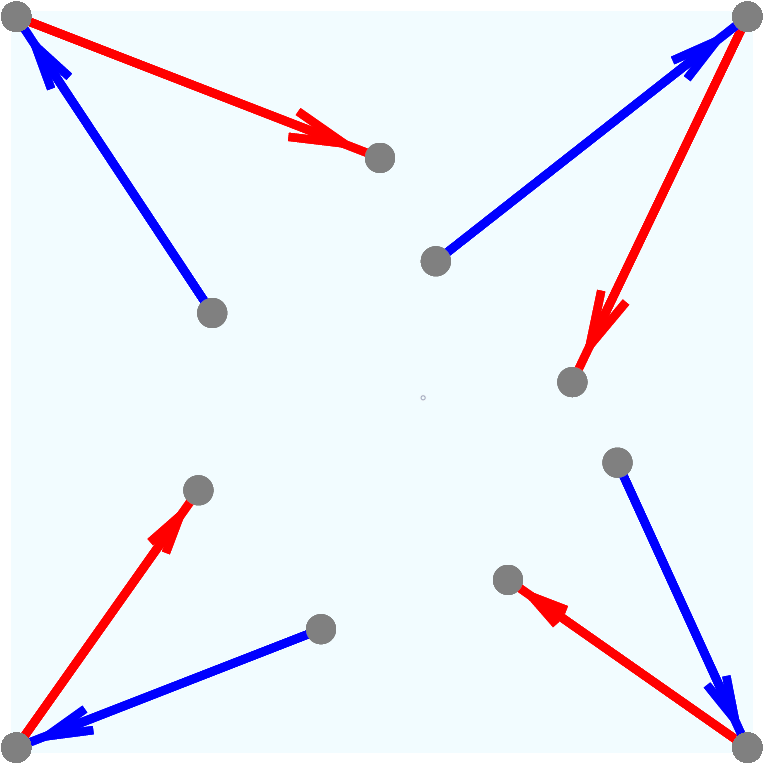}
    \end{center}
    \end{subfigure}
    \begin{subfigure}[b]{0.32\textwidth}
    \begin{center}
    \caption{Game-11}
     \includegraphics[scale=0.15]{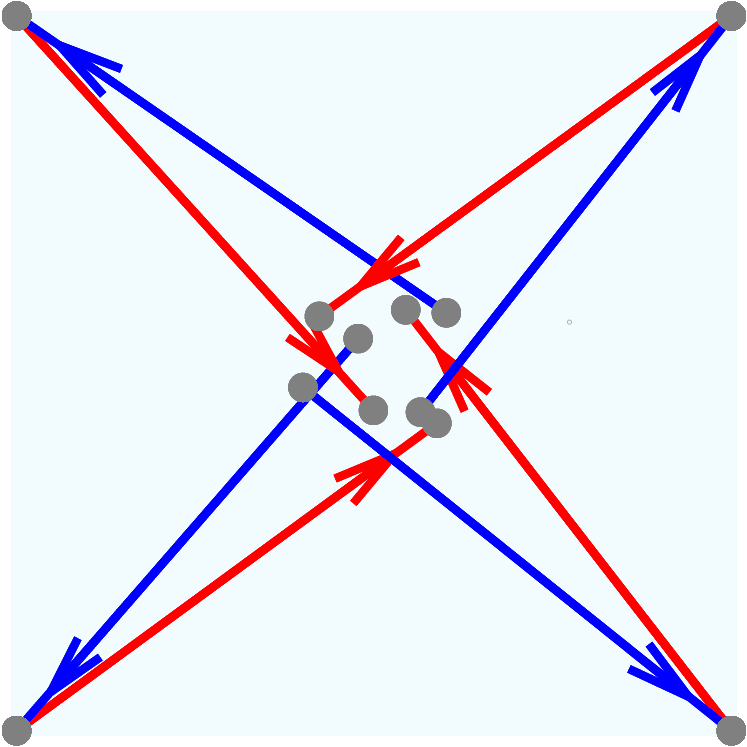}
     \end{center}
    \end{subfigure}
    \caption{The sub-figures are the schematic diagrams for aggregated backward (forward) transitions in blue arrows (in red arrows). These figures come from \cite{zhijian2012}, the data
    for which comes from \cite{erev2007}. }
    \label{fig:app_2012_vector}
\end{figure*}

Besides the cost (the payment given to student subjects and the time required of them, among other things) of the experiments, aspects to be considered when using the fixed-pair protocol include ensuring that this protocol is sufficient to test the predictable temporary deviation from equilibrium (PDTE). The references are as follows: 
\begin{itemize}
\item Human behavior in a fixed-pair, long-running game can be used to obtain PTDE. A representative example is O’Neill’s (1987) game \cite{ONeill1987}, which is firstly to determine the veracity and accuracy of the Nash equilibrium concept. In this game, the conditions dependent on human behavior are extensively studied \cite{2021Qinmei,Brown90,Binmore2001Minimax,zhijian2022,kandori2021}.  The protocol is a fixed-pair, two-person repeated game. 

However, in O'Neill 1987 data  \cite{ONeill1987}, the adjustment dynamics has been carefully analyzed  \cite{kandori2021}, and \cite{2021Qinmei}\cite{zhijian2020} have shown the eigencycle from the eigenvector in eigensystem of the replicator dynamics system. The eigencycle is a PTDE. 
This encourages the use of the fixed paired protocol.

\item A report published in 2012 \cite{zhijian2012} has shown that, in a fixed pair long running game, a cyclic dynamics pattern arises. This finding is based on the random parameter fixed pair two-player matrix games published in 2007 \cite{erev2007}. The transition vectors are significantly cyclic. Figure \ref{fig:app_2012_vector} gives an example of   the cyclic vectors, which also represents a PTDE.   
\end{itemize}

\subsection{Experiment procedure} 

\paragraph{Procedure}
The experiment was approved by the Experimental Social Science Laboratory of Zhejiang University and performed at Zhejiang University from August to September, 2018. The author confirms that this experiment was performed in accordance with the approved social experiment guidelines and regulations, as given in \cite{zhijian2014rps,zhijian2022,zhijian2013,zhijian2016}.

A total of 72 undergraduate students of Zhejiang University volunteered as the human subjects in this experiment. These students were openly recruited through a web registration system. Informed consent was obtained from all the participating subjects.

Each set of parameters was assigned 24 subjects. These subjects were divided into 12 groups of two, who played the game in a fixed-pair arrangement.

In a 1000 round experiment, a player can observe the strategy used ($X_1, X_2, ..., X_8$) by player X or ($Y_1, Y_2, ..., Y_8$) by player Y. 
However, each player can only observe their own payoffs, and the strategy used by their opponent.

The subjects were not provided with the payoff matrix, nor did we tell the subject that they were playing a zero–sum game, and no discussion was allowed. These considerations were set due to the effects of common knowledge \cite{samuelson1992}. 

The experiment sessions took 2 hours on average. The payoff of a subject, including their CNY 20 attendance fee and their rank fee, was CNY 120 on average. The experimental organization is shown in Table  \ref{tab:exp_org}. 
\begin{table} [!ht]
    \centering
    \begin{tabular}{|c|ccccc|}
    \hline
parameter&	SessionID&	Date&	Subjects&	Period & Share group\\
\hline
$A$&	80826z11&	20180826&	4&	1000& 1\\
(m,n)=(2,1)&	80826z12&	20180826&	4&	1000& 1\\
&	80901z21&	20180901&	4&	1000& 2\\
&	80901z22&	20180901&	4&	1000& 2\\
&	80901z23&	20180901&	2&	1000& 2\\
&	80902z21&	20180902&	4&	1000& 3\\
&	80902z22&	20180902&	2&	1000& 3\\
\hline
\hline
parameter&	SessionID&	Date&	Subjects&	Period& \\
\hline
$B$&	80901z11&	20180901&	4&	1000& 4\\
(m,n)=(3,2)&	80901z12&	20180901&	4&	1000& 4\\
&	80901z31&	20180901&	4&	1000&5 \\
&	80901z32&	20180901&	4&	1000&5 \\
&	80901z33&	20180901&	4&	1000& 5\\
&	80902z31&	20180902&	4&	1000& 6\\
\hline
\hline
parameter&	SessionID&	Date&	Subjects&	Period& \\
\hline
$C$&	80826z81&	20180826&	4&	1000& 7\\
(m,n)=(4,2)&	80826z83&	20180826&	6&	1000& 7\\
&	80902z11&	20180902&	4&	1000& 8\\
&	80902z12&	20180902&	4&	1000& 8\\
&	80902z32&	20180902&	4&	1000& 6\\
&	80902z33&	20180902&	2&	1000& 6\\
\hline
    
    \end{tabular}   
        \caption{The experiment organization}
    \label{tab:exp_org}
\end{table}

\section{SI --- Statics}

\subsection{IEDS order and survive strategy}\label{sec:app_IDES_servive}

This sub-section regards the analysis of the IEDS orders of the three treatments. The results relating to the main text are as follows:

\begin{itemize}
\item The main results are given in Table 1 in the main text as 'IEDS Round' and the 'Survive strategy'.
\item IEDS Round indicates   which strategy might yield  the pulse signal in collapse, see TS3 in the main text.  
\item Surviving strategy indicates where the cycle appears in a game's eigencycle spectrum, see TS2 in the main text. 
\end{itemize}

\paragraph{Treatment A --- ($m,n$)=(2,1)}
The order of elimination in the dominated strategy is as follows:
\begin{enumerate}
\item [Step 1] (As shown in Table \ref{tab:(2,1)matrix_step1}) 
    \begin{enumerate}
    \item For X population, \\
    $X_1$ was eliminated by $X_2$.\\
    $X_3$ was eliminated by $X_4$.\\
    $X_5$ was eliminated by $X_6$.\\
    $X_7$ was eliminated by $X_8$.
    
    \item For Y population, \\
    $Y_1,Y_3,Y_5,Y_6$, and $Y_7$ were  eliminated by $Y_2$.\\
    $Y_8$ was eliminated by $Y_4$.
 
    \end{enumerate}

\item [Step 2] (As shown in Table \ref{tab:(2,1)matrix_step2})  
$X_4$ and $X_8$ were eliminated by $X_2$.

\item [Result] -- Table \ref{tab:(2,1)matrix_step3} is the remaining strategy game after IEDS.

\end{enumerate}

\paragraph{Treatment B --- ($m,n$)=(3,2)}
The order of the eliminating dominated strategy is as follows:
\begin{enumerate}
\item [Step 1] (As shown in Table \ref{tab:(3,2)matrix_step1}) 
    \begin{enumerate}
    \item For X population, $X_3$ was eliminated by $X_6$.
    
    \item For Y population, $Y_1,Y_3,Y_5,Y_6$ and $Y_7$ were eliminated by $Y_2$. \\$Y_8$ was eliminated by $Y_4$.
 
    \end{enumerate}

\item [Step 2]
(As shown in Table \ref{tab:(3,2)matrix_step2}) $X_2,X_4,X_6,X_7$ and $X_8$ was eliminated by $X_1$.

\item [Step 3] (As shown in Table \ref{tab:(3,2)matrix_step3})
 $Y_2$ was eliminated by $Y_4$.

\item [Step 4] (As shown in Table \ref{tab:(3,2)matrix_step4})
 $X_5$ was eliminated by $X_1$.

\item [Result] -- Table \ref{tab:(3,2)matrix_step5} is the remaining strategy game after IEDS. 

\end{enumerate}

\paragraph{Treatment C --- ($m,n$)=(4,2)}
The order of the eliminating dominated strategy is as follows:
\begin{enumerate}
\item [Step 1] (As shown in Table \ref{tab:(4,2)matrix_step1})  
    \begin{enumerate}
    \item For X population,$X_3$ was eliminated by $X_6$.
    
    \item For Y population,$Y_1,Y_3,Y_5,Y_6$ and $Y_7$ were  eliminated by $Y_2$.\\$Y_8$ was eliminated by $Y_4$.
 
    \end{enumerate}

\item [Step 2] (As shown in Table \ref{tab:(4,2)matrix_step2})
$X_4, X_6, X_7$ and $X_8$ were eliminated by $X_1$.

\item [Result] -- Table \ref{tab:(4,2)matrix_step3} is the remaining strategy game after IEDS. 

\end{enumerate}

\begin{table}[!ht]
\begin{center}
\begin{tabular}{|c|c|c|c|c|c|c|c|c|}
  \hline
$(\textbf{m,n})\!=\!(2,1)$	&	$Y_1$ &	$Y_2$ &	$Y_3$ &	$Y_4$ &	$Y_5$ &	$Y_6$ &	$Y_7$ &	$Y_8$ \\ 
  \hline
$X_1$	&	0,~~0&0,~~0&0,~~0&0,~~0&0,~~0&0,~~0&0,~~0&0,~~0\\						
$X_2$	&	0,~~0&0,~~0&1,~~-1&1,~~-1&1,~~-1&1,~~-1&2,~~-2&2,~~-2\\	 
$X_3$	&	2,~~-2&-1,~~1&2,~~-2&-1,~~1&3,~~-3&0,~~0&3,~~-3&0,~~0\\	 
$X_4$	&	2,~~-2&-1,~~1&3,~~-3&0,~~0&4,~~-4&1,~~-1&5,~~-5&2,~~-2\\ 
$X_5$	&	4,~~-4&1,~~-1&1,~~-1&-2,~~2&4,~~-4&1,~~-1&1,~~-1&-2,~~2\\ 
$X_6$	&	4,~~-4&1,~~-1&2,~~-2&-1,~~1&5,~~-5&2,~~-2&3,~~-3&0,~~0\\ 
$X_7$	&	6,~~-6&0,~~0&3,~~-3&-3,~~3&7,~~-7&1,~~-1&4,~~-4&-2,~~2\\ 
$X_8$	&	6,~~-6&0,~~0&4,~~-4&-2,~~2&~~8,~~-8~~&2,~~-2&	~~6,~~-6~~&0,~~0\\	 
\hline
\end{tabular}
\end{center} 
\caption{The payoff matrix of the  (2, 1) game of treatment A. }
    \label{tab:(2,1)_matrix_2}
\end{table}

\begin{table}[!ht]
\begin{center}
\begin{tabular}{|c|c|c|c|c|c|c|c|c|}
  \hline
$(\textbf{m,n})\!=\!(2,1)$	&	\sout{$Y_1$} &	$Y_2$ &	\sout{$Y_3$ }&	$Y_4$ &	\sout{$Y_5$} &	\sout{$Y_6$} &	\sout{$Y_7$} &	\sout{$Y_8$} \\ 
  \hline
$X_1$	&	0,~~0&0,~~0&0,~~0&0,~~0&0,~~0&0,~~0&0,~~0&0,~~0\\	[-1.7ex]\hline\noalign{\vspace{\dimexpr 1.87ex-\doublerulesep}}			
$X_2$	&	\sout{0,~~0}&0,~~0&\sout{1,~~-1}&1,~~-1&\sout{1,~~-1}&\sout{1,~~-1}&\sout{2,~~-2}&\sout{2,~~-2}\\	 
$X_3$	&	2,~~-2&-1,~~1&2,~~-2&-1,~~1&3,~~-3&0,~~0&3,~~-3&0,~~0\\[-1.7ex]\hline\noalign{\vspace{\dimexpr 1.87ex-\doublerulesep}}	 
$X_4$	&	\sout{2,~~-2}&-1,~~1&\sout{3,~~-3}&0,~~0&\sout{4,~~-4}&\sout{1,~~-1}&\sout{5,~~-5}&\sout{2,~~-2}\\ 
$X_5$&	4,~~-4&1,~~-1&1,~~-1&-2,~~2&4,~~-4&1,~~-1&1,~~-1&-2,~~2\\ [-1.7ex]\hline\noalign{\vspace{\dimexpr 1.87ex-\doublerulesep}}
$X_6$	&	\sout{4,~~-4}&1,~~-1&\sout{2,~~-2}&-1,~~1&\sout{5,~~-5}&\sout{2,~~-2}&\sout{3,~~-3}&\sout{0,~~0}\\ 
$X_7$	&	6,~~-6&0,~~0&3,~~-3&-3,~~3&7,~~-7&1,~~-1&4,~~-4&-2,~~2\\ [-1.7ex]\hline\noalign{\vspace{\dimexpr 1.87ex-\doublerulesep}}
$X_8$	&	\sout{6,~~-6}&0,~~0&\sout{4,~~-4}&-2,~~2&\sout{~~8,~~-8~~}&\sout{2,~~-2}&	\sout{~~6,~~-6~~}&\sout{0,~~0}\\	 
\hline
\end{tabular}
\end{center} 
 \caption{
 $X_1$ was eliminated by $X_2$.
 $X_3$ was eliminated by $X_4$.
 $X_5$ was eliminated by $X_6$.
 $X_7$ was eliminated by $X_8$.
 $Y_1,Y_3,Y_5,Y_6$ and $Y_7$ were eliminated by $Y_2$.
 $Y_8$ was eliminated by $Y_4$.
 }
 \label{tab:(2,1)matrix_step1}
\end{table}

\begin{table}[!ht]
\begin{center}
\begin{tabular}{|c|c|c|c|c|c|c|c|c|}
  \hline
$(\textbf{m,n})\!=\!(2,1)$	&	 	$Y_2$ &		$Y_4$  \\ 
  \hline			
$X_2$	&	0,~~0&1,~~-1\\	  
$X_4$	&	-1,~~1&0,~~0\\[-1.7ex]\hline\noalign{\vspace{\dimexpr 1.87ex-\doublerulesep}} 
$X_6$	&	1,~~-1&-1,~~1\\ 
$X_8$	&	0,~~0&-2,~~2\\	[-1.7ex]\hline\noalign{\vspace{\dimexpr 1.87ex-\doublerulesep}} 
\hline
\end{tabular}
\end{center} 
 \caption{$X_4$ and $X_8$ were eliminated by $X_2$.}
  \label{tab:(2,1)matrix_step2}
\end{table}

\begin{table}[!ht]
\begin{center}
\begin{tabular}{|c|c|c|c|c|c|c|c|c|}
  \hline
$(\textbf{m,n})\!=\!(2,1)$	&	 	$Y_2$ &		$Y_4$  \\ 
  \hline			
$X_2$	&	0,~~0&1,~~-1\\	  
$X_6$	&	1,~~-1&-1,~~1\\ 
\hline
\end{tabular}
\end{center} 
 \caption{$X_2,X_6,Y_2$ and $Y_4$ are the remain strategies.}
 \label{tab:(2,1)matrix_step3}
\end{table}

\begin{table}[!ht] 
\begin{center}
\begin{tabular}{|c|c|c|c|c|c|c|c|c|}
  \hline
$(\textbf{m,n})\!=\!(3,2)$	&	$Y_1$ &	$Y_2$ &	$Y_3$ &	$Y_4$ &	$Y_5$ &	$Y_6$ &	$Y_7$ &	$Y_8$ \\
\hline
 $X_1$&	0,~~0&	0,~~0&	0,~~0&	0,~~0&	0,~~0&	0,~~0&	0,~~0&	0,~~0\\
$X_2$&	-2,~~2&	-2,~~2&	0,~~0&	0,~~0&	0,~~0&	0,~~0&	2,~~-2&	2,~~-2\\
$X_3$&	2,~~-2&	-2,~~2&	2,~~-2&	-2,~~2&	4,~~-4&	0,~~0&	4,~~-4&	0,~~0\\
$X_4$&	0,~~0&	-4,~~4&	2,~~-2&	-2,~~2&	4,~~-4&	0,~~0&	6,~~-6&	2,~~-2\\
$X_5$&	6,~~-6&	2,~~-2&	2,~~-2&	-2,~~2&	6,~~-6&	2,~~-2&	2,~~-2&	-2,~~2\\
$X_6$&	4,~~-4&	0,~~0&	2,~~-2&	-2,~~2&	6,~~-6&	2,~~-2&	4,~~-4&	0,~~0\\
$X_7$&	8,~~-8&	0,~~0&	4,~~-4&	-4,~~4&	10,~~-10&	2,~~-2&	6,~~-6&	-2,~~2\\
$X_8$&	6,~~-6&	-2,~~2&	4,~~-4&	-4,~~4&	10,~~-10&	2,~~-2&	~~8,~~-8~~&	0,~~0\\
 \hline
\end{tabular}
\end{center}
    \caption{The payoff matrix of the (3, 2)   game of treatment B.}
    \label{tab:(3,2)_matrix_2}
\end{table}

\begin{table}[!ht]
\begin{center}
\begin{tabular}{|c|c|c|c|c|c|c|c|c|}
  \hline
$(\textbf{m,n})\!=\!(3,2)$	&	\sout{$Y_1$} &	$Y_2$ &	\sout{$Y_3$} &	$Y_4$ &	\sout{$Y_5$} &	\sout{$Y_6$} &	\sout{$Y_7$} &	\sout{$Y_8$} \\
\hline
 $X_1$&	\sout{0,~~0}&	0,~~0&	\sout{0,~~0}&	0,~~0&	\sout{0,~~0}&	\sout{0,~~0}&	\sout{0,~~0}&	\sout{0,~~0}\\
$X_2$&	\sout{-2,~~2}&	-2,~~2&	\sout{0,~~0}&	0,~~0&	\sout{0,~~0}&	\sout{0,~~0}&	\sout{2,~~-2}&	\sout{2,~~-2}\\
$X_3$&	2,~~-2&	-2,~~2&	2,~~-2&	-2,~~2&	4,~~-4&	0,~~0&	4,~~-4&	0,~~0\\[-1.7ex]\hline\noalign{\vspace{\dimexpr 1.87ex-\doublerulesep}}
$X_4$&	\sout{0,~~0}&	-4,~~4&\sout{	2,~~-2}&	-2,~~2&	\sout{4,~~-4}&	\sout{0,~~0}&	\sout{6,~~-6}&	\sout{2,~~-2}\\
$X_5$&	\sout{6,~~-6}&	2,~~-2&	\sout{2,~~-2}&	-2,~~2&	\sout{6,~~-6}&	\sout{2,~~-2}&	\sout{2,~~-2}&	\sout{-2,~~2}\\
$X_6$&	\sout{4,~~-4}&	0,~~0&	\sout{2,~~-2}&	-2,~~2&	\sout{6,~~-6}&	\sout{2,~~-2}&	\sout{4,~~-4}&	\sout{0,~~0}\\
$X_7$&	\sout{8,~~-8}&	0,~~0&	\sout{4,~~-4}&	-4,~~4&\sout{	10,~~-10}&	\sout{2,~~-2}&	\sout{6,~~-6}&	\sout{-2,~~2}\\
$X_8$&	\sout{6,~~-6}&	-2,~~2&\sout{	4,~~-4}&	-4,~~4&\sout{	10,~~-10}&	\sout{2,~~-2}&	\sout{~~8,~~-8~~}&	\sout{0,~~0}\\
 \hline
 \end{tabular}
 \end{center}
 \caption{$X_3$ was eliminated by $X_6$. $Y_1,Y_3,Y_5,Y_6$ and $Y_7$ were eliminated by $Y_2$. $Y_8$ was eliminated by $Y_4$.}
  \label{tab:(3,2)matrix_step1}
\end{table}

\begin{table}[!ht]
\begin{center}
\begin{tabular}{|c|c|c|}
  \hline
$(\textbf{m,n})\!=\!(3,2)$	&	$Y_2$ &	$Y_4$  \\
\hline
 $X_1$&	0,~~0&	0,~~0\\
$X_2$&	-2,~~2&	0,~~0\\[-1.7ex]\hline\noalign{\vspace{\dimexpr 1.87ex-\doublerulesep}}
$X_4$&	-4,~~4&	-2,~~2\\[-1.7ex]\hline\noalign{\vspace{\dimexpr 1.87ex-\doublerulesep}}
$X_5$&	2,~~-2&	-2,~~2\\
$X_6$&	0,~~0&	-2,~~2\\[-1.7ex]\hline\noalign{\vspace{\dimexpr 1.87ex-\doublerulesep}}
$X_7$&	0,~~0&	-4,~~4\\[-1.7ex]\hline\noalign{\vspace{\dimexpr 1.87ex-\doublerulesep}}
$X_8$&	-2,~~2&	-4,~~4\\[-1.7ex]\hline\noalign{\vspace{\dimexpr 1.87ex-\doublerulesep}}
 \hline
 \end{tabular}
 \end{center}
 \caption{$X_2,X_4,X_6,X_7$ and $X_8$ was eliminated by $X_1$.}
 \label{tab:(3,2)matrix_step2}
\end{table}

\begin{table}
\begin{center}
\begin{tabular}{|c|c|c|}
  \hline
$(\textbf{m,n})\!=\!(3,2)$	&	\sout{$Y_2$} &	$Y_4$  \\
\hline
 $X_1$&	\sout{0,~~0}&	0,~~0\\
$X_5$&	\sout{2,~~-2}&	-2,~~2\\
 \hline
 \end{tabular}
 \end{center}
 \caption{$Y_2$ was eliminated by $Y_4$.}
 \label{tab:(3,2)matrix_step3}
\end{table}

\begin{table}[!ht]
\begin{center}
\begin{tabular}{|c|c|}
  \hline
$(\textbf{m,n})\!=\!(3,2)$	 &	$Y_4$  \\
\hline
 $X_1$&	0,~~0\\
$X_5$&	-2,~~2\\[-1.7ex]\hline\noalign{\vspace{\dimexpr 1.87ex-\doublerulesep}}
 \hline
 \end{tabular}
 \end{center}
 \caption{$X_5$ was eliminated by $X_1$.}
 \label{tab:(3,2)matrix_step4}
\end{table}

\begin{table}[!ht]
\begin{center}
\begin{tabular}{|c|c|}
  \hline
$(\textbf{m,n})\!=\!(3,2)$	 &	$Y_4$  \\
\hline
 $X_1$&	0,~~0\\
 \hline
 \end{tabular}
 \end{center}
 \caption{$X_1$ and $Y_4$  are the remain strategies.}
 \label{tab:(3,2)matrix_step5}
\end{table}

\begin{table}[!ht]
\begin{center}
\begin{tabular}{|c|c|c|c|c|c|c|c|c|}
  \hline
$(\textbf{m,n})\!=\!(4,2)$	&	$Y_1$ &	$Y_2$ &	$Y_3$ &	$Y_4$ &	$Y_5$ &	$Y_6$ &	$Y_7$ &	$Y_8$ \\
\hline
$X_1$	&	0,~~0&*0,~~0&0,~~0&*0,~~0&0,~~0&0,~~0&0,~~0&0,~~0\\						
$X_2$	&	-2,~~2&*-2,~~2&1,~~-1&1,~~-1&1,~~-1&1,~~-1&4,~~-4&4,~~-4\\					
$X_3$	&	2,~~-2&-3,~~3&2,~~-2&-3,~~3&5,~~-5&0,~~0&5,~~-5&0,~~0\\		
$X_4$	&	0,~~0&-5,~~5&3,~~-3&-2,~~2&6,~~-6&1,~~-1&9,~~-9&4,~~-4\\					
$X_5$	&	6,~~-6&*1,~~-1&1,~~-1&-4,~~4&6,~~-6&1,~~-1&1,~~-1&-4,~~4\\					
$X_6$	&	4,~~-4&-1,~~1&2,~~-2&-3,~~3&7,~~-7&2,~~-2&5,~~-5&0,~~0\\					
$X_7$	&	8,~~-8&-2,~~2&3,~~-3&-7,~~7&11,~~-11&1,~~-1&6,~~-6&-4,~~4\\				
$X_8$	&	6,~~-6&-4,~~4&4,~~-4&-6,~~6&12,~~-12&2,~~-2&10,~-10&0,~~0\\				 
\hline
\end{tabular}
\end{center} 
\caption{The payoff matrix of the  (4, 2) game of treatment C.}
    \label{tab:(4,2)_matrix_2}
\end{table}

\begin{table}[!ht]
\begin{center}
\begin{tabular}{|c|c|c|c|c|c|c|c|c|}
  \hline
$(\textbf{m,n})\!=\!(4,2)$	&	\sout{$Y_1$} &	$Y_2$ &	\sout{$Y_3$} &	$Y_4$ &	\sout{$Y_5$} &	\sout{$Y_6$} &	\sout{$Y_7$ }&	\sout{$Y_8$} \\
\hline
$X_1$	&	\sout{0,~~0}&*0,~~0&\sout{0,~~0}&*0,~~0&\sout{0,~~0}&\sout{0,~~0}&\sout{0,~~0}&\sout{0,~~0}\\						
$X_2$	&	\sout{-2,~~2}&*-2,~~2&\sout{1,~~-1}&1,~~-1&\sout{1,~~-1}&\sout{1,~~-1}&\sout{4,~~-4}&\sout{4,~~-4}\\				
$X_3$	&	2,~~-2&-3,~~3&2,~~-2&-3,~~3&5,~~-5&0,~~0&5,~~-5&0,~~0\\[-1.7ex]\hline\noalign{\vspace{\dimexpr 1.87ex-\doublerulesep}} 			
$X_4$	&	\sout{0,~~0}&-5,~~5&\sout{3,~~-3}&-2,~~2&\sout{6,~~-6}&\sout{1,~~-1}&\sout{9,~~-9}&\sout{4,~~-4}\\				
$X_5$	&	\sout{6,~~-6}&*1,~~-1&\sout{1,~~-1}&-4,~~4&\sout{6,~~-6}&\sout{1,~~-1}&\sout{1,~~-1}&\sout{-4,~~4}\\				
$X_6$	&	\sout{4,~~-4}&-1,~~1&\sout{2,~~-2}&-3,~~3&\sout{7,~~-7}&\sout{2,~~-2}&\sout{5,~~-5}&\sout{0,~~0}\\				
$X_7$	&	\sout{8,~~-8}&-2,~~2&\sout{3,~~-3}&-7,~~7&\sout{11,~~-11}&\sout{1,~~-1}&\sout{6,~~-6}&\sout{-4,~~4}\\				
$X_8$	&	\sout{6,~~-6}&-4,~~4&\sout{4,~~-4}&-6,~~6&\sout{12,~~-12}&\sout{2,~~-2}&\sout{10,~-10}&\sout{0,~~0}\\				 
\hline
\end{tabular}
\end{center} 
 \caption{$X_3$ was eliminated by $X_6$. $Y_1,Y_3,Y_5,Y_6$ and $Y_7$ were eliminated by $Y_2$. $Y_8$ was eliminated by $Y_4$.}
 \label{tab:(4,2)matrix_step1}
\end{table}

\clearpage
\begin{table}[!ht]
\begin{center}
\begin{tabular}{|c|c|c|}
  \hline
$(\textbf{m,n})\!=\!(4,2)$	&	$Y_2$ &		$Y_4$ \\
\hline
$X_1$	&	*0,~~0&*0,~~0\\		
$X_2$	&	*-2,~~2&1,~~-1\\		
$X_4$	&	-5,~~5&-2,~~2\\	[-1.7ex]\hline\noalign{\vspace{\dimexpr 1.87ex-\doublerulesep}} 		
$X_5$	&	*1,~~-1&-4,~~4\\		
$X_6$	&	-1,~~1&-3,~~3\\	[-1.7ex]\hline\noalign{\vspace{\dimexpr 1.87ex-\doublerulesep}} 		
$X_7$	&	-2,~~2&-7,~~7\\	[-1.7ex]\hline\noalign{\vspace{\dimexpr 1.87ex-\doublerulesep}} 		
$X_8$	&	-4,~~4&-6,~~6\\	[-1.7ex]\hline\noalign{\vspace{\dimexpr 1.87ex-\doublerulesep}} 		 
\hline
\end{tabular}
\end{center} 
 \caption{$X_4，X_6,X_7$ and $X_8$ were eliminated by $X_1$.}
  \label{tab:(4,2)matrix_step2}
\end{table}

\begin{table}[!ht]
\begin{center}
\begin{tabular}{|c|c|c|}
  \hline
$(\textbf{m,n})\!=\!(4,2)$	&	$Y_2$ &		$Y_4$ \\
\hline
$X_1$	&	*0,~~0&*0,~~0\\		
$X_2$	&	*-2,~~2&1,~~-1\\		
$X_5$	&	*1,~~-1&-4,~~4\\		 
\hline
\end{tabular}
\end{center} 
 \caption{$X_1，X_2,X_5,Y_2$ and $Y_4$ are the remain strategy.}
  \label{tab:(4,2)matrix_step3}
\end{table}

\subsection{Quadratic programming}\label{sec:nash_qp}
\paragraph{Method}  
Quadratic Programming (QP), used for determining the distribution of equilibrium, is applied here to predict this distribution.
 
We use the QP algorithm mainly because human game decisions are highly stochastic, operating via a trial-and-error process, according to which the mixed equilibrium is extremely important. The Quadratic Programming process is used to distinguish one mixed Nash equilibrium from the pure Nash equilibria that occur in a bimatrix or general sum two-person matrix game. It is based upon the algorithm presented by O.L. Mangasarian in 1964. It uses quadprog (a function in Matlab) to solve a quadratic programming problem for a two-person matrix game. For details of the algorithm, see \cite{mangasarian1964}. 
In practice, we use the Matlab function \textit{bimat.m} \cite{bimat2009} to calculate the equilibrium. 
\paragraph{Results}
According to the QP algorithm, the equilibrium distribution of 
the three treatments are shown in Table A2 in the main text as $\rho^T_{\text{QP}}$.

\section{SI --- Dynamics}
  
\subsection{Logit dynamics original form}\label{sec:dyn_logit_ori}
This study case is a zero–sum game. The dynamics system is a set of ordinary differential equations. The set contains two sub-sets corresponding to each of the two-player (X and Y) populations, respectively. Each population corresponds to eight equations used to describe the dynamic behaviors. We apply logit dynamics \cite{sandholm2010,dan2016}  to predict/interpret the evolution of the subjects’ game.

The logit dynamics system can be expressed as 
    \begin{eqnarray}
            \frac{dx(i)}{dt}&=& \frac{\exp\big(\lambda \cdot U_X (i)\big)}
            {\sum_{j=1}^8 
            \exp\big(\lambda \cdot U_X (j)\big)};\\ 
            \frac{dy(i)}{dt}&=& \frac{\exp\big(\lambda \cdot U_Y (i)\big)}
            {\sum_{j=1}^8 
            \exp\big(\lambda \cdot U_Y (j)\big)};	
    \end{eqnarray}
Where in 
\begin{itemize}
\item 
$i, j \in [1,2,...,8]$ represents the $i,j$-th strategy; 
\item 
$x(i)$ is the proportion of the $i$-th strategy in X population. Similarly,  $y(i)$ is the proportion of the $i$-th strategy in the  Y population;
\item $U_X$ and $U_Y$ are the payoff vector fields for the X and Y populations, respectively. They can be obtained by  
    \begin{eqnarray}\label{eq:logit_ori_set}
        U_X(i)&=&\sum_{j=1}^8 y(j)\cdot \text{M}_X(i,j);\\
        U_Y(i)&=&\sum_{j=1}^8 x(j) \cdot \text{M}_Y(i,j); \nonumber 
    \end{eqnarray}
    where $\text{M}_X(i,j)$ is the payoff of a player in population X who uses the $i$-th strategy and who meets a player in population Y who uses the $j$-th strategy;
\item
    $\lambda$ is the noise parameter in the logit dynamics model \cite{sandholm2010,dan2016}. 
\end{itemize}

\subsection{The unified dimension form}\label{sec:dyn_16dim}
We can unify the two population game into single population form via the following steps:
\begin{eqnarray}
   (x_1, x_2, ..., x_8) &\longrightarrow& (x_1, x_2, ..., x_8) \nonumber\\
    &\text{and} & \\
    (y_1, y_2, ..., y_8) &\longrightarrow& (x_9, x_{10}, ..., x_{16}).\nonumber
\end{eqnarray} 
The dynamics system in Equations set \ref{eq:logit_ori_set} can be presented as 
\begin{equation}
    \dot{x_i} =  \frac{x_i e^{\lambda U_i}}{\sum_{j=1+\gamma}^{8+\gamma} x_j e^{\lambda U_j}} 
\end{equation}
Wherein, the $\gamma$ is defined as follows:
\begin{itemize}
\item 
$\gamma = 0 ~~~\text{when} ~~~ i \in [1,2,...,8]$ for the X population; 
\item 
$\gamma = 8 ~~~\text{when} ~~~ i \in [9,10,...,16]$ for the Y population; 
\end{itemize}

 \subsection{The discrete time form}\label{sec:app_discrete_time_form}
In this case study, we have employed a time series generation approach. Then, we use the discrete time form of the logit dynamics equations:  
\begin{equation}\label{eq:logit_4_sim}
 x_i(t+1)=w \cdot \frac{x_i e^{\lambda U_i}}{\sum_{j} x_j e^{\lambda U_j}} + (1-w) \cdot x_i(t);
\end{equation} 
in which the following pertains:
\begin{itemize}
\item 
$x_i$ is the proportion of strategy $i$ used by a population; 
\item 
$\sum_j$ is the total proportion of strategies used by the population that uses the $j$-th strategy;   
\item 
$U_i$ is the payoff when the $i$-th strategy is used;  
\item 
$\lambda$ is the noise parameter;
\item 
$w$ is the time step or the weight of an agent to update its strategy, which is equivalent to $\Delta t$ in the main text.
\end{itemize}

\subsection{The specification and time series generation}\label{sec:app_specif_para_timeser_gen}
\paragraph{Parameter specification}
For parameter specification, we follow the framework of previous research \cite{dan2014,zhijian2013} to use the noise parameter ($\lambda$ = 50) and the adjustment time step ($\triangle t$ = 0.02).  

 The selected parameters ($\lambda$ = 50, $\triangle t$ = 0.02) are the most favorable in rendering the long-term distribution, 
 which is the most similar to the equilibrium distribution ($\rho^T_{\text{QP}}$, determined  by QP algorithm shown in section \ref{sec:nash_qp}) in  Euclidean distance measurement 
 (shown in section \ref{sec:si_euclidean}).

\paragraph{Time series generation}
To generate the numerical time series, we use the parameterized logit model applied to each treatment, with the following two specifications: 
\begin{enumerate}
\item 
 The initial conditions are set as uniformly random between the eight strategies used by each population; In the presented figures in main text, it is set as [1/8, 1/8, ..., 1/8; 1/8, ..., 1/8] over two 'populations'. 
\item 
2.	We generate the time series of 1000 rounds,  which is as long as the rounds in the laboratory human subject game experiment in this research.
\end{enumerate}

\paragraph{Algorithm for time series generation} ~\\
The algorithm used in the simulation is based on Eq. \ref{eq:logit_4_sim}. We use a similar form as applied in the time series simulation to study the consistency of dynamics theory and perform experiments on the price dispersion \cite{dan2005}.

\section{SI --- Distribution}\label{sec:si_distribution}

\subsection{Euclidean distance $\delta$ in experiment}\label{sec:si_euclidean}

In terms of Euclidean distance, the convergence to the equilibrium agrees with the theoretical expectations.  
This means that, across 1000 rounds, the game systems indeed go along the process to find Nash equilibrium. 

\paragraph{Method} 
The Euclidean distance is to measure 
the difference between the two strategy distributions vectors \cite{selten2008,dan2010,nowak2015}. The hypothesis is that, the smaller the distance, the closer the two distributions. We denote the two strategy distribution vectors of $N$ dimensions as
$\rho_1$ and $\rho_0$. The Euclidean distance from $\rho_1$ to  $\rho_0$ can be formulated as: 
\begin{equation}\label{eq:nash_distance}
   \delta = \Big(\sum_{i=1}^N(\rho_1(s_i)-\rho_0(s_i))^2\Big)^{1/2}
\end{equation}
We define the average of $\rho$ in the time interval $[t_1-t_0]$ as
\begin{equation}
    \rho(s_i,[t_0,t_1]) = \frac{1}{t_1-t_0} \sum_{t_j=t_0}^{t_1} \rho(s_i,t_j)
\end{equation}
Then the Euclidean distance of the time average is  determined via
\begin{equation}\label{eq:nash_distance2}
   \delta_{[t_0,t_1]} = \Big(\sum_{i=1}^N(\rho_1(s_i,[t_0,t_1])-\rho_0(s_i,[t_0,t_1]))^2\Big)^{1/2}
\end{equation}
Usually, we set $\rho_1$ = $\rho^E$. 
It is important to note the following:
\begin{enumerate}
\item In order to measure $\delta_{\textbf{QP}}$, we set $\rho_0 := \rho^T_{\textbf{QP}}$.  
\item In order to measure  $\delta_{\textbf{Dyn}}$, we set $\rho_0 := \rho^T_{\textbf{Dyn}}$.  
\item 
3.	In this study case, we unify the two-population game into a 
  16 dimensional game (for details, see Appendix section \ref{sec:dyn_16dim}), so the games have $N=$16 dimensions.  
\end{enumerate}

\paragraph{Numerical Results}
The numerical results can be calculated by  the following:
\begin{enumerate}
\item $\rho^T_{\textbf{QP}}$ and $\rho^T_{\textbf{Dyn}}$ are given in Table A2. 
\item The formula in Eq. \ref{eq:nash_distance}
\end{enumerate}
Details of the numerical results are shown in Table \ref{tab:nash_distance}, wherein, 
\begin{equation}
\triangle \delta_{\textbf{QP}}=\delta_{\textbf{QP[801,1000]}}-\delta_{\textbf{QP[1,200]}} <0
\end{equation}
means the evolution of $\rho^E$ to the expected Nash equilibrium from QP. 
\begin{equation}
\triangle \delta_{\textbf{Dyn}}=\delta_{\textbf{Dyn[801,1000]}}-\delta_{\textbf{Dyn[1,200]}} < 0
\end{equation}
means the evolution of the $\rho^E$ to the expected distribution by dynamics model. 
\begin{equation}
\triangle \delta =\delta_{\textbf{Dyn[801,1000]}}-\delta_{\textbf{QP[801,1000]}} < 0
\end{equation}
means  $\rho^E_{800,1000}$ is closer to that expected by the dynamics model than to the equilibrium distribution given by QP. 

\begin{table}[!ht]
    \centering
    \begin{tabular}{|l|c|c|c|}
    \hline
   &	 \multicolumn{3}{c|}{Treatment} \\\cline{2-4}
   &	A &	  B &	  C \\\hline
$\delta_{\textbf{QP[1,200]}}$&	~~0.4298&	~~0.5551&	~~0.5745\\\hline
$\delta_{\textbf{QP[801,1000]}}$&	~~0.3766&	~~0.4047&	~~0.4436\\\hline
~~~~~$\triangle \delta_{\textbf{QP}}$&	$-$0.0532&	$-$0.1504&	$-$0.1309\\\hline
$\delta_{\textbf{Dyn[1-200]}}$&	~~0.0781&	~~0.2088&	~~0.1442\\\hline
$\delta_{\textbf{Dyn[801-1000]}}$&	~~0.0059&	~~0.0097&	~~0.0072\\\hline
~~~~~$\triangle \delta_{\textbf{Dyn}}$&	$-$0.0722&	$-$0.1991&	$-$0.1370\\\hline
~~~~~~~$\triangle \delta$&	$-$0.3707&	$-$0.3950&	$-$0.4364\\\hline
    \end{tabular}
    \caption{The Euclidean distance between the experiment distribution $\rho^E$ of the three treatments.}
    \label{tab:nash_distance}
\end{table} 

\begin{figure}[!ht]
\begin{center}  
  \includegraphics[scale=0.8]{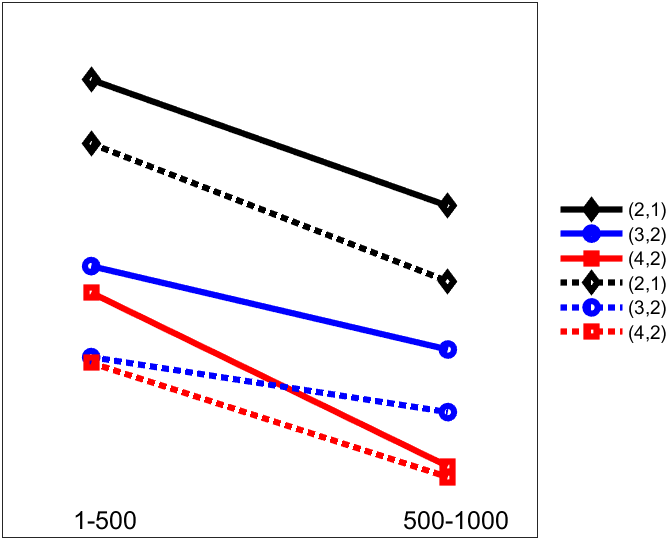}
\end{center}
\caption{ 
   Euclidean distance of the early periods (1-500 rounds) and the later periods (501-1000 rounds) in the game experiments. The Euclidean distance is between $\rho^E$ with respect to (1) the statics, $\rho^T_{\text{QP}}$, in solid line;  and (2) the  dynamics, $\rho^T_{\text{Dyn}}$, in dashed line.  
   \label{fig:result_distribution}
}
\end{figure}

\paragraph{Figurial Results}
 Using the numerical results, Figure \ref{fig:result_distribution} illustrates
 the convergence process. The following is made clear:
 
\begin{enumerate}
\item  For each of the treatment (A,B,C),
 the experiment distribution $\rho^E$ is closer to  $\rho^T_{\textbf{QP}}$ and $\rho^T_{\textbf{Dyn}}$  in terms of  time; This means that, in the 1000 rounds of the games, the systems indeed go
 along the process to find Nash equilibrium.  
\item For each of the treatment (A,B,C),  the dynamics model prediction gives a lower Euclidean value. As a result, the dynamics model performs better than the statics equilibrium model.
\end{enumerate}

\section{SI --- Cycle}
\subsection{Eigencycle spectrum definition}

The eigencycle spectrum presents the eigencycle of an $N$-dimensional dynamic system with a definite spectrum form.  

An eigencycle spectrum has two axes: the X-axis and the Y-axis. 
\begin{itemize}
\item 
X-axis value ($\text{N}_{\textbf{eic}}$): This is a natural number, called the eigencycle identifying number (ID). The eigencycle set has $N(N-1)/2$ elements \cite{zhijian2022,2021Qinmei,zhijian2020}, so the ID takes the form of 1 to  $N(N-1)/2$. 

The assignments are given in the order of  $m$ from 1 to $N-1$ first, and then $n$ from $m+1$ to $N$; This system is known as one-by-one mapping, where 
$$(m,n) \longrightarrow \text{N}_{eic} \in \Big[1,2,...,\frac{N(N-1)}{2}\Big],$$ 
The $\text{N}_{\textbf{eic}}$ is assigned as the X-axis value. 
For example, in this study case, $N$=16, then ID = [1, 2, ...,120]. For details of the assignment, see Table \ref{tab:ec_id_mn}; 

\item Y-axis value: This takes the eigencycle value of  $\sigma^{(mn)}$ as derived from the eigensystem analysis, or of $L^{mn}$ from time series measurement.  To see the method used to obtain an eigencycle from a time series, refer to the eigencycle measurement \cite{zhijian2022,2021Qinmei,zhijian2020}.  
\end{itemize}

\begin{table}[ht!]
    \centering
    \begin{tabular}{|c|c|c||c|c|c||c|c|c||c|c|c||c|c|c|}\hline
         $\text{N}_{\textbf{eic}}$ &$m$&$n$& $\text{N}_{\textbf{eic}}$ &$m$&$n$ &$\text{N}_{\textbf{eic}}$ &$m$&$n$ &$\text{N}_{\textbf{eic}}$ &$m$&$n$ &$\text{N}_{\textbf{eic}}$&$m$&$n$\\\hline
1&	1&	2&	25&	2&	12&	49&	4&	11&	73&	6&	14&	97&	9&	14\\
2&	1&	3&	26&	2&	13&	50&	4&	12&	74&	6&	15&	98&	9&	15\\
3&	1&	4&	27&	2&	14&	51&	4&	13&	75&	6&	16&	99&	9&	16\\
4&	1&	5&	28&	2&	15&	52&	4&	14&	76&	7&	8&	100&	10&	11\\
5&	1&	6&	29&	2&	16&	53&	4&	15&	77&	7&	9&	101&	10&	12\\
6&	1&	7&	30&	3&	4&	54&	4&	16&	78&	7&	10&	102&	10&	13\\
7&	1&	8&	31&	3&	5&	55&	5&	6&	79&	7&	11&	103&	10&	14\\
8&	1&	9&	32&	3&	6&	56&	5&	7&	80&	7&	12&	104&	10&	15\\
9&	1&	10&	33&	3&	7&	57&	5&	8&	81&	7&	13&	105&	10&	16\\
10&	1&	11&	34&	3&	8&	58&	5&	9&	82&	7&	14&	106&	11&	12\\
11&	1&	12&	35&	3&	9&	59&	5&	10&	83&	7&	15&	107&	11&	13\\
12&	1&	13&	36&	3&	10&	60&	5&	11&	84&	7&	16&	108&	11&	14\\
13&	1&	14&	37&	3&	11&	61&	5&	12&	85&	8&	9&	109&	11&	15\\
14&	1&	15&	38&	3&	12&	62&	5&	13&	86&	8&	10&	110&	11&	16\\
15&	1&	16&	39&	3&	13&	63&	5&	14&	87&	8&	11&	111&	12&	13\\
16&	2&	3&	40&	3&	14&	64&	5&	15&	88&	8&	12&	112&	12&	14\\
17&	2&	4&	41&	3&	15&	65&	5&	16&	89&	8&	13&	113&	12&	15\\
18&	2&	5&	42&	3&	16&	66&	6&	7&	90&	8&	14&	114&	12&	16\\
19&	2&	6&	43&	4&	5&	67&	6&	8&	91&	8&	15&	115&	13&	14\\
20&	2&	7&	44&	4&	6&	68&	6&	9&	92&	8&	16&	116&	13&	15\\
21&	2&	8&	45&	4&	7&	69&	6&	10&	93&	9&	10&	117&	13&	16\\
22&	2&	9&	46&	4&	8&	70&	6&	11&	94&	9&	11&	118&	14&	15\\
23&	2&	10&	47&	4&	9&	71&	6&	12&	95&	9&	12&	119&	14&	16\\
24&	2&	11&	48&	4&	10&	72&	6&	13&	96&	9&	13&	120&	15&	16\\
\hline
     \end{tabular}
    \caption{Eigencycle ID ($\text{N}_{\textbf{eic}}$) and the associated dimensional ID of the players' strategy space. In this study, in the $m$ and $n$ column, the dimensional IDs 1, 2, ..., 8 represent the dimensions $X_1, X_2, ..., X_8$ of player 1's strategy space, respectively, while the dimensional IDs 9, 10, ..., 16 represent dimensions  $Y_1, Y_2, ..., Y_8$ of player 2's strategy space, respectively.}
    \label{tab:ec_id_mn}
\end{table}

\subsection{Experimental and theoretical data} 
\begin{itemize}
\item 
The theoretical eigencycle is obtained from the time series, which is generated by the parameterized logit dynamics model with uniform random initial conditions. See Section \ref{sec:app_discrete_time_form} for the definition, 
and Section \ref{sec:app_specif_para_timeser_gen} 
describes the process of generating from the parameterized model.
\item 
The experimental eigencycle is obtained from the time series, which is derived from the experiments with human subjects. See SI --Experiment Section \ref{sec:app_experiment}.  
 \item  
The  theoretical and experimental 
eigencycle spectrum are shown in main text and its appendix. 
\end{itemize}

\subsection{Definition of a  closed loop  in eigencycle analysis}

In the context of eigencycle spectrum analysis, a \textbf{closed loop (or loop)} is defined as a cyclic structure formed by a sequence of strategy pairs \((m,n)\) that are connected end-to-end. More formally:

\begin{enumerate}
    \item \textbf{Nodes:} The strategy indices \(1, 2, \dots, 16\), where indices \(1\) to \(8\) correspond to strategies \(X_1\) to \(X_8\), and indices \(9\) to \(16\) correspond to strategies \(Y_1\) to \(Y_8\).
    
    \item \textbf{Edges:} Each \(x\)-value in the eigencycle spectrum corresponds to an unordered strategy pair \((m,n)\). The associated strength \(\theta\) indicates the extent to which that strategy pair participates in cyclic motion.
    
    \item \textbf{Loop structure:} A set of edges \((m_1,n_1), (m_2,n_2), \dots, (m_k,n_k)\) forms a \textit{closed loop} if they can be arranged into a directed cyclic sequence:
    \[
    m_1 \to n_1 = m_2 \to n_2 = \cdots \to m_k \to n_k = m_1.
    \]
    wherein the signal of the $m_i \to n_i$ having the same signal, $+$ or $-$. Herein, $(m_i \to n_i)$ = $-1 \times (n_i \to m_i)$ referring to the definition of the eigencycle. 
\end{enumerate}

Mathematically, a loop \(L\) is represented as an ordered sequence of nodes:
\[
L = (v_1, v_2, \dots, v_k, v_1),
\]
such that each edge \((v_i, v_{i+1})\) is observed in the eigencycle spectrum with a non-negligible strength, and the sequence returns to its starting point. The smallest closed loop is three nodes. 

\paragraph{Empirical verification}~~ In this study, as shown in main text section 3.1 and A.2.1, the theoretically predicted loop structures (e.g., those in treatments A and C) are specified by the IEDS/dynamic model. Empirical validation consists of testing whether the edges of these predicted loops exhibit significantly high strengths in the experimental spectrum (e.g., appearing in the top 12 reported  in section A2.1, Table~A2).

A loop is considered  empirically verified  if the following criteria are satisfied:
\begin{itemize}
    \item All edges of the loop   appear among the top 12 eigencycles in the experimental spectrum (as reported in Table~A2);
    \item The mean strength of the loop edges is significantly higher than that of non-loop edges, as confirmed by a Wilcoxon rank-sum test.
\end{itemize}

This definition provides a rigorous and reproducible basis for identifying and validating cyclic structures in both theoretical predictions and experimental observations.

\section{SI --- Collapse}

\subsection{Definition: strategy space collapse} 
\paragraph{Player strategy space} 
Assume $i$ is a player in a game (denoted as $G$) with $N$ players, where player $i$'s strategy is a $d_i$ dimensional vector $v_i$,
$$ v_i := (s_i^1, s_i^2, ..., s_i^{d_i}), $$
in which $s_i^j \in \textbf{R}$ is a real number. Assume $v_i \in S_i$, $S_i$ is the strategy space 
of player $i$ in the game $G$. 

\paragraph{Game strategy space}
Then, the \textbf{game strategy space} is defined  as: 
$$ S_G = \prod_{i=1}^N S_i, $$ 
where $S_i$ is player $i$'s strategy space. With this definition, 
we can determine that
\begin{itemize}
\item At any given time $t_0$, game $G$ is located at a point  in $\Sigma$;
\item In a long running  repeated game, the game's evolution process is represented as a trajectory within the game strategy space $\Sigma$. 
\item As $S_i$ has dimension $d_i$, the dimension of game ($d_G$) is 
$$d_G = \sum_{i=1}^{N} d_i.$$
\end{itemize}
 
\paragraph{Game strategy space collapse}
Following the definition of the game strategy space, 
we can define its collapse. 
If the observed   player $i$'s space is reduced from $d_i$ to $d_i'$ dimension and 
$$d_i' < d_i$$
which means that, the freedom of the $i$-th player is reduced, and as such, the dimension of game $D_G$ is reduced. 
We refer to this dimensional decline as a  \textbf{strategy space collapse}.

\subsection{Experimental and Theoretical data}

The theoretical accumulated curve, crossover point and  pulse signal are obtained from the time series, which is generated by the parameterized logit dynamics models (see Section \ref{sec:app_discrete_time_form} for the definition. See Section \ref{sec:app_specif_para_timeser_gen} for the time series.  

 \section{SI --- Statistics}

To establish that the observed pulses and crossover points are mutually corroborating manifestations of the same underlying dynamic process, we conducted a series of cross-variable correlation tests. Each test pairs a specific empirical or theoretical measurement with another, with sample sizes and data sources clearly identified below.

\subsection{Test for Statistical Significant of Pulse Signal \label{sec:comm_stat}}
\subsubsection{Exact binomial test on pulse (logit dynamics, all treatments)}

\paragraph{Question: } Is the pulse significance an artifact?  In a huge sample space, the significance can be a random effect. To convince the significance is not fake, so multiple-testing correction is necessary.\\

We conducted a pooled exact binomial test to evaluate the predictive power of the logit dynamics across all three treatments combined, using the pulse candidates specified in Table~A5 as the theoretical target set and the statistically significant pulses  in Table~A4 as the empirical observations.

\paragraph{Sample space} 
The total testing space \(M\) is (the sum over all treatments of the number of possible strategy-pair) \(\times\) (time-blocks):
\begin{equation}\label{eq:sample6500}
    M = 100 \times (12+12+7+7+15+12) = 100 \times 65 = 6500.
\end{equation}
Here, time block = 1000 round, and 1 block has 10 rounds. The components of testing space (pulse candidate) are shown in Table \ref{tab:M_65_decomp},
\begin{table}[ht!]
\centering

\begin{tabular}{c c c c c}
\toprule
\textbf{Treat-} & \textbf{Popul-} & \textbf{Dominating} & \textbf{Dominated} & \textbf{Size} \\
ment&ation&set&set&\\
\midrule
\multirow{2}{*}{A} & X & \(\{2,6\}\) & \(\{1,3,4,5,7,8\}\) & \(2 \times 6 = 12\) \\
                   & Y & \(\{2,4\}\) & \(\{1,3,5,6,7,8\}\) & \(2 \times 6 = 12\) \\
\midrule
\multirow{2}{*}{B} & X & \(\{1\}\) & \(\{2,3,4,5,6,7,8\}\) & \(1 \times 7 = 7\) \\
                   & Y & \(\{4\}\) & \(\{1,2,3,5,6,7,8\}\) & \(1 \times 7 = 7\) \\
\midrule
\multirow{2}{*}{C} & X & \(\{1,2,5\}\) & \(\{3,4,6,7,8\}\) & \(3 \times 5 = 15\) \\
                   & Y & \(\{2,4\}\) & \(\{1,3,5,6,7,8\}\) & \(2 \times 6 = 12\) \\
\bottomrule
\end{tabular} \\
\medskip
\footnotesize
\caption{Components of the total possible pulse $(65)$ by treatment and population}
\label{tab:M_65_decomp}
\end{table}
 
The theoretical pulses candidate set size is \(K = 18\)  from Table~A5. The empirical \(n = 9\) significant pulses in (Table~A4). Among these, \(k = 5\) pulses \textbf{exactly} hit the theoretical predictions in terms  (i.e., the 5 key columns of Table~A4 and Table~A5) of the 
\begin{eqnarray}
   &&\text{ (1) treatment}, \\
   &&\text{ (2) population}, \\
   &&\text{ (3) dominated strategy}, \\
   &&\text{ (4) dominating strategy} \\
   &&\text{ (5) time block}
\end{eqnarray}
 Among the 18 targets shown as the theoretical expectation in Table A5, 5 are hit, with details shown in Table \ref{tab:5hit18logit} 

The parameters for exact binomial test  (logit dynamics, all treatments) are shown in Table \ref{tab:ebt_logit}.

\begin{table}[!ht]
    \centering

    \begin{tabular}{cccc}
    \hline
    Treat- & Domin- & Domin- & time     \\
    ments & ated &  ation   &  block   \\
          & $s_j$   & $s_i$ &  $[t_0,t_1]$     \\
    \hline
     	 A & $X_8$ & $X_2$ & 11-20   \\
     	 A & $X_8$ & $X_6$ & 11-20 \\
     	 A & $X_8$ & $X_2$ & 21-30   \\
     	 A & $X_4$ & $X_6$ & 41-50 \\ 
     	 C & $X_8$ & $X_5$ & 11-20\\
    \hline
    \end{tabular}
 \caption{The theoretical targets in Table A5 are hit by experimental pulses  in Table A4. }
\label{tab:5hit18logit}
\end{table}

\begin{table}[ht!]
\centering

\begin{tabular}{l c c c}
\toprule
\textbf{Parameter} & \textbf{Symbol} & \textbf{Value}  & Note\\
\midrule
Total testing space (all possible positions) & \(M\) & 6500 & Eq. \ref{eq:sample6500} \\
Theoretically predicted candidate set size & \(K\) & 18 & Table A5\\
Random hit probability under the null & \(p_0 = K / M\) & \(0.00276923\) \\
Total number of empirically detected signals & \(n\) & 9 & Table A4 \\
Number of hits (falling within candidate set) & \(k\) & 5 & Table \ref{tab:5hit18logit} \\
\bottomrule
\end{tabular}
\caption{Exact binomial test parameters (logit dynamics, all treatments)} 
\label{tab:ebt_logit}
\end{table}

\paragraph{Null hypotheses}
 
    \textbf{\(H_0\):} The theory has no predictive power. Each observed signal falls into the candidate set with probability \(p_0 = 18/6500 \approx 0.002769\), independently across observations.
      \textbf{The alternative hypothesis \(H_1\):} The theory has predictive power; the observed hit rate is significantly higher than the random probability.

\paragraph{Exact binomial test}
Under the null, the number of hits \(X\) follows a binomial distribution: \(X \sim \text{Binomial}(n=9, p_0 = 18/6500)\). We observed \(k = 5\) hits. The \(p\)-value is the probability of obtaining at least 5 hits:

\[
p\text{-value} = P(X \ge 5) = \sum_{i=5}^{9} \binom{9}{i} p_0^i (1-p_0)^{9-i} \\
\approx 2.03 \times 10^{-11}.
\]

    To guard against inflated Type I error from multiple comparisons, we apply the Bonferroni correction, with \(M = 65\) and the nominal significance level \(\alpha = 0.05\), the corrected threshold is:
\[
\alpha_{\text{Bonferroni}} = \frac{0.05}{6500} \approx 7.69 \times 10^{-6}.
\]

\begin{table}[ht!]
\centering
\begin{tabular}{l c}
\toprule
\textbf{Indicator} & \textbf{Value} \\
\midrule
Random hit probability \(p_0\) & 0.277\% \\
Observed hit rate & 5/9 = 55.6\% \\
Expected number of hits under \(H_0\) & \(9 \times p_0 \approx 0.0249\) \\
Enrichment factor (observed / expected) & 	
200.8 \\
Exact \(p\)-value & \(\mathbf{2.03 \times 10^{-11}}\) \\
Bonferroni-corrected threshold & \(7.69 \times 10^{-6}\) \\
Significance & \(p \ll 0.001\) (highly significant) \\
\bottomrule
\end{tabular}
\end{table}

\paragraph{Conclusion}

Under the null hypothesis of no predictive power, the probability of observing at least 5 hits out of 9 empirically detected pulses is only about \(2.03 \times 10^{-11}\),  below the stringent Bonferroni-corrected threshold. Therefore, the theoretical predictions exhibit statistically significant predictive power for the pulse signals across all treatments. 

\subsubsection{Exact binomial test on pulse (logit dynamics, (A,B,C) treatments)}

\paragraph{Sample space} We conducted separate exact binomial tests for treatments A and C to evaluate the predictive power of the dynamic theory within each treatment. The theoretical candidate sets are derived from Table~A5, and the empirical signals are the statistically significant pulses reported in Table~A4.

The total testing space for treatments is given by Eq. \ref{eq:sample6500}. Theoretical pulse candidate set size and total observed signals  refer to Table A5 and A4, respectively.

\begin{table}[ht!]
\centering

\begin{tabular}{l c c c c}
\toprule
\textbf{Parameter} & \textbf{Symbol} & \textbf{Treatment} & \textbf{Treatment} & \textbf{Treatment} \\
\textbf{ } & \textbf{ } & \textbf{  A} & \textbf{  B} & \textbf{  C} \\
\midrule
Total testing space & \(M\) & 2400 & 1400 & 2700 \\
Theoretical candidate set size & \(K\) & 12 & 0 & 5 \\
Random hit probability & \(p_0 = K/M\) & \(0.005\) & -- & \(0.001852\) \\
Total observed signals & \(n\) & 6 & 0 & 3 \\
Number of hits & \(k\) & 4 & 0 & 1 \\
\bottomrule
\end{tabular}
\medskip
\footnotesize
\textit{Note}: For Treatment B, no theoretical candidate set is predicted (\(K=0\)), hence \(p_0\) is undefined.
\caption{Exact binomial test parameters across treatments}
\label{tab:binomial_params}
\end{table}

\paragraph{Null hypotheses}

For each treatment, the null hypothesis \(H_0\) states that the theory has no predictive power: each observed signal falls into the candidate set with probability \(p_0 = K/M\), independently across observations. The alternative hypothesis \(H_1\) states that the observed hit rate is significantly higher than random.

\paragraph{Exact binomial test} 
 
Under the null, \(X \sim \text{Binomial}(n, p_0)\). The \(p\)-value is the probability of observing at least \(k\) hits:
\[
p\text{-value} = P(X \ge k) = \sum_{i=k}^{n} \binom{n}{i} p_0^i (1-p_0)^{n-i}.
\]

\textbf{Treatment A} (\(n=6\), \(p_0=0.005\), \(k=4\)):
\[
p\text{-value} = \sum_{i=4}^{6} \binom{6}{i} (0.005)^i (0.995)^{6-i} \approx 9.30 \times 10^{-9}.
\]

\textbf{Treatment C} (\(n=3\), \(p_0=5/2700\), \(k=1\)):
\[
p\text{-value} = 1 - (1 - 0.001852)^3 \approx 0.0055453.
\]

\textbf{Multiple-Testing Correction}
Applying Bonferroni correction for the total testing space in each treatment:
\[
\alpha_{\text{Bonf, A}} = \frac{0.05}{2400} \approx 2.08 \times 10^{-5}, \qquad
\alpha_{\text{Bonf, C}} = \frac{0.05}{2700} \approx 1.85 \times 10^{-5}.
\]

\begin{table}[ht!]
\centering
\begin{tabular}{l c c c}
\toprule
\textbf{Indicator} & \textbf{Treatment A} & \textbf{Treatment B}& \textbf{Treatment C} \\
\midrule
Random hit probability \(p_0\) & 0.5\% &&  0.185\% \\
Observed hit rate & \(4/6 = 66.7\%\) &&  \(1/3 = 33.3\%\) \\
Exact \(p\)-value & \(\mathbf{9.30 \times 10^{-9}}\) & &  \(\mathbf{5.5453 \times 10^{-3}}\) \\
Bonferroni-corrected threshold & \(2.08 \times 10^{-5}\) & &  \(1.85 \times 10^{-5}\) \\
Significance (uncorrected) & \(p < 0.001\) &-- &  \(p < 0.05\) \\
Significance (Bonferroni) & Significant & -- &  Not significant \\
\bottomrule
\end{tabular}
\caption{Summary of exact binomial test by treatments} \label{tab:ebt_by_treatment}
\end{table}

\paragraph{Conclusion}
In the summary Table \ref{tab:ebt_by_treatment}, 
in Treatment A, the probability of observing at least 4 hits out of 6 detected pulses under the null is only \(9.30 \times 10^{-9}\), which remains highly significant even after Bonferroni correction. In Treatment C, the observed hit rate (\(1/3 = 33.3\%\)) yields an uncorrected \(p\)-value of \(0.00554\), significant at the conventional \(\alpha = 0.05\) level, but it does not survive the stringent Bonferroni correction.  Overall, the results support the predictive validity of the dynamic framework, with stronger evidence in Treatment A.

\subsection{Exact binomial test on pulse (IEDS, all treatments)}
\paragraph{Question}
Does the classical IEDS possess predictive power at the strategy level   to generate pulse signals during the collapse phase? 

To address this, we conduct a pooled exact binomial test across all three treatments, using the surviving strategies predicted by IEDS (Table~1) as the theoretical candidate set and the experimental pulses reported in Table~A4 as the empirical observations.

\paragraph{Sample space}

The total testing space \(M\) is the total number of possible pulse combinations across all treatments:
\[
M = (12+12) + (7+7) + (15+12) = 65.
\]
Table \ref{tab:M_65_decomp} provides the component of the total possible pulse $M$. 

The theoretical candidate set size \(K\) is the number of strategies, which being deleted at later rounds, rather than in the 1st round, in Table~1, and total number of candidate are:
\[
K = 4 + 0 + 1 + 1 + 12 + 0 = 18.
\]
Table \ref{tab:K_IEDS_decomp} provides the decomposition of the total target.

\begin{table}[ht!]
\centering

\begin{tabular}{c c c c c}
\toprule
\textbf{ } & \textbf{ } & \textbf{Dominating} & \textbf{Dominated} & \textbf{} \\
\textbf{Treatment} & \textbf{Population} & \textbf{  set} & \textbf{  set} & \textbf{Size} \\
\midrule
\multirow{2}{*}{A} & X & \(\{2,6\}\) & \(\{4,8\}\) & \(2 \times 2 = 4\) \\
                   & Y & \(\{2,4\}\) & \(0\) & \(2 \times 0 = 0\) \\
\midrule
\multirow{2}{*}{B} & X & \(\{1\}\) & \(\{5\}\) & \(1 \times 1 = 1\) \\
                   & Y & \(\{4\}\) & \(\{2\}\) & \(1 \times 1 = 1\) \\
\midrule
\multirow{2}{*}{C} & X & \(\{1,2,5\}\) & \(\{4,6,7,8\}\) & \(3 \times 4 = 12\) \\
                   & Y & \(\{2,4\}\) & \(0\) & \(2 \times 0 = 0\) \\
\bottomrule
\end{tabular}\\
\medskip
\footnotesize
\textit{Note}:  Empty sets \(0\) indicate that no candidate in the population expected by IEDS.
\caption{Component of the total target $K= 18$ by treatment and population}
\label{tab:K_IEDS_decomp}
\end{table}

The empirical pulses \(n = 9\) are shown in (Table~A4). Among these, \(k = 8\) pulses exactly hit at the strategy-pair level, as determined by comparing the multi-round eliminates final period strategy frequencies in Table~A1. 

\begin{table}[ht!]
\centering
\begin{tabular}{l c c}
\toprule
\textbf{Parameter} & \textbf{Symbol} & \textbf{Value} \\
\midrule
Total testing space (all possible positions) & \(M\) & 65 \\
Theoretically predicted candidate set size & \(K\) & 18 \\
Random hit probability under the null & \(p_0 = K / M\) & \(0.2769\) \\
Total number of empirically detected signals & \(n\) & 9 \\
Number of hits (falling within candidate set) & \(k\) & 8 \\
\bottomrule
\end{tabular}
\end{table}

\paragraph{Null hypotheses\(H_0\):} IEDS has no predictive power at pulse signal. \textbf{Alternative hypothesis \(H_1\):} The theory has predictive power; the observed hit rate is significantly higher than the random probability.

\paragraph{Exact Binomial Test}
Under the null, the number of hits \(X\) follows a binomial distribution: \(X \sim \text{Binomial}(n=9, p_0 = 18/65)\), observed \(k = 8\) hits. The \(p\)-value:

\[
p\text{-value} = P(X \ge 8) = \sum_{i=8}^{9} \binom{9}{i} p_0^i (1-p_0)^{9-i}.\\
               \approx 2.346 \times 10^{-4}.
\]

\begin{table}[ht!]
\centering
\begin{tabular}{l c}
\toprule
\textbf{Indicator} & \textbf{Value} \\
\midrule
Random hit probability \(p_0\) & 32.3\% \\
Observed hit rate & 8/9 = 88.9\% \\
Expected number of hits under \(H_0\) & \(9 \times p_0 \approx 2.91\) \\
Exact \(p\)-value & \(\mathbf{2.346 \times 10^{-4}}\) \\
Bonferroni-corrected threshold & \( 0.05/M = 7.69 \times 10^{-4}\) \\
Significance & \(p < 0.001\) (significant) \\
\bottomrule
\end{tabular}
\caption{Summary of IEDS test} \label{tab:IEDS_test_pulse}
\end{table}

\paragraph{Conclusion}

Referring to Table \ref{tab:IEDS_test_pulse}, Under the null hypothesis of no predictive power of IEDS, the probability of at least 8 hits out of 9 targets is only about \(2.346 \times 10^{-4}\). This probability is below the conventional significance level and the Bonferroni correction for \(M = 65\) comparisons (\(p < 7.69 \times 10^{-4}\)). We therefore reject the null hypothesis and conclude that the IEDS  exhibits  predictive power for pulse signals.

\subsection{Exact binomial test  on pulse (replicator dynamics, all treatments)}

\begin{table}[ht!]
    \centering
    \begin{tabular}{cccrcrcc}
    \hline 
    Treat- & Domin- & Domin- & time   & paired- &   Surplus& Sample&Pulse  \\
    ments & ated &  ation   &  block  &  ttest($p$) &    & size &observed    \\
    & $s_j$   & $s_i$ &  $[t_0,t_1]$ &        & $\psi^T$  &  \textbf{N}&(Table A4)    \\
    \hline 
A&	$X_	4$&	$X_	2$&	1-10&	0&	2.68&	120\\
A&	$X_	4$&	$X_	2$&	11-20&	0&	2.64&	120\\
A&	$X_	4$&	$X_	6$&	1-10&	0&	0.20&	120\\
A&	$X_	4$&	$X_	6$&	11-20&	0&	2.85&	120\\
A&	$X_	4$&	$X_	6$&	21-30&	0&	7.22&	120\\
A&	$X_	4$&	$X_	6$&	31-40&	0&	10.48&	120\\
A*&	$X_	4$&	$X_	6$&	41-50&	0&	10.98&	120 & \textbf{Hit}\\
A&	$X_	4$&	$X_	6$&	51-60&	0&	9.17&	120\\
A&	$X_	4$&	$X_	6$&	61-70&	0&	6.28&	120\\
A&	$X_	4$&	$X_	6$&	71-80&	0&	3.21&	120\\
A*&	$X_	4$&	$X_	6$&	81-90&	0&	0.35&	120 & \textbf{Hit}\\
A&	$X_	7$&	$X_	2$&	1-10&	0&	1.40&	120\\
A&	$X_	8$&	$X_	2$&	1-10&	0&	4.74&	120\\
A*&	$X_	8$&	$X_	2$&	11-20&	0&	1.40&	120 & \textbf{Hit}\\
A&	$X_	8$&	$X_	6$&	1-10&	0&	2.26&	120\\
A*&	$X_	8$&	$X_	6$&	11-20&	0&	1.61&	120 & \textbf{Hit}\\
A&	$Y_	8$&	$Y_	2$&	1-10&	0&	0.35&	120\\
   \hline
B&	$X_	3$&	$X_	1$&	1-10&	0&	0.93&	120\\
B&	$X_	4$&	$X_	1$&	1-10&	0&	0.56&	120\\
B&	$X_	5$&	$X_	1$&	1-10&	0&	4.61&	120\\
B&	$X_	5$&	$X_	1$&	11-20&	0&	4.99&	120\\
B&	$X_	6$&	$X_	1$&	1-10&	0&	3.72&	120\\
B&	$X_	6$&	$X_	1$&	11-20&	0&	2.30&	120\\
B&	$X_	7$&	$X_	1$&	1-10&	0&	4.25&	120\\
B&	$X_	8$&	$X_	1$&	1-10&	0&	4.52&	120\\
   \hline
C&	$X_	4$&	$X_	1$&	1-10&	0&	2.43&	120\\
C&	$X_	4$&	$X_	5$&	1-10&	0&	2.31&	120\\
C&	$X_	4$&	$X_	5$&	11-20&	0&	1.00&	120\\
C&	$X_	6$&	$X_	1$&	1-10&	0&	2.89&	120\\
C&	$X_	6$&	$X_	5$&	1-10&	0&	2.77&	120\\
C&	$X_	6$&	$X_	5$&	11-20&	0&	4.69&	120\\
C&	$X_	6$&	$X_	5$&	21-30&	0&	1.68&	120\\
C&	$X_	8$&	$X_	1$&	1-10&	0&	1.56&	120\\
C*&	$X_	8$&	$X_	5$&	1-10&	0&	1.44&	120 & \textbf{Hit}\\

   \hline
    \end{tabular}
    \caption{ Theoretical pulse from replicator dynamics($\psi^T$).  The all 34 samples  ordered by  the treatments (A,B,C). Note: the time series from the dynamics model are smooth, meaning that more significant pulse signals can be obtained from this model, when compared with those from the highly stochastic experimental time series.The vainlue of $\psi^T$  is multiplied by 12, referring to 12 session in experiment.  The starred entries are \textbf{exactly} equal the corresponding values for treatments, Population, Dominated ($s_j$), Domination ($s_i$) and time‑block ($[t_0,t_1]$) reported in Table A4 in the main text.}
    \label{tab:Replicator_dyn_Pulse_significant}
\end{table}  

\paragraph{The question.} whether the target of logit dynamics comes from parameter tuning?

In detail, The theoretical candidate windows used in the previous binomial test were derived from the logit dynamics model with specific parameter values ($\lambda=50$, $\Delta t=0.02$). One might suspect that the predictive success could be an artifact of the particular parameterization rather than a genuine feature of the underlying dynamics.

To address this concern, we employ    fully parameter-free and mostly applied model, \textit{replicator dynamics} as an independent benchmark. 

\paragraph{Sample space} 
We generate simulated time series of 1,000 rounds from the replicator dynamics equations, using uniform random initial conditions. From 1000 times repeated simulation trajectories, we obtain the average, then identify the theoretical pulse candidates following exactly the same definition as in the main text (see Appendix~A.3.1). 

\begin{figure}[!ht]
    \centering
    \includegraphics[width=0.5\linewidth]{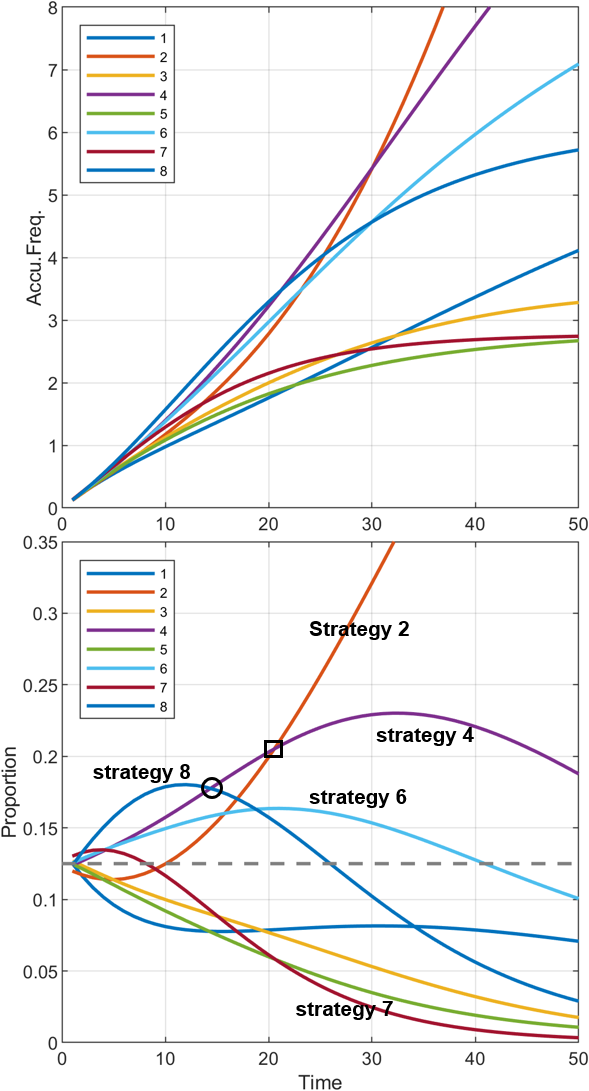}
    \caption{Theoretical pulse candidates predicted by the replicator dynamics for treatment A. The shaded regions indicate the time intervals $t: [0,40]$ where pulse ($X_8$ or $X_4$ is exceeds $X_2$ or $X_6$) is expected.}
    \label{fig:replicator_pulses}
\end{figure}

Table~\ref{tab:Replicator_dyn_Pulse_significant} is the list of all of the potential candidate pulse. Entries marked with an asterisk ($^*$) indicate pulses that exactly match the experimental pulse in Table~A4.

We evaluate the predictive power of the replicator-dynamics-based theory using the same exact binomial test as in Section~\ref{sec:comm_stat}. The sample specifications are summarized in Table~\ref{tab:replicator_binomial}.

\begin{table}[ht!]
\centering

\begin{tabular}{l c c c}
\toprule
\textbf{Parameter} & \textbf{Symbol} & \textbf{Value} & \textbf{Note}\\
\midrule
Total testing space & \(M\) & 6500  & Eq. \ref{eq:sample6500} \\
Theoretical candidate set size & \(K\) & 34 & Table \ref{tab:Replicator_dyn_Pulse_significant} \\
Random hit probability under the null & \(p_0 = K / M\) & \(0.005231\) \\
Total empirical detected signals & \(n\) & 9 & Table A4\\
Number of hits (within candidate set) & \(k\) & 5 &Table \ref{tab:Replicator_dyn_Pulse_significant} \\
\bottomrule
\end{tabular}
\caption{Exact binomial test parameters for replicator dynamics predictions}
\label{tab:replicator_binomial}
\end{table}

\paragraph{Null hypothesis and Exact binomial test}

Under the null hypothesis that the theory has no predictive power, the probability of observing $k$ or more hits out of $n$ signals follows a binomial distribution:
\[
p\text{-value} = \sum_{x=k}^{n} \binom{n}{x} p_0^x (1-p_0)^{n-x}.
\]

With $n=9$, $k=5$, and $p_0 = 34/6500 \approx 0.005231$, we obtain:
\[
p\text{-value} = \sum_{x=5}^{9} \binom{9}{x} (0.005231)^x (0.994769)^{9-x} \approx 4.8485 \times 10^{-10}.
\]

\begin{table}[ht!]
\centering

\begin{tabular}{l c}
\toprule
\textbf{Indicator} & \textbf{Value} \\
\midrule
Random hit probability \(p_0\) & 0.523\% \\
Observed hit rate & 55.6\% (5/9) \\
Exact \(p\)-value & \(\mathbf{4.8485 \times 10^{-10}}\) \\
Significance & \(p \ll 0.001\) (highly significant) \\
\bottomrule
\end{tabular}
\caption{Summary of exact binomial test results for replicator dynamics predictions}
\end{table}

\paragraph{Conclusion.} The replicator dynamics, a fully parameter-free model, independently predicts pulse signals that match the experimental data at a highly significant level ($p = 4.8485 \times 10^{-10}$). This result reinforces our earlier findings from the logit dynamics and addresses the concern that the predictive success might be an artifact of parameter tuning. 

\subsection{Exact binomial test on pulse (By session)} 

\paragraph{Question: }
is the classical-game-theoretic prediction that every dominating strategy is used with strictly higher probability than every dominated strategy empirically supported, or is it rejected by the observed prevalence of pulse signals?

\paragraph{Null Hypothesis \(H_0\)}
The classical prediction holds: for every session and any time interval, the minimum usage rate among the dominating  \(\{X_2, X_6\}\) is  larger than the maximum usage rate among the dominated \(\{X_4, X_8\}\):
\[
H_0: \min(X_2, X_6) > \max(X_4, X_8) \quad \text{for all sessions}.
\]
Here, $(X_4, X_8)$ are the expected pulse signal by the two dynamics (logit dynamics Table A5 in main text and replicator dynamics Table \ref{tab:Replicator_dyn_Pulse_significant} in SI)

{Alternative Hypothesis \(H_1\)}
The pulse phenomenon is present: there exists  noticeable  number of sessions in which a dominated strategy temporarily exceeds a dominating strategy. Equivalently, in a significant number of sessions,
\[
H_1: \min(X_2, X_6) < \max(X_4, X_8).
\]

\paragraph{Testing Logic}
The presence of a statistically significant number of sessions where \(\min(X_2, X_6) < \max(X_4, X_8)\) constitutes empirical evidence \emph{against} the classical theory and \emph{in favor} of the pulse phenomenon predicted by the game dynamics framework.

\paragraph{Sample construction}
The following analysis compares, for each session, the minimum usage count of strategies \(X_2\) and \(X_6\) against the maximum usage count of strategies \(X_8\) and \(X_4\). The per-session usage counts for all eight strategies are given in Table~\ref{tab:strategy_counts} (rows = sessions, columns = strategies \(X_1\) to \(X_8\)), counted frequency of $X_1, X_2,...,X_8$ between $T=1,2,...,40$ time interval.

For each session, we compute the pair \((\min(X_2, X_6),\ \max(X_8, X_4))\). These pairs are displayed alongside the corresponding rows  in Table~\ref{tab:strategy_counts}.

\begin{table}[ht!]
\centering

\begin{tabular}{c|cccccccc|cc}
\toprule
Session & \(X_1\) & \(X_2\) & \(X_3\) & \(X_4\) & \(X_5\) & \(X_6\) & \(X_7\) & \(X_8\) & \(\min\) & \(\max\) \\
&&&&&&&&&\((X_2,X_6)\)&\((X_8,X_4)\) \\
\midrule
1  & 2  & 3  & 11 & 6  & 7  & 2  & 8  & 1  & 2 & 6 \\
2  & 2  & 27 & 0  & 0  & 0  & 11 & 0  & 0  & 11 & 0 \\
3  & 4  & 4  & 4  & 5  & 4  & 6  & 1  & 12 & 4 & 12 \\
4  & 3  & 4  & 5  & 8  & 7  & 5  & 5  & 3  & 4 & 8 \\
5  & 4  & 2  & 4  & 6  & 5  & 8  & 7  & 4  & 2 & 6 \\
6  & 5  & 2  & 2  & 4  & 1  & 8  & 6  & 12 & 2 & 12 \\
7  & 3  & 5  & 9  & 4  & 12 & 2  & 3  & 2  & 2 & 4 \\
8  & 9  & 2  & 3  & 3  & 4  & 3  & 8  & 8  & 2 & 8 \\
9  & 3  & 5  & 3  & 3  & 3  & 5  & 2  & 16 & 5 & 16 \\
10 & 5  & 7  & 5  & 9  & 5  & 1  & 4  & 4  & 1 & 9 \\
11 & 1  & 6  & 11 & 3  & 6  & 5  & 2  & 6  & 5 & 6 \\
12 & 4  & 1  & 4  & 6  & 1  & 4  & 5  & 15 & 1 & 15 \\
\bottomrule
\end{tabular}
\caption{Per-session strategy counts and derived paired measures}
\label{tab:strategy_counts}
\end{table}

\paragraph{Exact binomial test (sign test)}
Under the null hypothesis \(H_0\) that the two strategy groups are equally likely to dominate, the probability of observing a positive sign is \(p = 0.5\). This is a sign test, equivalent to an exact binomial test.

We conduct a one-sided test against the alternative hypothesis \(H_1: p > 0.5\), because dynamics, both logit or replicator, predicts that the maximum of \(X_8, X_4\) ought to be larger than the minimum of \(X_2, X_6\). Among the 12 sessions, 11 pairs yield a positive sign and 1 pair yields a negative sign. The one-sided \(p\)-value is:
\[
P(X \ge 11) = \binom{12}{11}(0.5)^{11}\times (1-0.5)^{1} + \binom{12}{12}(0.5)^{12} = \frac{13}{4096} \approx 0.00317.
\]

\paragraph{Paired \(t\)-test:} over the difference between the maximum of \(X_8, X_4\) and the minimum of \(X_2, X_6\) for session
\[
\begin{aligned} 
p &= 0.0183, \quad 95\%\ \text{CI} = [1.0425,\ 9.1242].
\end{aligned}
\]

\paragraph{Wilcoxon signed-rank test:}
\[
p = 0.0215.
\]

\begin{table}[ht!]
\centering

\begin{tabular}{l l c c}
\toprule
Variable&\textbf{Test} & \textbf{Statistic} & \textbf{p-value} \\
\midrule
$\min\{X_2,X_6\}$ vs $\min\{X_4,X_8\}$ &Exact sign test &   & \(0.0032\) \\
&Paired \(t\)-test & \(t = 2.77\) & \(0.0183\) \\
&Wilcoxon signed-rank & \(W = 71\) & \(0.0215\) \\
\bottomrule
\end{tabular}
\caption{Summary of paired statistical tests by session}
\label{tab:strategy_freq_test}
\end{table}

\paragraph{Conclusion} 
All the 3 tests reject the null hypothesis. The results indicate that the maximum of strategies \(X_8\) and \(X_4\) is significantly larger than the minimum of strategies \(X_2\) and \(X_6\) (\(p \leq 0.0032\) in the sign tests). This statistical evidence, in the experimental sessions level, consists with the dynamics predictions on   pulse signal.

\subsection{Experimental pulse intensity vs. experimental crossover timing (E–E)}

\paragraph{Question:} Are stronger pulses associated with later crossover points within the experimental data?

\paragraph{Sample construction:} The 9 statistically significant pulses strategy from Table~A4 corresponds to their   crossover points from Table~6. According to Table~A4 (pulse intensity $\psi^E$) and Table~A6 (crossover timing $\tau^E$), the performance of the three strategies in treatment A is summarised as follows:

\begin{table}[ht!]
\centering

\begin{tabular}{c|c|c|c|c|c}
\toprule
Treatment & Strategy & Rank by & Total intensity   &   Crossover & Crossover \\
(Tab.~A4)   & (Tab.~A4)   &   (observ.) &  $\sum \psi^E$ (Tab.~A4) &   $\tau^E$ (Tab.~A6)  &timing rank \\
\midrule
A & $X_8$ & 1~(3) & $17\!+\!19\!+\! 15 = 51$ & 186 ($X_8$ vs $X_6$) & 1 \\
A & $X_4$ & 2~(2) & $13 + 15 = 28$ & 169 ($X_4$ vs $X_6$) & 2\\
A & $X_1$ & 3~(1) & 10 & 106 ($X_1$ vs $X_6$) & 4 \\
\bottomrule
\end{tabular}
\caption{Ranking of strategies by pulse intensity and crossover timing}
\label{tab:pulse_cross_ranking}
\end{table}

There are 12 candidates in total (6 dominated $\times$ 2 dominatiing) in treatment A. The crossover timing (see Table A6) predicts that 4 specific items, $X_4, X_4,X_5,X_1$.  will occupy the 1st, 2nd,  3rd and 4th positions, respectively. 

In the actual experiment, $X_8$ is indeed the 1st, $X_4$ is the 2nd, and $X_1$ appears within the top four positions (i.e., either 3rd or 4th). We adopt a relaxed criterion: we do not require $X_1$ to be exactly 3rd; it suffices that the top two positions are perfectly predicted and that $X_1$ falls in the top four.

\paragraph{Null hypothesis.} 
Under the null hypothesis, the ordering of the top four positions is uniformly random among all possible permutations of the 12 items.

 \paragraph{Method: Exact permutation test for ranking accuracy}

{Total number of possible orderings for the top four positions:
\[
12 \times 11 \times 10 \times 9 = 11880.
\]

\textbf{Number of favourable orderings:}
To satisfy the relaxed condition:
\begin{itemize}
    \item The 1st position is fixed to $A$: 1 way.
    \item The 2nd position is fixed to $B$: 1 way.
    \item The 3rd and 4th positions must contain $X_1$ in either the 3rd or 4th slot (2 choices), and the other slot is filled by any of the remaining 9 items (excluding $X_8,X_4,X_1$): 9 choices.
\end{itemize}
Thus,
\[
\text{Favourable orderings} = 1 \times 1 \times 2 \times 9 = 18.
\]

\textbf{Probability under the null hypothesis:}
\[
p = \frac{\text{Favourable orderings}}{\text{Total number of possible
ordering}}  \approx 0.001515.
\]

\paragraph{Conclusion:} 
The probability of observing the relaxed event (top two perfectly correct and the third predicted item within the top four) under random ordering is only about 0.0015. We therefore reject the null hypothesis of random ranking between crossover measurement and pulse measurement. 

The results provide  an example showing the \textbf{internal consistency} among various experimental measurements.

\subsection{Experimental vs. theoretical cycle intensity
(E–T)}
\subsubsection{Correlation between experimental vs. theoretical cycle intensity} 

\paragraph{Question:} For the eigencycles that are successfully predicted by the theory (i.e., those that appear in both the experimental and theoretical top-12 lists), does the strength ordering match between theory and experiment? In other words, are the relative intensities of the identified cycles consistent with theoretical predictions?

\paragraph{Sample construction:} The sample consists of the matched eigencycle indices across treatments. For each treatment, we extract the \(x\) values that appear in both the experimental top-12 
Table A3 in main text, 
left panel) and the theoretical top-12 
Table A3 in main text, 
right panel). For each matched \(x\), we record its experimental strength \(\theta^E\) and its theoretical strength \(\theta^T\). The matched pairs are summarised in Table~\ref{tab:eic_matched_strength}.

\begin{table}[ht!]
\centering

\begin{tabular}{c|c|c|c}
\toprule
Treatment & Matched $x$ & Experimental $\theta^E$ & Theoretical $\theta^T$ \\
\midrule
\multirow{8}{*}{A} 
& 25  & -1.000 & -1.000 \\
& 23  &  0.830 &  0.995 \\
& 71  &  0.700 &  0.928 \\
& 69  & -0.660 & -0.923 \\
& 59  & -0.110 & -0.068 \\
& 9   & -0.104 & -0.021 \\
& 11  &  0.104 &  0.021 \\
& 61  &  0.081 &  0.068 \\
\midrule
\multirow{5}{*}{B} 
& 11  & -0.269 & -0.046 \\
& 101 & -0.148 & -0.024 \\
& 59  & -0.143 & -0.030 \\
& 23  &  0.126 &  0.084 \\
& 29  & -0.081 & -0.037 \\
\midrule
\multirow{6}{*}{C} 
& 23  &  0.272 &  0.150 \\
& 25  & -0.153 & -0.107 \\
& 101 & -0.094 & -0.060 \\
& 11  &  0.086 & -0.168 \\
& 61  &  0.056 &  0.256 \\
& 59  & -0.045 & -0.117 \\
\bottomrule
\end{tabular}
\caption{Summary of matched eigencycle indices and their strength values across treatments, source: Table A3 in main text}
\label{tab:eic_matched_strength}
\end{table}

\paragraph{Null hypothesis:} Under the null hypothesis that the theoretical strength ordering has no predictive power, the ranks of \(\theta^E\) and \(\theta^T\) are independent of each other, i.e., the Spearman rank correlation coefficient \(\rho_s = 0\).

\paragraph{Method:} For each treatment, we compute the Spearman rank correlation coefficient between the experimental strengths \(\theta^E\) and the theoretical strengths \(\theta^T\) among the matched \(x\) values. The \(p\)-value is obtained from the test of whether \(\rho_s\) is significantly different from zero. The results are reported in Table~\ref{tab:eic_hit_summary}.

\begin{table}[ht!]
\centering

\begin{tabular}{c|c|c|c|c}
\toprule
Treatment & Hits $k$ & Hit rate & Spearman $\rho_s$ & $p$-value \\
\midrule
A & 8/12 & 66.7\% & 0.976 & 0.0004 \\
B & 5/12 & 41.7\% & 0.600 & 0.3500 \\
C & 6/12 & 50.0\% & 0.143 & 0.8028 \\
\bottomrule
\end{tabular}
\caption{Summary of eigencycle prediction hits and strength correlation across treatments}
\label{tab:eic_hit_summary}
\end{table}

\paragraph{Results:} The Spearman correlation tests yield the following results. In Treatment A, we observe a near-perfect rank correlation between experimental and theoretical strengths (\(\rho_s = 0.976\), \(p = 0.0004\)), indicating that the model accurately predicts not only which eigencycles appear but also their relative intensities. In Treatment B, the correlation is moderate but not statistically significant (\(\rho_s = 0.600\), \(p = 0.3500\)). In Treatment C, the correlation is weak and not significant (\(\rho_s = 0.143\), \(p = 0.8028\)).

\subsubsection{Enrichment of experimental eigencycles in theoretical candidate set}
\paragraph{Sample space} 
For each of the three treatments, the total number of possible eigencycle indices is 120, which comes from the definition of eigencycle (see SI). The theoretical model preselects the 12 indices with the largest predicted strengths as the target set (see main text Table A5 $\theta^T$ column), while the experimental data select the 12 eigencycles with the largest observed strengths (see main text Table A5 $\theta^E$ column). The numbers of hits (i.e., experimental indices that fall within the theoretical target set) are 8, 5, and 6 for treatments A, B, and C, respectively, see Table \ref{tab:eic_matched_strength}.

\paragraph{Null hypothesis}
For each treatment, the theoretical top-\(K\) indices and the experimental top-\(n\) indices are independently and uniformly drawn from the \(N\) possible indices. Under this null, the number of hits \(k\) follows a hypergeometric distribution.

\paragraph{Method}
We performed a one-sided hypergeometric exact test for each treatment, with parameters \(N=120\), \(K=12\), \(n=12\), and observed hits \(k\) as reported. The \(p\)-value is \(P(X \ge k)\). A pooled test combines the three treatments (\(N=360\), \(K=36\), \(n=36\), \(k=19\)). All \(p\)-values were computed using MATLAB's \texttt{hygecdf}. To adjust for multiple comparisons, a Bonferroni correction was applied to the three individual tests (\(\alpha = 0.05/3 \approx 0.0167\)).

\paragraph{Result}
Similarly, hpergeometric exact test results for enrichment of theoretical eigencycle predictions are reported in Table~\ref{tab:hypergeo_eigencycle}.

\paragraph{Conclusion:} The strength correlation analysis reveals a clear distinction across treatments. Treatment A provides strong evidence that the theoretical model captures both the identity and the relative ordering of the strongest eigencycles. Treatments B and C show weaker or non-significant strength correlations. It is not surprising, as both statics and dynamics expect, the cycle signals are weak and noise will outperform.

The hypergeometric tests show that the theoretical eigencycle predictions are significantly enriched across all treatments (A: \(p=2.56\times10^{-7}\), B: \(p=2.27\times10^{-3}\), C: \(p=1.76\times10^{-4}\)). Referring to the Bonferroni-corrected threshold (\(\alpha=0.05/120\approx 4.1667\times 10^{-4}\)), treatment A and C remain significant. 

The pooled result is highly significant (\(p=1.69\times10^{-12}\)) which remains below the Bonferroni-corrected threshold (\(\alpha=0.05/360\approx 1.3889\times 10^{-4}\)), confirming that the observed matches are unlikely to be due to chance.

 \begin{table}[ht!]
\centering

\begin{tabular}{c c c c c c c}
\toprule
\textbf{Level} & \(N\) & \(K\) & \(n\) & \(k\) & \textbf{\(p\)-value} & \textbf{Significance} \\
\midrule
Treatment A & 120 & 12 & 12 & 8 & \(2.56 \times 10^{-7}\) & \(p < 0.001\) \\
Treatment B & 120 & 12 & 12 & 5 & \(2.27 \times 10^{-3}\) & \(p < 0.01\) \\
Treatment C & 120 & 12 & 12 & 6 & \(1.76 \times 10^{-4}\) & \(p < 0.001\) \\
\midrule
Overall (pooled) & 360 & 36 & 36 & 19 & \(1.69 \times 10^{-12}\) & \(p < 0.001\) \\
\bottomrule
\end{tabular}
\medskip
\footnotesize
~\\
\textit{Notes}: \(N\) = total possible positions, referring to the definition of eigencycle; \(K\) = number of theoretically predicted top-12 candidates; \(n\) = number of experimental top-12 indices; 
\(k\) = number of hits (see Table \ref{tab:eic_matched_strength}). All \(p\)-values are from the hypergeometric test, computed as \(p = P(X \ge k \mid N, K, n)\). The overall test pools the three treatments together (\(N = 3 \times 120\), \(K = 3 \times 12\), \(n = 3 \times 12\), \(k = 8+5+6 = 19\)).
\caption{Hypergeometric exact test results for enrichment of theoretical eigencycle predictions}
\label{tab:hypergeo_eigencycle}
\end{table}

\clearpage

\bibliographystyle{unsrt}
\bibliography{references}

@article{huyck2001,
  author = {Raymond Battalio and Larry Samuelson and John Van~Huyck},
  title = {Optimization incentives and coordination failure in laboratory stag hunt games},
  journal = {Econometrica},
  volume = {69},
  number = {3},
  pages = {749--764},
  year = {2001}
}

@article{hans2021,
  author = {Volker Benndorf and Ismael Mart{\'\i}nez-Mart{\'\i}nez and Hans-Theo Normann},
  title = {Games with coupled populations: An experiment in continuous time},
  journal = {Journal of Economic Theory},
  volume = {195},
  pages = {105281},
  year = {2021}
}

@book{binmore2007game,
  author = {Ken Binmore},
  title = {Game theory: a very short introduction},
  volume = {173},
  publisher = {Oxford University Press},
  year = {2007}
}

@book{colin2003,
  author = {Colin F. Camerer},
  title = {Behavioral game theory: Experiments in strategic interaction},
  publisher = {Princeton University Press},
  year = {2003}
}

@article{dan2010,
  author = {Timothy N. Cason and Daniel Friedman and Ed Hopkins},
  title = {Testing the {TASP}: An experimental investigation of learning in games with unstable equilibria},
  journal = {Journal of Economic Theory},
  volume = {145},
  number = {6},
  pages = {2309--2331},
  year = {2010}
}

@article{dan2014,
  author = {Timothy N. Cason and Daniel Friedman and Ed Hopkins},
  title = {Cycles and instability in a rock--paper--scissors population game: a continuous time experiment},
  journal = {Review of Economic Studies},
  volume = {81},
  number = {1},
  pages = {112--136},
  year = {2014}
}

@article{dan2021,
  author = {Timothy N. Cason and Daniel Friedman and Ed Hopkins},
  title = {An experimental investigation of price dispersion and cycles},
  journal = {Journal of Political Economy},
  volume = {129},
  number = {3},
  pages = {789--841},
  year = {2021}
}

@article{dan1991,
  author = {Daniel Friedman},
  title = {Evolutionary games in economics},
  journal = {Econometrica: Journal of the Econometric Society},
  pages = {637--666},
  year = {1991}
}

@article{dan1998,
  author = {Daniel Friedman},
  title = {Evolutionary economics goes mainstream: A review of the theory of learning in games},
  journal = {Journal of Evolutionary Economics},
  volume = {8},
  number = {4},
  pages = {423--432},
  year = {1998}
}

@book{dan2016,
  author = {Daniel Friedman and Barry Sinervo},
  title = {Evolutionary games in natural, social, and virtual worlds},
  publisher = {Oxford University Press},
  year = {2016}
}

@misc{mit2014,
  author = {{Emerging Technology from the arXiv}},
  title = {Best of 2014: How to win at rock-paper-scissors --- {MIT} technology review},
  year = {2014},
  note = {[Online; accessed 07-January-2023]}
}

@book{fudenberg1998,
  author = {Drew Fudenberg and David K. Levine},
  title = {The theory of learning in games},
  volume = {2},
  publisher = {MIT Press},
  year = {1998}
}

@article{levine2016,
  author = {Drew Fudenberg and David K. Levine},
  title = {Whither game theory? towards a theory of learning in games},
  journal = {Journal of Economic Perspectives},
  volume = {30},
  number = {4},
  pages = {151--70},
  year = {2016}
}

@book{gintis2000,
  author = {Herbert Gintis},
  title = {Game theory evolving: A problem-centered introduction to modeling strategic behavior},
  publisher = {Princeton University Press},
  year = {2000}
}

@article{Hofbauer2011Survival,
  author = {Josef Hofbauer and William H. Sandholm},
  title = {Survival of dominated strategies under evolutionary dynamics},
  journal = {Theoretical Economics},
  volume = {6},
  number = {3},
  pages = {341--377},
  year = {2011}
}

@article{Hofbauer1996Evolutionary,
  author = {Josef Hofbauer and Jörgen W. Weibull},
  title = {Evolutionary selection against dominated strategies},
  journal = {JEL},
  volume = {71},
  number = {2},
  pages = {558--573},
  year = {1996}
}

@article{nowak2015,
  author = {Moshe Hoffman and Sigrid Suetens and Uri Gneezy and Martin A. Nowak},
  title = {An experimental investigation of evolutionary dynamics in the rock-paper-scissors game},
  journal = {Scientific Reports},
  volume = {5},
  number = {1},
  pages = {1--7},
  year = {2015}
}

@misc{kandori2021,
  author = {Michihiro Kandori},
  title = {Adjustment dynamics for human players},
  note = {Nobel Symposium on One Hundred Years of Game Theory, December 17--19, 2021},
  year = {2021}
}

@article{lax1960,
  author = {Melvin Lax},
  title = {Fluctuations from the nonequilibrium steady state},
  journal = {Reviews of Modern Physics},
  volume = {32},
  number = {1},
  pages = {25},
  year = {1960}
}

@article{mangasarian1964two,
  author = {Olvi L. Mangasarian and H. Stone},
  title = {Two-person nonzero-sum games and quadratic programming},
  journal = {Journal of Mathematical Analysis and Applications},
  volume = {9},
  number = {3},
  pages = {348--355},
  year = {1964}
}

@article{mertikopoulos2016learning,
  author = {Panayotis Mertikopoulos and William H. Sandholm},
  title = {Learning in games via reinforcement and regularization},
  journal = {Mathematics of Operations Research},
  volume = {41},
  number = {4},
  pages = {1297--1324},
  year = {2016}
}

@article{stratego2022,
  author = {Julien Perolat and others},
  title = {Mastering the game of stratego with model-free multiagent reinforcement learning},
  journal = {Science},
  volume = {378},
  number = {6623},
  pages = {990--996},
  year = {2022}
}

@book{plott2008,
  editor = {Charles R. Plott and Vernon L. Smith},
  title = {Handbook of experimental economics results},
  volume = {1},
  publisher = {Elsevier},
  year = {2008}
}

@article{rapoport2000mixed,
  author = {Amnon Rapoport and Wilfred Amaldoss},
  title = {Mixed strategies and iterative elimination of strongly dominated strategies: An experimental investigation of states of knowledge},
  journal = {Journal of Economic Behavior \& Organization},
  volume = {42},
  number = {4},
  pages = {483--521},
  year = {2000}
}

@article{samuelson2016,
  author = {Larry Samuelson},
  title = {Game theory in economics and beyond},
  journal = {Journal of Economic Perspectives},
  volume = {30},
  number = {4},
  pages = {107--30},
  year = {2016}
}

@book{sandholm2010,
  author = {William H. Sandholm},
  title = {Population games and evolutionary dynamics},
  publisher = {MIT Press},
  year = {2010}
}

@article{saran2016,
  author = {Rene Saran},
  title = {Bounded depths of rationality and implementation with complete information},
  journal = {Journal of Economic Theory},
  volume = {165},
  pages = {517--564},
  year = {2016}
}

@article{selten2008,
  author = {Reinhard Selten and Thorsten Chmura},
  title = {Stationary concepts for experimental 2x2-games},
  journal = {American Economic Review},
  volume = {98},
  number = {3},
  pages = {938--66},
  year = {2008}
}

@article{stephenson2019,
  author = {Daniel Stephenson},
  title = {Coordination and evolutionary dynamics: When are evolutionary models reliable?},
  journal = {Games and Economic Behavior},
  volume = {113},
  pages = {381--395},
  year = {2019}
}

@article{huyck1999,
  author = {John Van~Huyck and Frederick Rankin and Raymond Battalio},
  title = {What does it take to eliminate the use of a strategy strictly dominated by a mixture?},
  journal = {Experimental Economics},
  volume = {2},
  number = {2},
  pages = {129--150},
  year = {1999}
}

@article{wang2017,
  author = {Yijia Wang and Xiaojie Chen and Zhijian Wang},
  title = {Testability of evolutionary game dynamics based on experimental economics data},
  journal = {Physica A: Statistical Mechanics and its Applications},
  volume = {486},
  pages = {455--464},
  year = {2017}
}

@misc{zhijian2012,
  author = {Zhijian Wang and Bin Xu},
  title = {Evolutionary rotation in switching incentive zero-sum games},
  year = {2012},
  note = {arXiv:1203.2591}
}

@article{zhijian2014rps,
  author = {Zhijian Wang and Bin Xu and Hai-Jun Zhou},
  title = {Social cycling and conditional responses in the rock-paper-scissors game},
  journal = {Scientific Reports},
  volume = {4},
  pages = {5830},
  year = {2014}
}

@article{zhijian2022,
  author = {Zhijian Wang and Shujie Zhou and Qinmei Yao and Yijia Wang},
  title = {Dynamic structure in four-strategy game: Theory and experiment},
  journal = {Contributions to Game Theory and Management},
  volume = {15},
  pages = {365--385},
  year = {2022}
}

@misc{wiki:Gravitational_collapse,
  author = {{Wikipedia}},
  title = {Gravitational collapse},
  note = {Wikipedia, the free encyclopedia, [Online; accessed 07-January-2023]},
  year = {2023}
}

@misc{wiki:High-frequency_trading,
  author = {{Wikipedia}},
  title = {High frequency trading},
  note = {Wikipedia, the free encyclopedia, [Online; accessed 07-January-2023]},
  year = {2023}
}

@article{zhijian20142x2,
  author = {Bin Xu and Shuang Wang and Zhijian Wang},
  title = {Periodic frequencies of the cycles in 2 $\times$ 2 games: evidence from experimental economics},
  journal = {The European Physical Journal B},
  volume = {87},
  number = {2},
  pages = {46},
  year = {2014}
}

@article{zhijian2011,
  author = {Bin Xu and Zhijian Wang},
  title = {Evolutionary dynamical pattern of 'coyness and philandering': Evidence from experimental economics},
  journal = {Unifying Themes in Complex Systems},
  volume = {8},
  year = {2011}
}

@article{zhijian2013,
  author = {Bin Xu and Hai-Jun Zhou and Zhijian Wang},
  title = {Cycle frequency in standard rock--paper--scissors games: evidence from experimental economics},
  journal = {Physica A: Statistical Mechanics and its Applications},
  volume = {392},
  number = {20},
  pages = {4997--5005},
  year = {2013}
}

@misc{2021Qinmei,
  author = {Qinmei Yao},
  title = {Theoretical analysis and experiment of dynamic structure of high dimensional game},
  year = {2021},
  note = {\url{https://cdmd.cnki.com.cn/Article/CDMD-10335-1021626407.htm}, doi=10.27461/d.cnki.gzjdx.2021.000847}
}

@misc{zhijian2020,
  author = {Zhijian Wang and Qingmei Yao},
  title = {Human social cycling spectrum},
  year = {2020},
  note = {arXiv preprint arXiv:2012.03315}
}

@article{Walunj2026,
  author = {Tushar Shankar Walunj and Shiksha Singhal and Veeraruna Kavitha and Jayakrishnan Nair},
  title = {Equilibrium cycle: A "Dynamic" equilibrium},
  journal = {Games and Economic Behavior},
  volume = {157},
  pages = {286--297},
  year = {2026}
}

@misc{johnston2026,
  author = {Tom Johnston and Michael Savery and Alex Scott and Bassel Tarbush},
  title = {Game Connectivity and Adaptive Dynamics},
  year = {2026},
  note = {arXiv preprint arXiv:2309.10609}
}

@misc{wang2026eigen,
  author = {Zhijian Wang and Qingmei Yao and Lixia Shan and Yijia Wang and Chuchen Zhang},
  title = {Eigenmanifold in Game: Evidence from human continuous strategy game experiments},
  year = {2026},
  note = {arXiv preprint arXiv:2607.09782}
}

@article{Dan2026,
  author = {Gary Charness and Daniel Friedman and Weinan Gong and Lones Smith},
  title = {Cyclical Behavior in Large Location Games},
  journal = {Journal of Economic Behavior \& Organization},
  volume = {248},
  pages = {107609},
  year = {2026}
}

@article{Newton2018,
  author = {Jonathan Newton},
  title = {Evolutionary Game Theory: A Renaissance},
  journal = {Games},
  volume = {9},
  number = {2},
  pages = {31},
  year = {2018}
}

@article{ONeill1987,
  author = {B. O'Neill},
  title = {Nonmetric test of the minimax theory of two-person zerosum games},
  journal = {Proceedings of the National Academy of Sciences},
  year = {1987}
}

@article{Binmore2001Minimax,
  author = {Ken Binmore and Joe Swierzbinski and Chris Proulx},
  title = {Does Minimax work? an experimental study},
  journal = {The Economic Journal},
  year = {2001}
}

@article{Brown90,
  author = {James N. Brown and Robert W. Rosenthal},
  title = {Testing the Minimax Hypothesis: A Re-examination of O'Neill's Game Experiment},
  journal = {Econometrica},
  volume = {58},
  number = {5},
  pages = {1065--1081},
  year = {1990}
}

@article{erev2007,
  author = {Ido Erev and Alvin E. Roth and Robert L. Slonim and Greg Barron},
  title = {Learning and equilibrium as useful approximations: Accuracy of prediction on randomly selected constant sum games},
  journal = {Economic Theory},
  volume = {33},
  number = {1},
  pages = {29--51},
  year = {2007}
}

@article{zhijian2016,
  author = {Zhijian Wang and Yanran Zhou and Jaimie W. Lien and Jie Zheng and Bin Xu},
  title = {Extortion can outperform generosity in the iterated prisoner's dilemma},
  journal = {Nature Communications},
  volume = {7},
  number = {1},
  pages = {1--7},
  year = {2016}
}

@article{samuelson1992,
  author = {Larry Samuelson},
  title = {Dominated strategies and common knowledge},
  journal = {Games and Economic Behavior},
  volume = {4},
  number = {2},
  pages = {284--313},
  year = {1992}
}

@article{mangasarian1964,
  author = {Olvi L. Mangasarian and H. Stone},
  title = {Two-person nonzero-sum games and quadratic programming},
  journal = {Journal of Mathematical Analysis and Applications},
  volume = {9},
  number = {3},
  pages = {348--355},
  year = {1964}
}

@misc{bimat2009,
  author = {Bapi Chatterjee},
  title = {Bimatrix Game},
  year = {2009},
  note = {MATLAB Central File Exchange}
}

@article{dan2005,
  author = {Timothy N. Cason and Daniel Friedman and Florian Wagener},
  title = {The dynamics of price dispersion, or Edgeworth variations},
  journal = {Journal of Economic Dynamics and Control},
  volume = {29},
  number = {4},
  pages = {801--822},
  year = {2005}
}

\end{document}